\documentclass[aps,groupedaddress,a4paper,showpacs,twocolumn,prb,superscriptaddress]{revtex4-1}
\usepackage{graphicx}
\usepackage{amsmath}
\usepackage{amssymb}
\usepackage{verbatim}
\usepackage{color}
\usepackage{dsfont}
\usepackage{bm}
\usepackage{hyperref}
\usepackage{mathtools}
\usepackage{pdfpages}

\definecolor{violet}{rgb}{0.7,0,0.5}
\definecolor{newgreen}{rgb}{0,0.6,0.0}
\definecolor{grey}{rgb}{0.4,0.4,0.4}

\usepackage{amsthm}

 \theoremstyle{definition}

 \theoremstyle{remark}

\begin{document}

\title{Monte Carlo studies of the properties of the Majorana quantum error correction code: is self-correction possible during braiding?}
\author{Fabio L. Pedrocchi}
\address{JARA Institute for Quantum Information, RWTH Aachen University, D-52056 Aachen, Germany}
\author{N. E. Bonesteel}
\address{Department of Physics and National High Magnetic Field Laboratory, Florida State University, Tallahassee, Florida 32310, USA}
\author{David P. DiVincenzo}
\address{JARA Institute for Quantum Information, RWTH Aachen University, D-52056 Aachen, Germany}

\begin{abstract}
The Majorana code is an example of a stabilizer code where the quantum information is stored in a system supporting well-separated Majorana Bound States (MBSs).   We focus on one-dimensional realizations of the Majorana code, as well as networks of such structures, and investigate their lifetime when  coupled to a parity-preserving thermal environment. We apply the Davies prescription,  a standard method that describes the basic aspects of a thermal environment,  and derive a master equation in the Born-Markov limit. We first focus on a single wire with immobile MBSs  and perform error correction to annihilate thermal excitations. In the high-temperature limit,  we show both analytically and numerically that the lifetime of the Majorana qubit grows logarithmically with the size of the wire. We then study a trijunction with four MBSs  when braiding is executed. We study the occurrence of \emph{dangerous} error processes that prevent the lifetime of the Majorana code from growing with the size of the trijunction. The origin of the dangerous processes is the braiding itself, which separates pairs of excitations and renders the noise nonlocal; these processes arise from the basic constraints of moving MBSs in 1D structures.  We confirm our predictions with Monte Carlo simulations in the low-temperature regime, i.e. the regime of practical relevance. Our results put a restriction on the degree of self-correction of this particular 1D topological quantum computing architecture. 
\end{abstract}

\pacs{03.65.Yz, 05.30.Pr, 03.67.Pp, 03.67.Lx}

\maketitle

\section{Introduction}
Topological quantum computation (TQC) describes the general idea of storing and processing quantum information in topological states of matter.\cite{KitaevToric, Review2008} The most appealing aspects of TQC reside in the intrinsic protection of the ground-state subspace against local (static) perturbations; topological ground states are thus viewed as a good place to hide quantum information. Furthermore, quantum gates are executed by performing highly non-local operations that consist in the exchange (or braiding) of quasi-particles in the form of non-abelian anyons. While it is difficult for the environment to induce such exchanges, an external observer is able to do it by adiabatically controlling the parameters of the system. Also, the applied quantum gates depend only on the \emph{topology} of the exchange and are thus insensitive to local imperfections.

In the last decade,  it has appeared that Ising anyons are the non-abelian particles most likely to occur in physical systems in the laboratory, see Ref.~\onlinecite{Review2008} and references therein.  Although their braiding statistics is not rich enough to generate a universal set of gates, they allow the implementation of the Clifford group in a topologically protected fashion and are thus of strong interest for quantum computation.
In this context, the so-called Kitaev wire \cite{LiebSchulzMattis, Kitaev2001} has recently attracted tremendous attention. In fact, unpaired Majorana modes appear in this model and, when braided in a network of one-dimensional wires, they behave as Ising anyons. \cite{Alicea2011} Considerable theoretical \cite{SauPRL, AliceaPRB2010, LutchynPRL2010, OregPRL2010} and experimental \cite{MourikScience2012, DengNano, DasNat, RokhinsonNatPhys, FinckPRL, ChurchillPRB} efforts have been invested to investigate semiconducting hybrid structures that could realize the Kitaev wire.

Although Majorana qubits exhibit many favorable properties, more and more studies have focused on the fragility of such topological qubits. In particular, several sources of noise that limit the applicability of such setups have been reported. \cite{GoldsteinPRB, BudichPRB, BravyiKoenig, LossRainis, Schmidt, ZyuzinPRL, Hassler, KlinovajaLoss,Cheng2011,Karzig, Scheurer,Karzig2015,Karzig20152,Mazza2013,Campbell2015} 

In this paper we start from a microscopic model and study the lifetime of the Majorana code, see Refs.~\onlinecite{BravyiKoenig,TerhalNJP,short}, as well as Sec.~\ref{sec:ECEC} for a definition,  when coupled to a \emph{parity-preserving} thermal environment. We apply the Davies prescription to derive a Born-Markov master equation. We first focus on a single wire with immobile Majorana Bound States (MBSs)  and discuss how to perform error correction to counteract the effect of the environment. We demonstrate in the high-temperature limit, both analytically and with Monte Carlo methods, that the lifetime grows logarithmically with the system size. This result is not unexpected as similar behavior was observed by Bravyi and Koenig for a closed system with Hamiltonian perturbations.\cite{BravyiKoenig}  As a next step, we study a trijunction with moving MBSs. Our main result is the investigation of details of \emph{dangerous} error processes that prevent the lifetime of the system from increasing with the system size. The origin of dangerous errors is the braiding itself that renders a local error source highly non local by dragging excitations across the trijunction. In particular, we demonstrate that performing error correction at the end of the braid does not allow the dangerous errors to be cured. We confirm our predictions with a Monte Carlo simulation. Our work is an extension of Ref.~\onlinecite{short}. Here, we present additional physical results as well as the technical details leading to the main results of Ref.~\onlinecite{short}. 

In the context of a full quantum computing protocol, where several braids are executed, our results imply that error correction at the end of all the braids, i.e. purely passive,  is not enough. Our results show also that a more active scheme, in which error correction is executed at the end of each braid, is also too weak to solve the problem of dangerous errors. We are led to the view that error correction will only be successful if it is fully active, i.e., where several error correction steps are executed during each braid, to counteract the decoherence effects of dangerous errors. Therefore, our  work brings additional evidence that even in non-abelian topological codes active error correction, in the same sense as for ordinary quantum error correction codes, is necessary. \cite{WoottonLoss, BrellPRX, Hutter, WoottonHutter}

The paper is organized as follows. In Sec.~\ref{sec:SingWire} we present the main aspects of a single Kitaev wire that carries MBSs at the junction between topological and nontopological segments. In Sec.~\ref{sec:Box} we introduce a box representation of the wire that turns out to be useful to understand the phenomenology of the wire as well as the way we simulate it. In Sec.~\ref{sec:ECEC} we define the Majorana code, i.e. a stabilizer code that encodes a logical qubit in the ground-state subspace of the Kitaev wire. In Sec.~\ref{sec:EC} we define string operators that create, annihilate, and move excitations in the wire. The string operators give us a rigorous way to perform error correction. In Sec.~\ref{sec:BathWire} we study the coupling between the Kitaev wire and a bosonic bath. We follow the Davies prescription and derive a Markovian master equation in Sec.~\ref{sec:Davies}. In Sec.~\ref{sec:analytical} we focus on the lifetime of the single wire Majorana code at high temperatures. We derive an analytical formula for the lifetime in Sec.~\ref{sec:Analytical2} and confirm it with Monte Carlo simulations in Sec.~\ref{sec:MonteCarlo}. In Sec.~\ref{sec:Trij} we introduce the trijunction setup used to braid MBSs and in Sec.~\ref{sec:TrijEncoding} we show how the logical qubit is encoded in four well separated MBSs. In Sec.~\ref{sec:Unitary} we study in detail the unitary evolution arising when MBSs are moved. In particular, we focus on the behavior of excitations. In Sec.~\ref{sec:ad}, we present a rigorous definition of what adiabaticity means in our study. In Sec.~\ref{sec:TrijDavies} we show how the master equation for the time-independent Kitaev wire generalizes to the time-dependent trijunction setup in the adiabatic limit. In Sec.~\ref{sec:ErrorCorr} we present the algorithm we use to perform error correction in the trijunction. Section~\ref{sec:Dangerous} contains our main results; we identify dangerous error processes that prevent the lifetime of the trijunction logical qubit from increasing with system size when braiding is executed. Finally we confirm our predictions with Monte Carlo simulations in Sec.~\ref{sec:MonteCarloTrij}. The Appendices contain additional information and details about the derivations.

\section{Single wire}\label{sec:SingWire}
We review here the basic aspects of the physical model considered here and already exposed in our previous work, see Ref.~\onlinecite{short}. 

We start our study with a single wire of size $L$ supporting immobile MBSs. The wire  Hamiltonian is \cite{LiebSchulzMattis, Kitaev2001}
\begin{eqnarray}\label{eq:Hamiltonian}
H_{\text{W}}&=&-\sum_{j=1}^{L}\mu_{j}a_{j}^{\dagger}a_{j}-\sum_{j=1}^{L-1}t(a_{j}^{\dagger}a_{j+1}+a_{j+1}^{\dagger}a_{j})\nonumber\\
&&+\sum_{j=1}^{L-1}(\Delta a_{j}a_{j+1}+\Delta^{*}a_{j+1}^{\dagger}a_{j})\,,
\end{eqnarray}
where $a_{j}^{\dagger}$ and $a_{j}$ are fermionic creation and annihilation operators at site $j$.
The first term describes a site-dependent chemical potential $\mu_{j}\leqslant0$. The second and third terms describe respectively nearest-neighbor hopping with $t>0$ and superconducting pairing with $\Delta=\vert\Delta\vert e^{i\theta}$. 

To understand the physics of $H_{\text{W}}$ in simple terms, it is useful to go to a representation in terms of Majorana operators,  $a_{j}=\frac{e^{-i\theta/2}}{2}(\gamma_{2j-1}+i\gamma_{2j})$ with $\{\gamma_{i},\gamma_{j}\}=2\delta_{ij}$ and $\gamma_{i}^{\dagger}=\gamma_{i}$. For the case $t=\vert\Delta\vert$ and $\mu_{j}=0\,\forall j$, we obtain the simplified expression
\begin{equation}\label{eq:HStop}
H_{\text{W}}^{\text{top}}=-\vert\Delta\vert\sum_{j=1}^{L-1}i\gamma_{2j+1}\gamma_{2j}\,.
\end{equation}

The first Majorana mode $\gamma_{1}$ as well as the last Majorana mode $\gamma_{2L}$ are decoupled from the rest of the chain and $\left[H_{\text{W}}^{\text{top}},\gamma_{1}\right]=\left[H_{\text{W}}^{\text{top}},\gamma_{2L}\right]=0$. This allows one to define a zero-energy delocalized fermionic mode with annihilation operator
\begin{equation}
d_{0}=\frac{1}{2}(\gamma_{1}+i\gamma_{2L})\,.
\end{equation}
Using the eigenmode operators $d_{j}=\frac{1}{2}(\gamma_{2j}+i\gamma_{2j+1})$, the wire Hamiltonian takes the fully diagonal form
\begin{equation}\label{eq:HSdiag}
H_{\text{W}}^{\text{top}}=\sum_{j=0}^{L-1}\epsilon_{j}\,(2d_{j}^{\dagger}d_{j}-1)\,,
\end{equation}
where $\epsilon_{0}=0$ and $\epsilon_{j}=\vert\Delta\vert$ for $j=1,\ldots,L-1$.
As originally proposed by Kitaev, \cite{Kitaev2001} it is tempting to encode a qubit in the ground-state subspace of $H_{\text{W}}$. The reason is that local (static) perturbations lead to a ground-state splitting exponentially small in $L$. Therefore, the decoherence induced by such undesirable splitting can be exponentially suppressed by increasing a parameter that is easy to control, namely the size of the wire.

\begin{figure}[h!]
	\centering
		\includegraphics[width=0.47\textwidth]{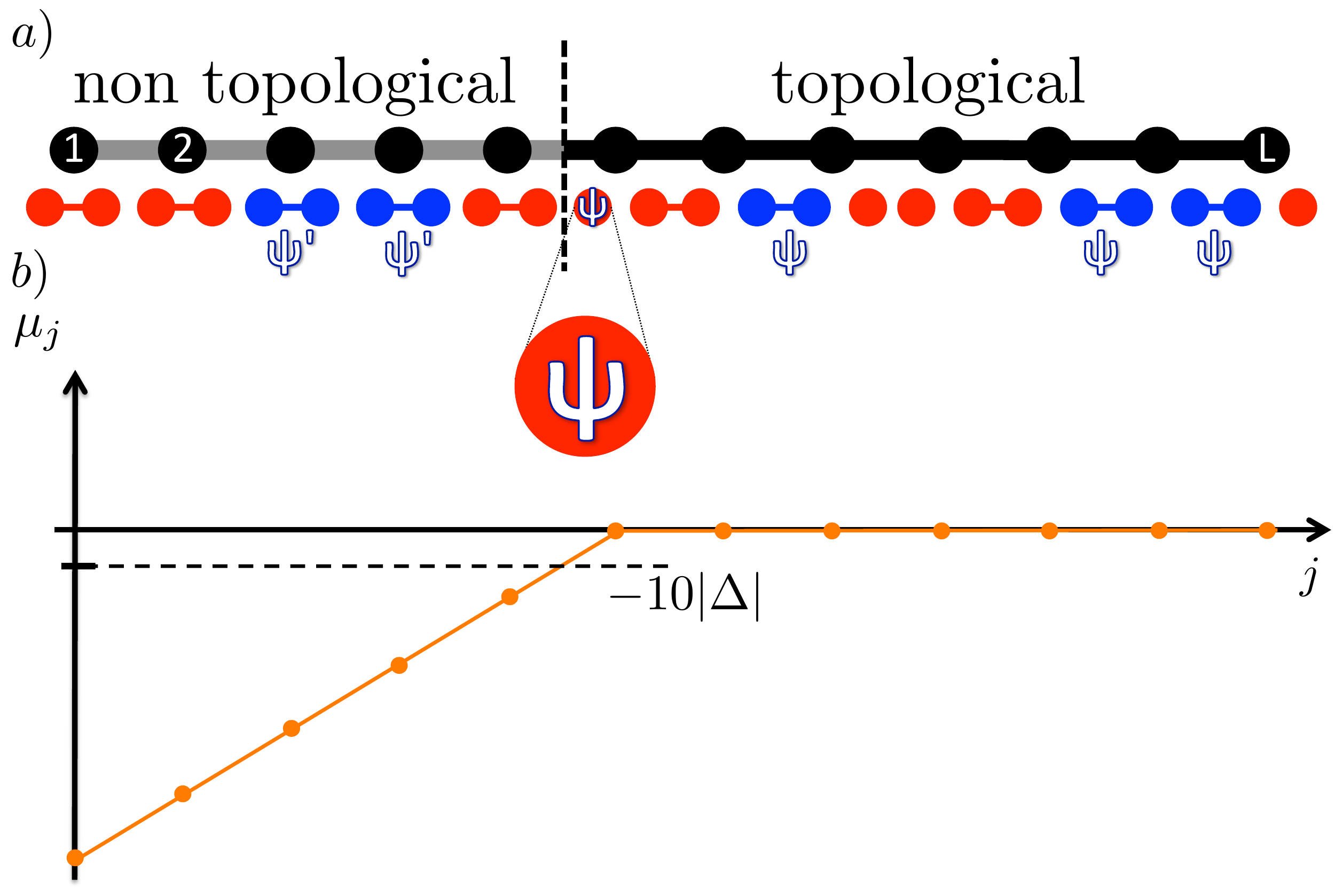}
	\caption{$a)$ Pictorial representation of the Kitaev wire with topological (black) and nontopological (gray) segments. The large black dots describe the fermionic sites, while the line in-between describe hopping and superconducting pairing. The smaller dots below represent the Majorana modes whose pairing are depicted by lines connecting the dots. Pairings in the topological segment are shifted as compared to pairings in the nontopological segment. A possible pattern of $\psi$- and $\psi^{\prime}$-excitations is shown. $b)$ Value of the chemical potentials $\mu_{j}$ corresponding to the situation in $a)$. The chemical potentials in the nontopological segment have a gradient in order to localize the $\psi^{\prime}$-excitations. }
	\label{fig:TopNonTop}
\end{figure}

Away from the limit $\mu_{j}=0$, MBSs  localized at the end of the chain persist as long as $\vert\mu_{j}\vert\leqslant 2t$; we call this the \emph{topological} phase. However, when $\mu_{j}\neq0$ the MBSs  are not localized anymore on a single site but have support in the bulk of the chain; the amplitude of the MBS  wave function decreases exponentially away from the end sites. For $\vert\mu\vert\geqslant 2t$ the localized modes disappear; this characterizes the \emph{nontopological} phase. Deep in the nontopological phase, with $\vert\mu_{j}\vert\gg\vert\Delta\vert=0$, the Majorana Hamiltonian approaches
\begin{equation}\label{eq:HWnontop}
H_{\text{W}}^{\text{nontop}}=-\frac{i}{2}\sum_{j=1}^{L}\mu_{j}\,\gamma_{2j-1}\gamma_{2j}\,.
\end{equation}
In Eq.~(\ref{eq:HWnontop}), the Majorana modes  are paired in a shifted way as compared to the topological case, see Eq.~(\ref{eq:HStop}). We present a pictorial representation of these two different pairings in Fig.~\ref{fig:TopNonTop}a.

Having in mind the Majorana pairing pattern in the topological and non topological segments, it is straightforward to see that MBSs  appear at the junctions between topological and nontopological segments of the wire, see Fig.~\ref{fig:TopNonTop}a. By varying the chemical potential, one can increase or decrease the size of the nontopological segments and thus move the position of the localized MBSs.\cite{Alicea2011} This idea will be used later in Sec.~\ref{sec:Trij} to braid the MBSs. 
\begin{figure}[h!]
	\centering
		\includegraphics[width=0.45\textwidth]{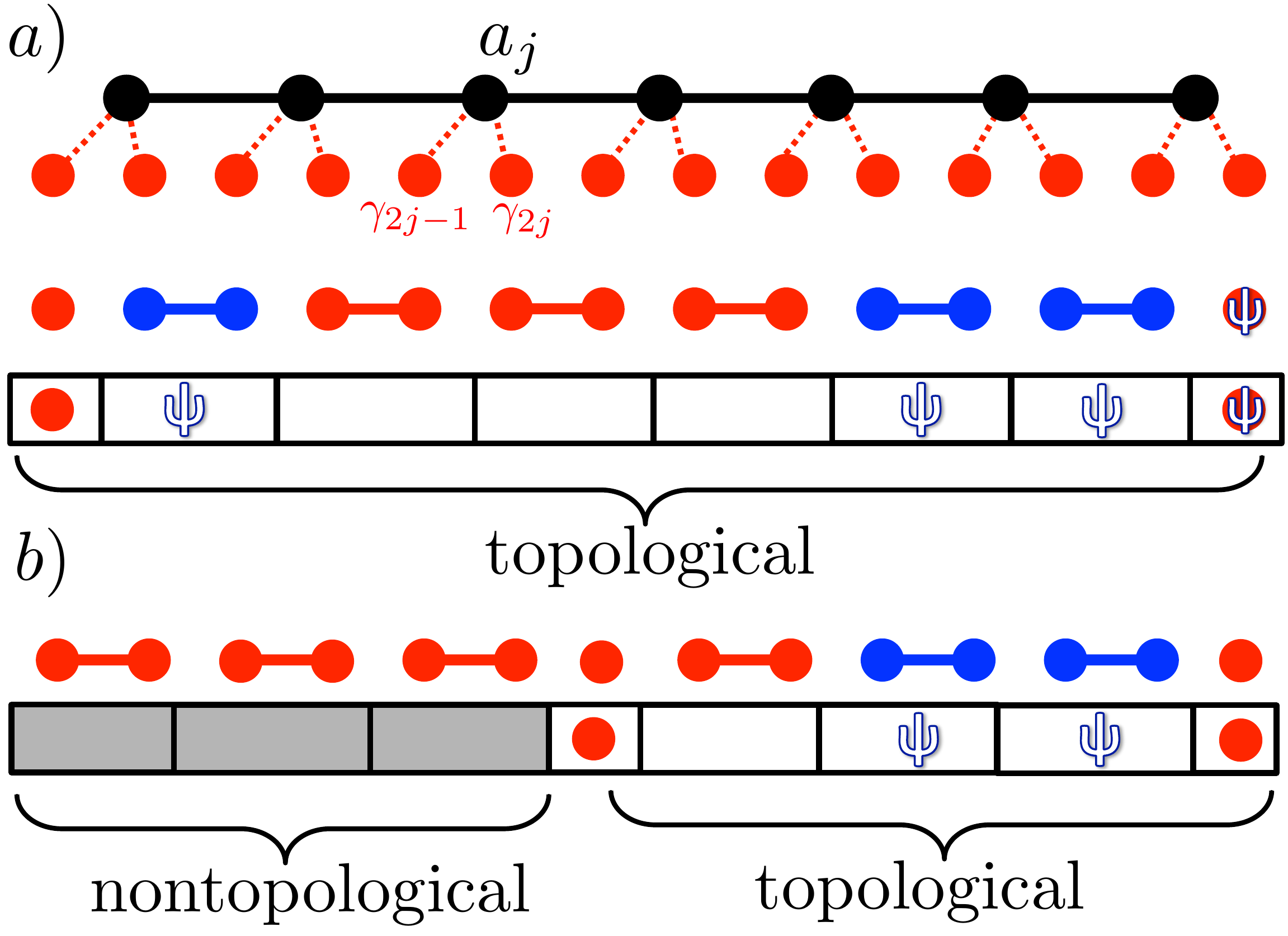}
	\caption{Box representation of the Kitaev wire. A wire of length $L$ is represented by $L+1$ boxes. A box represents either a full fermionic mode or a Majorana mode. Here and below, the boxes in the non topological sections are colored gray. The blue links carry an excitation and we draw a $\psi$ inside the corresponding box. A $\psi$ inside an MBS  describes a flip of the logical parity $S_{0}$. a) The wire is completely topological with two MBSs at the ends and four excitations. $b)$ The wire carries a topological and a nontopological segment, with two MBSs  at the boundary of the topological segment. Here the topological segment hosts two excitations.}
	\label{fig:Box}
\end{figure}

\subsection{Box representation of the wire}\label{sec:Box}
It is useful to use a box representation of the wire to understand the phenomenology of the model and the way we will simulate it. In Fig.~\ref{fig:Box} we present the details of our representation; if the wire contains $L$ sites, its box representation contains $L+1$ boxes. It is worth pointing out that a box is used to represent either a Majorana mode or a fermionic mode. While the size of the boxes vary in Fig.~\ref{fig:Box} for clarity, the size of each box has no meaning. In the rest of this work, all the boxes will have the same size. 

\subsection{Majorana Code}\label{sec:ECEC}
Following the approach of our previous work Ref.~\onlinecite{short},  it is convenient to take an information-theoretical approach to the encoding of logical qubits into the ground states of the Kitaev wire. \cite{BravyiKoenig, TerhalNJP} The ground-state subspace of $H_{\text{W}}^{\text{top}}$ forms a stabilizer code \cite{NielsenChuang,BarbaraReview}  with stabilizer operators $S_{j}=i\gamma_{2j+1}\gamma_{2j}$ i.e., the terms in the Hamiltonian Eq.~(\ref{eq:HStop}). As usual for a stabilizer quantum error correcting code, two logical qubit states $\vert\bar{0}\rangle$ and $\vert\bar{1}\rangle$ have the property
\begin{equation}
S_{j}\vert\bar{0}\rangle=\vert\bar{0}\rangle\,\quad\text{and}\quad S_{j}\vert\bar{1}\rangle=\vert\bar{1}\rangle\,.
\end{equation}
The Majorana code can then be interpreted as a one-dimensional version of Kitaev's  toric code.\cite{KitaevToric,KayPRL} Excitations above the ground states are localized and defined through the conditions $S_{j}=-1$. These excitations are denoted as quasi-particles $\psi$. We represent the code with boxes where the first and last boxes host the MBSs, while the other sites support either vacuum ($S_{j}=+1$) or a $\psi$ ($S_{j}=-1$), see Fig.~\ref{fig:Box}a. We represent a flip of the logical parity $S_{0}=i\gamma_{1}\gamma_{2L}\rightarrow -S_{0}$ by drawing a $\psi$ inside the left or right MBS. A $\psi$ inside an MBS   does not correspond to a real excitation since it does not cost any energy to be created, rather it signifies that the logical qubit has been flipped. A $\psi$ inside the left MBS  is proportional to a $X$ Pauli, a $\psi$ inside the right MBS  is proportional to a $Y$ Pauli, and a $\psi$ inside both MBSs is proportional to a $Z$ Pauli.

\subsection{String Operators, Fusion, and Error Correction}\label{sec:EC}
\begin{figure}[h!]
	\centering
		\includegraphics[width=0.47\textwidth]{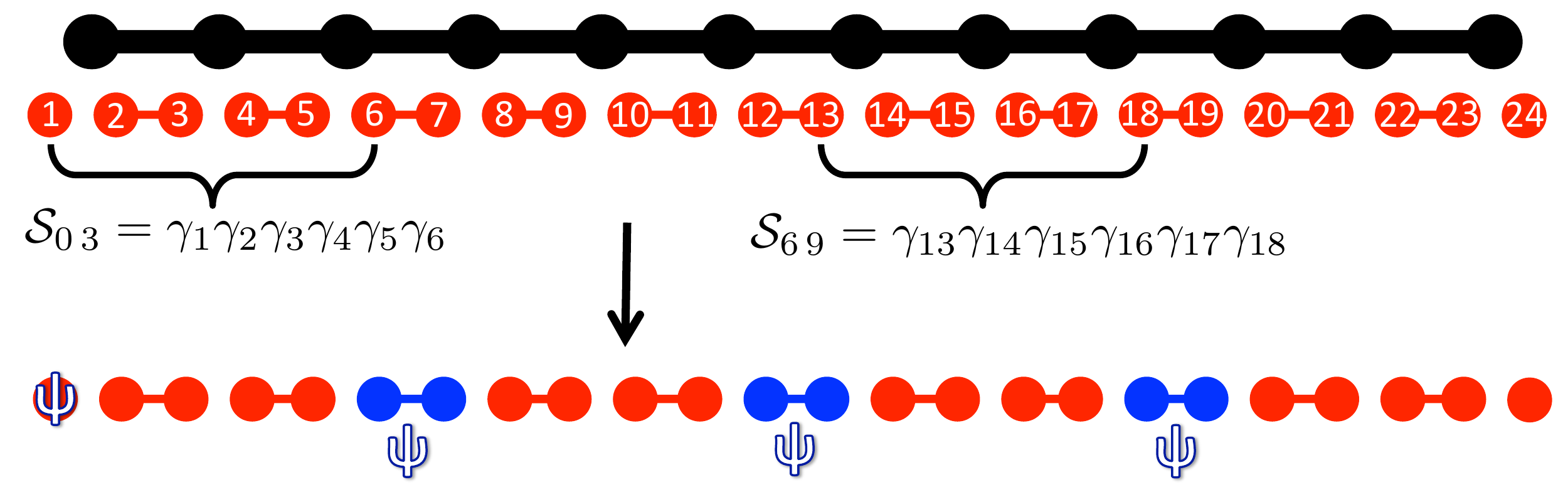}
	\caption{Excitations created by application of string operators $\mathcal{S}_{j\,k}$ defined in Eq.~(\ref{eq:String}).}
	\label{fig:Figure2_3}
\end{figure}

We solely consider parity-conserving perturbations and $\psi$ particles are thus always created in pairs. Pairs of excitations are generated by string operators; a string operator creating excitations $S_{j}=-1$ and $S_{k}=-1$ reads
\begin{equation}\label{eq:String}
\mathcal{S}_{jk}=\gamma_{2j+1}\gamma_{2j+2}\cdots\gamma_{2k}\,,
\end{equation}
see Fig.~\ref{fig:Figure2_3}.  We define the weight of a string operator $\mathcal{S}_{j\, k}$ as $\vert j-k\vert$. 
It is then clear that 
\begin{equation}
Z\propto \mathcal{S}_{0\, L}=\gamma_{1}\gamma_{2}\cdots\gamma_{2L}\,
\end{equation} 
maps the ground-state subspace into itself creating a $\psi$ inside the left MBS  and a $\psi$ inside the right MBS. Since $X$ and $Y$ logicals are generated by an odd number of $\psi$ particles, they cannot be implemented in a parity-preserving scenario with immobile MBSs. As we will see, when MBSs are braided the situation changes drastically.

String operators give us a way to fuse excitations and thus to perform error correction. Two $\psi$ particles $S_{a}=-1$ and $S_{b}=-1$ are fused by applying $\mathcal{S}_{ab}$. The effect is to bring back the system into its ground state by annihilating the quasi-particles. Similarly, a $\psi$ particle $S_{a}=-1$ can be fused to the left (right) MBS  by applying $\mathcal{S}_{0\,a}$ ($\mathcal{S}_{a\,L}$). 

In light of the above considerations, it is clear that the phenomenology of the Majorana wire  is the same as the Ising anyon model, \cite{BondersonThesis,PachosBook}
\begin{equation}\label{eq:fusion}
\psi\times\psi=1\,,\quad\sigma\times\psi=\psi\,,\quad\text{and}\quad\sigma\times\sigma=1+\psi\,,
\end{equation}
where $\sigma$ is the standard label for an Ising anyon and $1$ for vacuum. Here $\sigma$ particles are identified with the MBSs.  The second Eq.~(\ref{eq:fusion}) indicates that, as we have seen, a $\psi$ inside an MBS  is invisible to an external observer. Also it is clear that two MBSs give us a two-dimensional Hilbert space,  $\sigma\times\sigma=1+\psi$; as we have seen $1$ corresponds to an empty delocalized mode with $d_{0}^{\dagger}d_{0}=0$ and $\psi$ to a filled delocalized mode $d_{0}^{\dagger}d_{0}=1$.

In the following, we assume that the wire is in contact with a thermal bath that generates excitations. In order to conserve the information stored in the ground states of $H_{\text{W}}^{\text{top}}$, one needs to define a protocol for error correction based on the knowledge of the positions of $\psi$ in the bulk (recalling that $\psi$ inside an MBS  is invisible), the so-called error syndrome. If a pair of $\psi$'s is created in the bulk of the chain and not annihilated, then one particle can diffuse to the left end, while the second one diffuses to the right end. The operation performed on the logical qubit is then proportional to $Z$.

\begin{figure}[h!]
	\centering
		\includegraphics[width=0.5\textwidth]{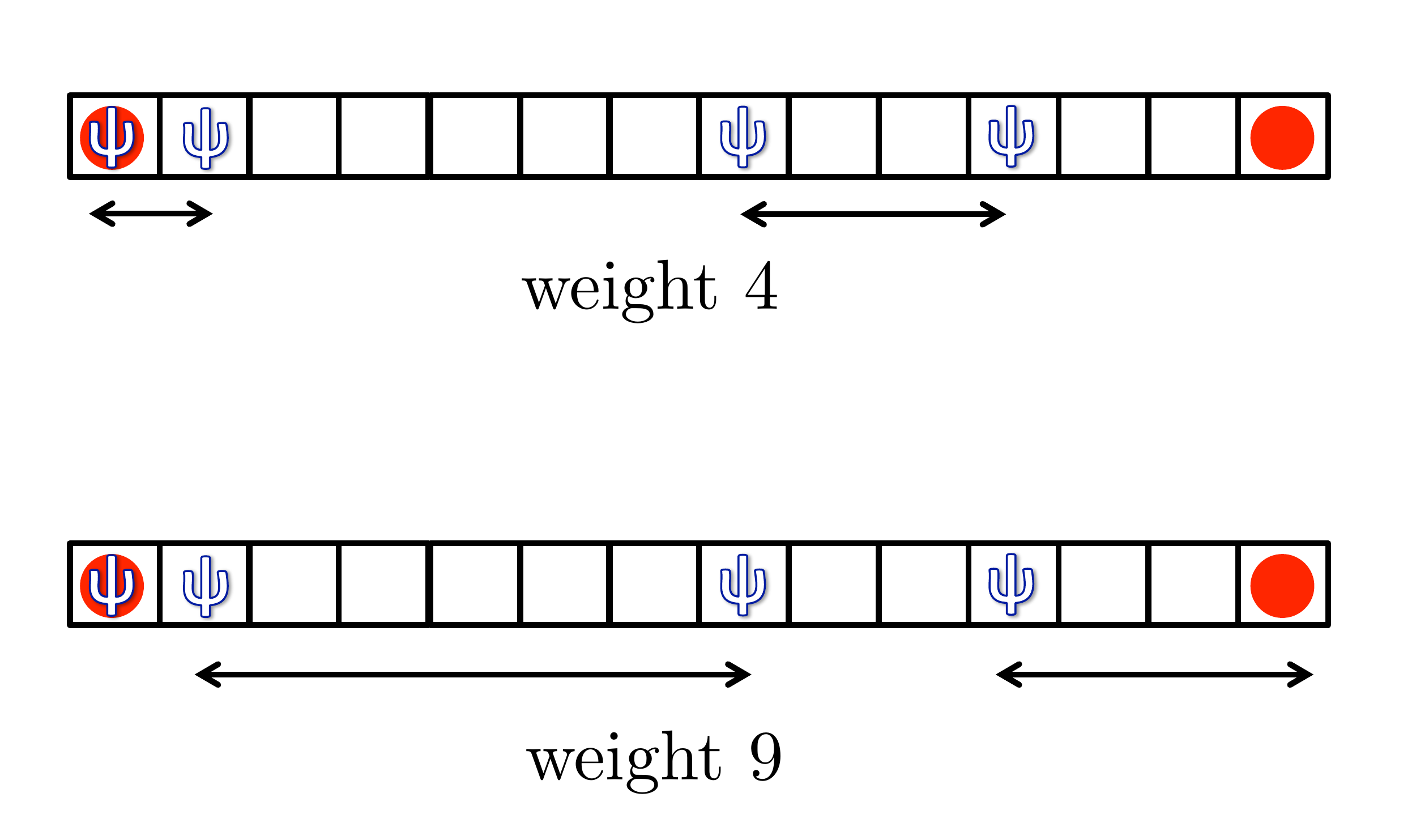}
	\caption{ Two examples of the Minimal Weight Perfect Matching (MWPM) algorithm. The total weight of the algorithm is defined as the sum of the weights of all the strings operators applied to remove the excitations in the bulk of the chain, see Eq.~(\ref{eq:String}). Top: The first $\psi$ is fused with the left MBS. The total weight is $4$. It leads to a successful recovery of the encoded logical qubit. Bottom: The first $\psi$ is fused to the second $\psi$. The total weight is $9$. This leads to a faulty procedure and the logical qubit is lost. }
	\label{fig:MWPM}
\end{figure}

The goal of error correction is to counteract the effect of the environment by finding a procedure that annihilates the excitations in a definite manner such that the stored quantum information is retrieved. Since it is reasonable to assume that nearby $\psi$'s originate from the same error event (as is the case at small times), we annihilate them following a Minimal Weight Perfect Matching (MWPM) algorithm for the single wire.  In one dimension there are only two possibilities to perform such pairings. \cite{BravyiKoenig}  One of them will lead to a successful logical qubit recovery, while the second one will introduce a $Z$ error, see Fig.~\ref{fig:MWPM}. Which of the two schemes is chosen depends on the total number of moves to be applied. We choose the scheme with minimal weight. Note that we will eventually use a different algorithm when we study the trijunction, see Sec.~\ref{sec:ErrorCorr}. 
\subsection{Coupling to thermal bath}\label{sec:BathWire}
The total Hamiltonian for the wire and the thermal bath is the one considered in Ref.~\onlinecite{short}, 
\begin{equation}\label{eq:Hamiltonian}
H=H_{S}+H_{B}+H_{SB}\,.
\end{equation}
Here we choose the system Hamiltonian $H_{S}=H_{\text{W}}^{\text{top}}$, see Eq.~(\ref{eq:HStop}). This choice  ensures that all the errors originate purely from thermal fluctuations. Bravyi and Koenig have considered the opposite regime where errors are solely due to Hamiltonian imperfections with $\mu_{j}\neq0$ and $t\neq\vert\Delta\vert$. \cite{BravyiKoenig} In their scenario, they showed that the lifetime of the Majorana code increases logarithmically with $L$. As we will see, this is also true for our thermal-bath model at large temperatures.

The bath Hamiltonian is bosonic and take the generic form
\begin{equation}
H_{B}=\sum_{j}\mathfrak{B}_{j}\,,
\end{equation}
where $\mathfrak{B}_{j}$ are local bosonic operators associated with fermionic site $j$.

The last term $H_{SB}$ in Eq.~(\ref{eq:Hamiltonian}) stands for the bath-wire interaction that we assume to be \emph{parity conserving},
\begin{equation}\label{eq:SB}
H_{SB}=-\sum_{j}B_{j}\otimes (2a_{j}^{\dagger}a_{j}-1)=-i \sum_{j}B_{j}\otimes \gamma_{2j-1}\gamma_{2j} \,.
\end{equation}
This form of the coupling seems quite natural since it corresponds to quantum fluctuations of the chemical potential. Note that $\psi$ excitations are created in pairs by the bath since $\left[\gamma_{2j-1}\gamma_{2j}, S_{k}\right]=0$ for $k\neq j, j-1$ and 
\begin{equation}
\{\gamma_{2j-1}\gamma_{2j}, S_{j-1}\}=\{\gamma_{2j-1}\gamma_{2j},S_{j}\}=0\,,
\end{equation}
with $j=1,\ldots,L$ and $S_{L}=S_{0}=i\gamma_{1}\gamma_{2L}$  is the parity of the logical qubit.

\subsubsection{Davies Prescription}\label{sec:Davies}
Following the prescription of Davies, \cite{Davies1974, BreuerPetruccione} that has become standard in many quantum information problems,\cite{ChesiNJP, ChesiPRA, HaahPRL, HutterPRA} we derive  the master equation for the wire in the memoryless (Markov) limit,
\begin{equation}\label{eq:MasEq}
\dot{\rho}_{S}(t)=-i[H_{S},\rho_{S}(t)]+\mathcal{D}(\rho_{S}(t))\,.
\end{equation}
The first term describes unitary evolution while the second one the exchange of energy between the bath and the wire. The so-called dissipator is
\begin{eqnarray}\label{eq:MasterEq}
\mathcal{D}(\rho_{S}(t))&=&\sum_{i,j}\sum_{\omega}\gamma^{ij}(\omega)\left(A^{i}(\omega)\rho_{S}(t)(A^{j}(\omega))^{\dagger}\right.\nonumber\\
&&\hspace{1cm}\left.-\frac{1}{2}\{(A^{j}(\omega))^{\dagger}A^{i}(\omega),\rho_{S}(t)\}\right)\,,
\end{eqnarray}
where $\gamma^{ij}(\omega)=\int_{-\infty}^{\infty}ds\,e^{i\omega s}\,\langle B_{i}^{\dagger}(s)B_{j}(0)\rangle$ are the bath spectral functions. Here, $\langle\cdots\rangle=\text{Tr}\left(\cdots e^{-\beta H_{B}}\right)$ is the thermal expectation value at inverse temperature $\beta$. 

In Appendix ~\ref{app:Davies}, we derive explicit expressions for the \emph{jump operators} $A^{i}(\omega)$. Importantly, they are local and satisfy detailed balance. The Davies prescription ensures that the steady state of Eq.~(\ref{eq:MasEq}) is the Gibbs state $\rho_{\text{Gibbs}}=\exp\left(e^{-\beta H_{\text{W}}^{\text{top}}}\right)/\text{Tr}\left(e^{-\beta H_{\text{W}}^{\text{top}}}\right)$. 

\begin{figure}[h!]
	\centering
		\includegraphics[width=0.45\textwidth]{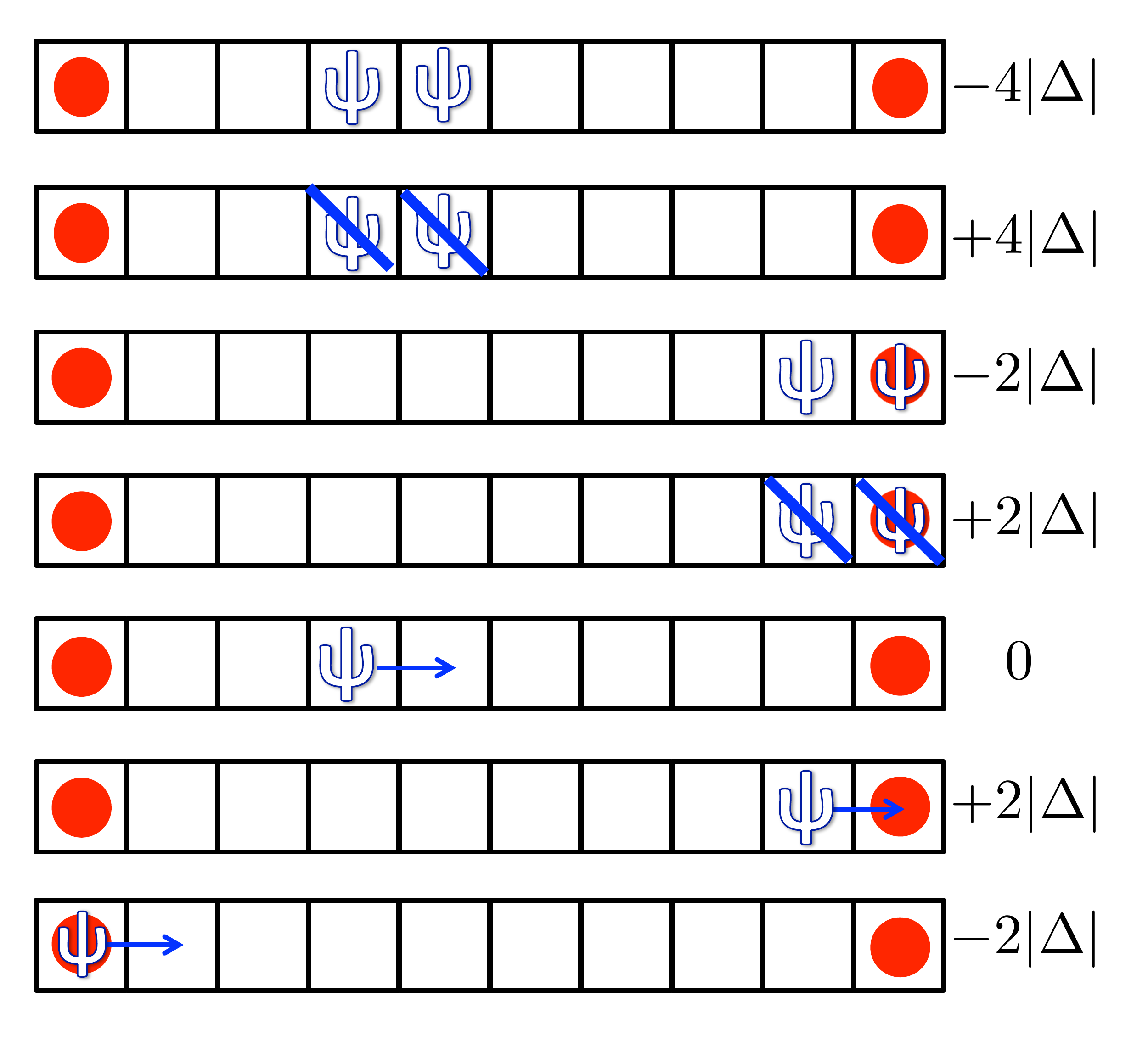}
	\caption{Pictorial box representation of all the possible error processes generated by the environment. Here we consider the simple time-independent problem where the MBSs  are immobile at the end of the wire. On the right is shown the energy costs of the processes. A positive energy means that energy is transferred from the wire to the bath.}
	\label{fig:Processes_Time_Ind}
\end{figure}

The jump operators $A_{i}(\omega)$ cause transitions between eigenstates of $H_{\text{S}}$, with energy difference $\omega$. We distinguish between the following categories of transitions, see Fig.~\ref{fig:Processes_Time_Ind}:
\begin{itemize}
\item Pair creation (annihilation) of $\psi$ in the bulk, with $\omega=-4\vert\Delta\vert$ ($\omega=+4\vert\Delta\vert$).
\item Pair creation (annihilation) of $\psi$ at the boundary, with $\omega=-2\vert\Delta\vert$ ($\omega=+2\vert\Delta\vert$). More precisely, one $\psi$ is created (annihilated) and the eigenvalue of $S_{0}=i\gamma_{1}\gamma_{2L}$ changes sign.
\item Hopping of a $\psi$ to a nearest-neighbor site inside the bulk, with $\omega=0$.
\item Hopping of a $\psi$ into a nearest-neighbor MBS, with $\omega=+2\vert\Delta\vert$.
\item Hopping of a $\psi$ out from an MBS  to a nearest-neighbor site of the bulk, with $\omega=-2\vert\Delta\vert$.
\end{itemize}
Here we use the convention that a negative sign of $\omega$ describes an energy transfer from the bath to the wire.

The time evolution of the diagonal elements of $\rho_{S}(\tau)$ decouple from the off-diagonal elements, see Appendix~\ref{app:Davies}, and the Pauli master equation for the population $P(n,\tau)$ in an eigenstate $\vert n\rangle$ of $H_{S}$ satisfies
\begin{equation}\label{eq:Populations}
\frac{dP(n,\tau)}{d\tau}=\sum_{m}\left[W(n\vert m)P(m,\tau)-W(m\vert n)P(n,\tau)\right]\,,
\end{equation}
with transition rates
\begin{equation}\label{eq:W}
W(n|m)=\gamma(\omega_{mn})\vert\langle m\vert A^{i_{mn}}(\omega_{mn})\vert n\rangle\vert^{2}\,.
\end{equation}
Here $\omega_{mn}$ is the energy difference between the eigenstates $\vert m\rangle$ and $\vert n\rangle$. Note that $\vert m\rangle$ and $\vert n\rangle$ can be degenerate, with $\omega_{mn}=0$.  We have removed the superscripts on the spectral function $\gamma(\omega)$ as it does not depend on the position; we assume that the sites are coupled to identical and independent baths.

In this work we consider an Ohmic bath where the rates $\gamma(\omega)$ are
\begin{equation}\label{eq:spectral}
\gamma(\omega)=\kappa\left\vert\frac{\omega}{1-\exp(-\beta\omega)}\right\vert\,,
\end{equation}
with coupling constant $\kappa$ and inverse temperature $\beta$. 

\subsection{Lifetime of Majorana Code: Infinite Temperature}\label{sec:analytical}
We focus here on the infinite-temperature limit, where we derive transparent analytical results for the lifetime of the Majorana code. As mentioned in Ref.~\onlinecite{short},   the Majorana code represents a useful quantum memory with a lifetime that grows with the wire's size $L$. A similar scaling behavior was discovered by Bravyi and Koenig in Ref.~\onlinecite{BravyiKoenig} and Kay in Ref.~\onlinecite{KayPRL}. However, these references considered unitary evolution, while we focus here on dissipative dynamics.  Unfortunately, the scaling is logarithmic and thus very modest.  Here we present an analytical proof of this result.

\subsubsection{Analytic derivation}\label{sec:Analytical2}
It is convenient to map the problem to spins via a Jordan-Wigner transformation
\begin{equation}\label{eq:w}
a_{j}=\left(\prod_{k=1}^{j-1}S_{k}^{z}\right)S_{j}^{+}\,\quad\text{and}\quad S_{j}^{z}=2a_{j}^{\dagger}a_{j}-1\,.
\end{equation} 
In spin language $H_{\text{W}}^{\text{top}}$ takes the simple form
\begin{equation}
H_{\text{W}}^{\text{top}}=-4\vert\Delta\vert\sum_{j=1}^{L-1}S_{j}^{x}S_{j+1}^{x}\,.
\end{equation}
In spin language the logical states $\vert\bar{0}\rangle$ and $\vert\bar{1}\rangle$ are recognized as the states with all spins pointing along $x$ and $-x$.
The logical $Z$ Pauli is then obtained by application of the parity operator
\begin{equation}
Z\propto\prod_{j=1}^{L}S_{j}^{z}\,.
\end{equation}

We model the error process taking place on the spin chain as follows. All spins point initially along $-x$. After a time step $\tau$, we assume that a number $n$ of spin flips has been applied on randomly chosen sites, where $n$ is taken from a Poisson distribution with mean $N_{0}=\tau\, W_{\text{tot}}$. Here $W_{\text{tot}}$ is the total rate of all allowed error processes. We assume that $W_{\text{tot}}$ is state independent; this corresponds to an infinite temperature scenario where $\gamma(0)=\gamma(\pm2\vert\Delta\vert)=\gamma(\pm 4\vert\Delta\vert)$, see Eq.~(\ref{eq:spectral}) in the limit $\beta\rightarrow0$. For simplicity we choose $\gamma(0)=\gamma(\pm2\vert\Delta\vert)=\gamma(\pm 4\vert\Delta\vert)=W_{\text{tot}}/L=:w_{\text{tot}}$.  Since events in a Poisson process are \emph{independent}, one can simplify the problem by just considering a single spin. We have
\begin{equation}
\langle S_{j}^{x}\rangle=-\sum_{k=0}^{\infty}(-1)^{k}\frac{(\tau\,w_{\text{tot}})^{k}}{k!}e^{-\tau w_{\text{tot}}}=-e^{-2\tau w_{\text{tot}}}\,.
\end{equation}
Similarly, the standard deviation is given by
\begin{equation}
\sigma_{S}=\sqrt{\langle (S_{j}^{x})^{2}\rangle-\langle S_{j}^{x}\rangle^2}=\sqrt{1-e^{-4 \tau w_{\text{tot}}}}\,.
\end{equation}
We thus have
\begin{equation}\label{eq:mean}
\mu_{\text{tot}}=\langle S^{x}_{\text{tot}}\rangle=\left\langle \sum_{i=1}^{L} S_{i}^{x}\right\rangle=L\langle S_{j}^{x}\rangle=-L e^{-2\tau w_{\text{tot}}}\,.
\end{equation}
From the central limit theorem we derive the probability distribution $g(s)$ of the random variable $s=S_{x}^{\text{tot}}$ with standard deviation $\sigma_{\text{tot}}=\sqrt{L}\sqrt{1-e^{-4\tau\vert\Delta\vert}}$, namely
\begin{equation}
g(s,\tau,L)=\frac{1}{\sigma_{\text{tot}}\sqrt{2\pi}}e^{-\frac{(s-\mu_{tot})^{2}}{2\sigma_{\text{tot}}^{2}}}\,.
\end{equation}

Error correction in the spin chain is performed straightforwardly: one flips either all the spins pointing along $-x$ or along $x$, such that all the spins point along the same direction after the error correcting step. The choice of which spins to flip is done according to the total number of spins that need to be addressed; we choose to flip the minimal number of spins.  In the original fermionic language, this is the Minimal Weight Perfect Matching algorithm of Sec.~\ref{sec:EC}, and the two choices correspond to the possibility to fuse the first $\psi$ with the left MBS  or not. As a direct consequence, the question whether error correction is successful after time $\tau$ maps to the question whether the majority of spins still points in the same direction as the initial state. If, for example, the initial state has all spins pointing along the $-x$-direction, then the condition for successful recovery after time $\tau$ is that (with high probability)
\begin{equation}
\sum_{j=1}^{L}S_{j}^{x}<0\,.
\end{equation}
The times $\tau$ for which error correction has a high probability of success are thus those for which
\begin{equation}
\langle S^{x}_{\text{tot}}\rangle+\sigma_{\text{tot}}\lesssim0
\end{equation}
and so
\begin{equation}
L\gtrsim e^{4\tau w_{\text{tot}}}-1\,.
\end{equation}
Stated in other terms, the Majorana code has \emph{no threshold}; the lifetime of the memory increases only logarithmically with $L$. A similar scaling behavior was discovered by Bravyi and Koenig in Ref.~\onlinecite{BravyiKoenig} and Kay for the surface code with one-dimensional Hamiltonian perturbations. \cite{KayPRL} Note that these references considered unitary evolution, while we focus here on dissipative dynamics.  The infinite temperature limit is a special case which allows
for straightforward analytic treatment.  However, because it is in a
sense a worst case scenario for error correction we expect the "no
threshold" result to hold for more general error models. 
\begin{figure}[h!]
	\centering
		\includegraphics[width=0.5\textwidth]{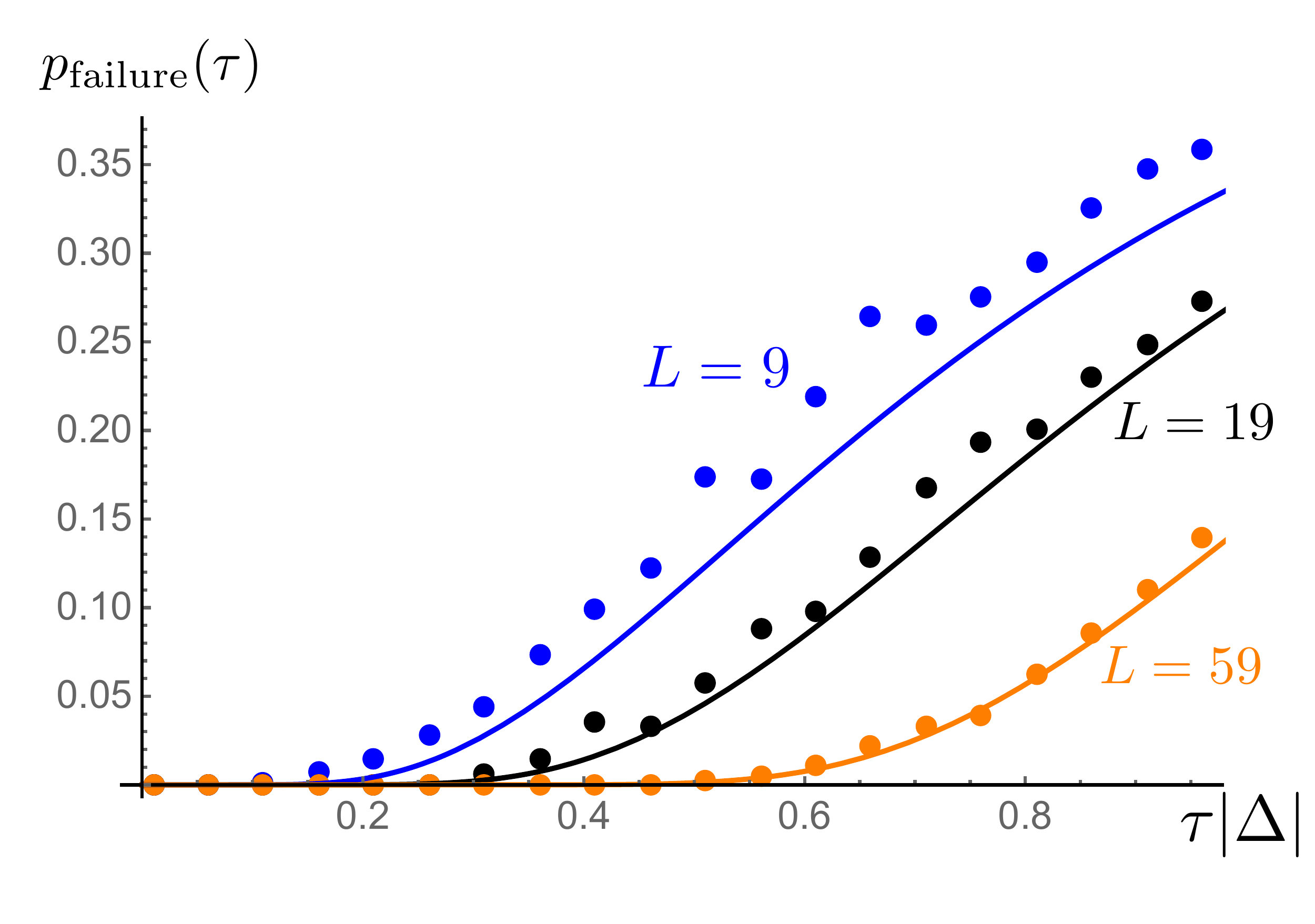}
	\caption{Probability of failure $p_{\text{failure}}(\tau)$ as function of time for Kitaev wires of lengths $L=9, 19, 59$. The results are obtained in the high-temperature regime with $W_{\text{tot}}=L\vert\Delta\vert$. The two MBSs  are immobile and we perform the MWPM algorithm to remove excitations. The solid lines represent the analytical result of Eq.~(\ref{eq:pfailure}) while the dots are obtained via Monte Carlo simulation. We find good agreement between analytics and numerics; the improvement becomes better for larger $L$, as expected.}
	\label{fig:MWPM_Linear}
\end{figure}

Finally, the probability that error correction fails after time $\tau$ is then simply given by
\begin{equation}\label{eq:pfailure}
p_{\text{failure}}(\tau,L)=\int_{0}^{\infty}g(s,\tau,L)ds=\frac{1}{2}\left(1+\text{Erf}\left(\frac{\mu_{\text{tot}}}{\sqrt{2}\sigma_{\text{tot}}}\right)\right)\,.
\end{equation}
In Sec.~\ref{sec:MonteCarlo}, we compare our analytical result with a Monte Carlo simulation and find very good agreement.

\subsubsection{Monte Carlo Simulation for the wire}\label{sec:MonteCarlo}
We have applied standard Monte Carlo methods to sample  Eq.~(\ref{eq:Populations}). We use the box representation of Fig.~\ref{fig:Box} to describe the simulation. 
An error caused by a system operator $\gamma_{2j-1}\gamma_{2j}$, see Eq.~(\ref{eq:SB}), is implemented in the simulation by adding a $\psi$ in boxes $2j-1$ and $2j$. An even number of $\psi$ in a box is identical to vacuum. Note that a $\psi$ and an MBS  can coincide in the same box. To be more precise, we implement the effect of the error operator $\gamma_{1}\gamma_{2}$ ($\gamma_{2L-1}\gamma_{2L}$) by adding a $\psi$ in boxes $1$ and $2$ ($L$ and $L+1$), although box $1$ ($L+1$) carries an MBS.  This is just to signify that the parity of the logical qubit has been flipped, $S_{0}\rightarrow -S_{0}$. A logical $Z$ Pauli occurs when two $\psi$ only are present in the chain, namely in boxes $1$ and $L+1$, see Fig.~\ref{fig:Errors_2}. 

An iteration of the simulation decomposes into the following steps. i) We register all the relevant parameters of the system, in particular the actual configuration of excitations. ii) For a given time interval $\delta\tau$, if $\tau+\delta \tau\leqslant \tau_{\text{sim}}$, we update the time to $\tau+\delta \tau$ and go to step iii). If $\tau+\delta \tau> \tau_{\text{sim}}$ we go directly to step v). The time $\tau_{\text{sim}}$ is the simulation time and describes how long the wire and the thermal bath have been in contact with each other. iii) We draw the number $n$ of error processes from a Poisson distribution with mean $W_{\text{tot}}\, \delta \tau$. It is worth pointing out that $W_{\text{tot}}$ is a state-dependent quantity; for a given eigenstate $\vert n\rangle$ of $H_{S}$, the total transition rate is
\begin{equation}\label{eq:Wtot}
W_{\text{tot}}(n)=\sum_{m}W(m\vert n)\,,
\end{equation}
where $\vert m\rangle$ are eigenstates of $H_{S}$.

However, in the infinite-temperature limit considered here, the total error rate becomes state independent. iii) We apply $n$ error processes randomly according to their relative rates\cite{BrellPRX} and go back to step i). v) We perform the MWPM algorithm described in Sec.~\ref{sec:ECEC} and finally record whether the error correction was successful or not. 

To obtain reliable statistics we perform these five steps on several thousands of samples for each $\tau_{\text{sim}}$. In Fig.~\ref{fig:MWPM_Linear} we plot the probability of failure as function of time for different lengths of the wire. 
The solid lines describe the analytical results (\ref{eq:pfailure}), while the dots are obtained from the Monte Carlo simulation just described. We see that both results coincide very well  (with the agreement improving for bigger $L$) and the logarithmic lifetime of the memory is confirmed.

At low temperatures (i.e. $\beta\gg 1/\vert\Delta\vert$), the dominant processes leading to faulty error correction is the diffusion of a single pair of $\psi$ particles. The reason is that, at low temperature, it is not favorable to create quasi-particle pairs and it costs much less energy for an existing pair to diffuse than for a new pair to be created. We will treat this case in the following sections, when we consider the trijunction.

\section{Trijunction}\label{sec:Trij}
In this section we follow our earlier approach \cite{short} and present the main aspects of the trijunction setup. 

As a one-dimensional wire does not have enough space to exchange MBSs, Ref.~\onlinecite{Alicea2011} proposed to use a trijunction and to move MBSs  by tuning locally the different chemical potentials. Reference \onlinecite{Alicea2011} demonstrated that, when MBSs are exchanged in the trijunction, they obey the same non-abelian braiding statistics as Ising anyons. It is thus very important to understand their properties when coupled to a thermal environment. In particular, below we will determine how the induced thermal noise affects braiding in a nontrivial way.

\begin{figure}[h!]
	\centering
		\includegraphics[width=0.45\textwidth]{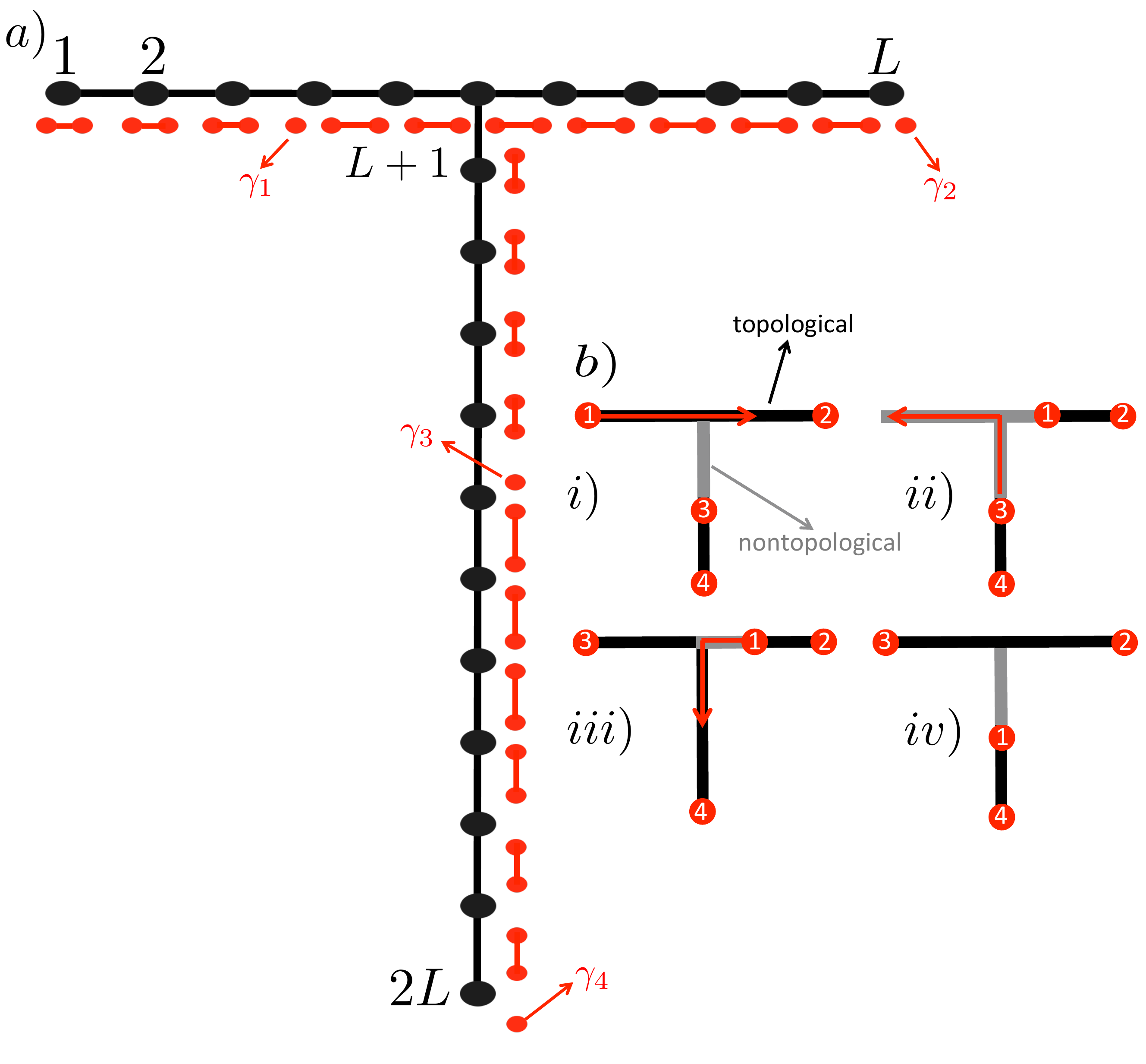}
	\caption{$a)$ Pictorial representation of the trijunction setup. It is composed of a horizontal wire coupled to a vertical wire through hopping and superconducting pairing. The black (larger) dots represent the fermionic sites and the black bonds represent hopping and superconducting pairing. The red (smaller) dots describe the Majorana modes and the lines in-between their pairings. The four Majorana modes $\gamma_{1,2,3,4}$ are depicted. $b)$  Representation of the braiding motion considered in this work. MBSs $2$ and $4$ remains immobile, while MBSs  $1$ and $3$ are exchanged. The black and gray regions represent respectively topological and non topological segments of the trijunction.}
	\label{fig:Figure1_Long}
\end{figure}

The trijunction Hamiltonian is taken to be time dependent,
\begin{eqnarray}
&&H_{\text{trij}}(\tau)=-\sum_{j=1}^{2L}\mu_{j}(\tau)a_{j}^{\dagger}a_{j}-\sum_{j=1}^{2L-1}t(a_{j}^{\dagger}a_{j+1}+a_{j+1}^{\dagger}a_{j}^{\dagger})\nonumber\\
&&+\sum_{j=1}^{2L-1}(\Delta a_{j}a_{j+1}+\Delta^{*}a_{j+1}^{\dagger}a_{j})-t(a_{L/2}^{\dagger}a_{L+1}+a_{L+1}^{\dagger}a_{L/2})\nonumber\\
&&+(\Delta a_{L/2}a_{L+1}+\Delta^{*}a_{L+1}^{\dagger}a_{L/2}^{\dagger}).
\end{eqnarray}
The sites $1,\ldots,L$ correspond to the horizontal wire and the sites $L+1,\ldots, 2L$ to the vertical wire, see Fig.~\ref{fig:Figure1_Long}. The last two terms in $H_{\text{trij}}(\tau)$ describe the coupling between the vertical and horizontal wires of the trijunction. In terms of Majorana operators, the trijunction Hamiltonian becomes
\begin{eqnarray}
H_{\text{trij}}(\tau)&=&-\frac{i}{2}\sum_{j=1}^{2L}\mu_{j}(\tau)\gamma_{2j-1}\gamma_{2j}+i\vert\Delta\vert\sum_{j=1}^{2L-1}\gamma_{2j}\gamma_{2j+1}\nonumber\\
&&+i\vert\Delta\vert\gamma_{L}\gamma_{2L+1}\,,
\end{eqnarray}
where the last term describes the coupling at the trijunction point. Note that the time-dependent chemical potentials satisfy $\mu_{j}(\tau)\leqslant0$ at all times $\tau$.  

The set of chemical potentials $\mu_{j}(\tau)$ is controlled externally such that the full braid depicted in Fig.~\ref{fig:Figure1_Long}b is implemented. In the following, topological segments are characterized by $\mu_{j}=0$  with Hamiltonian $H_{\text{W}}^{\text{top}}$, see Eq.~(\ref{eq:HStop}). Nontopological segments have $\vert\mu_{j}\vert\gg \vert\Delta\vert$, such that they are well approximated by $H_{\text{W}}^{\text{nontop}}$, see Eq.~(\ref{eq:HWnontop}).

\begin{figure}[h!]
	\centering
		\includegraphics[width=0.45\textwidth]{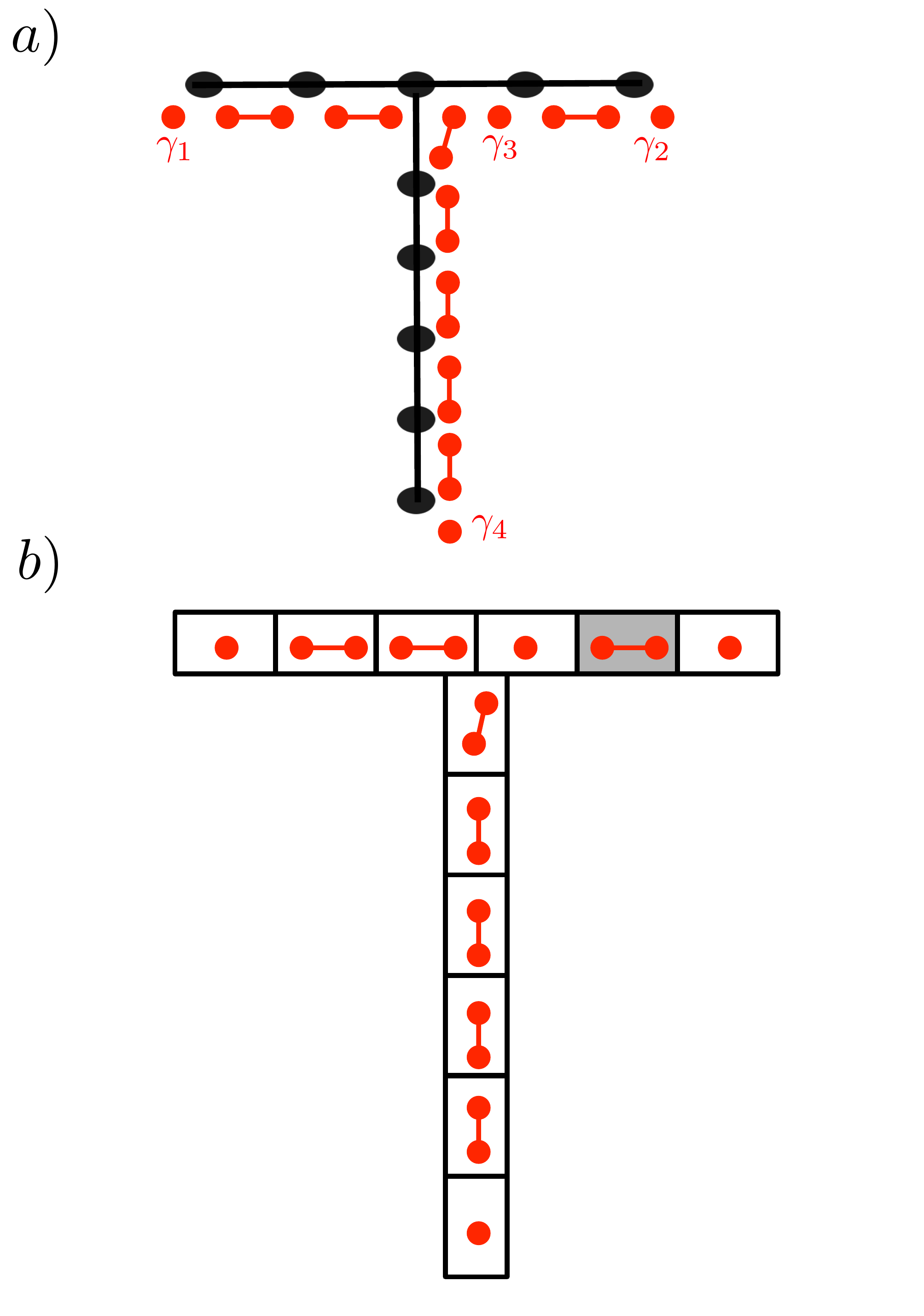}
	\caption{$a)$ Pictorial representation of the trijunction during braiding stage $iii)$ of Fig.~\ref{fig:Figure1_Long}b; three MBSs lie on the horizontal wire. $b)$ Box representation of $a)$. }
	\label{fig:Box_Trij}
\end{figure}

For clarity, Fig.~\ref{fig:Box_Trij} shows the box representation of the full trijunction when the horizontal wire carries three MBSs  i.e., in braiding stage $iii)$ of Fig.~\ref{fig:Figure1_Long}b.  Indeed, one must pay attention that a Majorana mode of the horizontal wire will be paired with a Majorana mode of the vertical wire during this braiding and this is reflected in the box representation as shown in Fig.~\ref{fig:Box_Trij}.  

\subsection{Encoding}\label{sec:TrijEncoding}
Consider four MBSs $\gamma_{1,2,3,4}$ as in Fig.~\ref{fig:Figure1_Long}. Following the procedure of Ref.~\onlinecite{BravyiPRA}, we encode the logical qubit in a fixed-parity sector, say $i\gamma_{1}\gamma_{2}\,i\gamma_{3}\gamma_{4}=+1$. Thus, while the ground-state subspace is fourfold degenerate, we use only two states to encode the qubit. This is required since the overall parity is fixed by the superconducting pairing terms, so that gate operations can only be performed within a fixed-parity sector. The logical qubit states satisfy $i\gamma_{1}\gamma_{2}\vert\bar{0}\rangle=i\gamma_{3}\gamma_{4}\vert\bar{0}\rangle=\vert\bar{0}\rangle$  and $i\gamma_{1}\gamma_{2}\vert\bar{1}\rangle=i\gamma_{3}\gamma_{4}\vert\bar{1}\rangle=-\vert\bar{1}\rangle$. 

\begin{figure}[h!]
	\centering
		\includegraphics[width=0.45\textwidth]{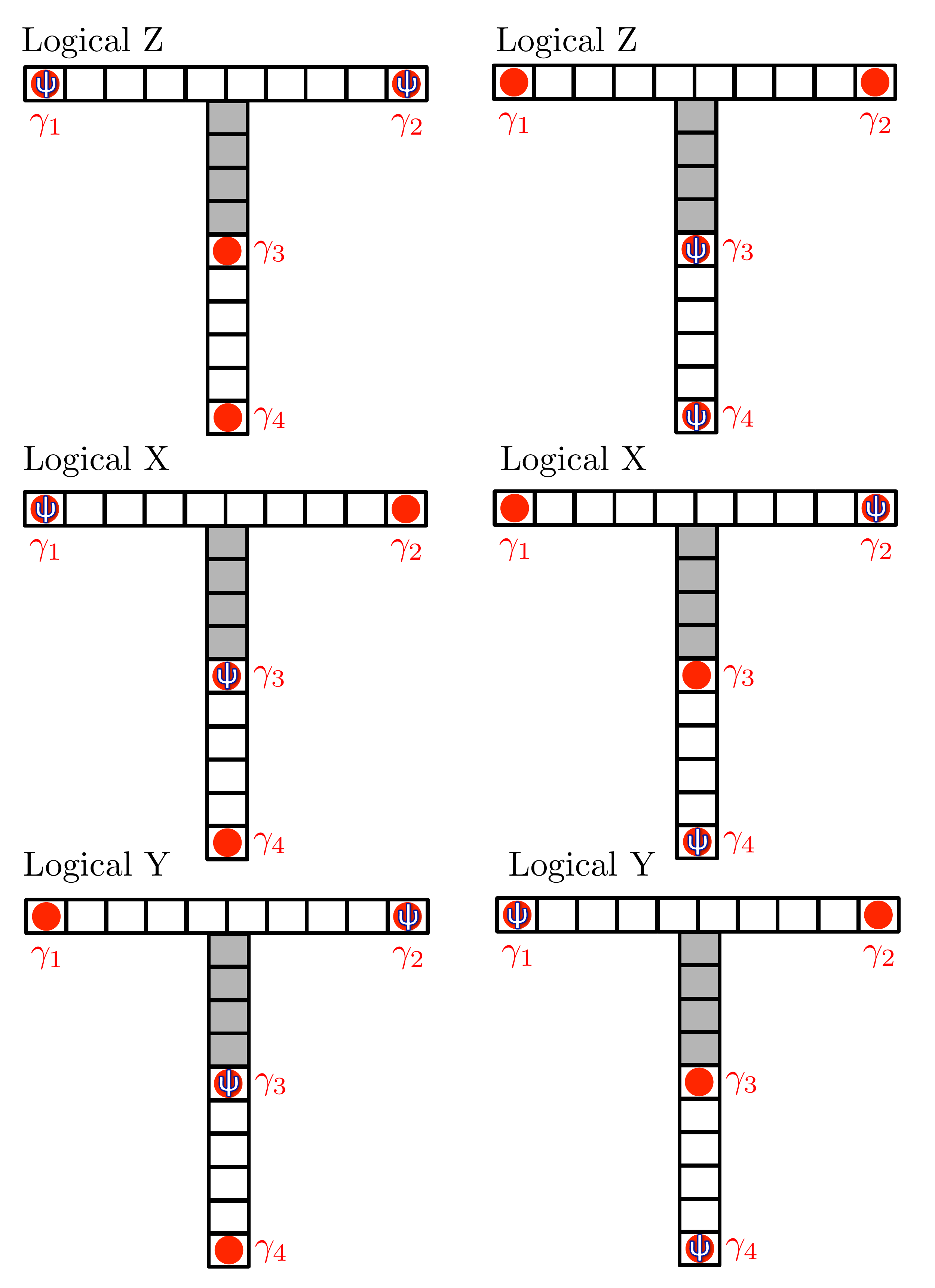}
	\caption{Pictorial box representation of logical $Z$-, $X$-, and $Y$- errors in terms of $\psi$-particles inside MBSs.}
	\label{fig:Errors_2}
\end{figure}

Again, the logical $X$, $Y$, and $Z$ Pauli operators are represented in terms of $\psi$ particles inside MBSs, see Fig.~\ref{fig:Errors_2}.

\section{Unitary Evolution and MBS Motion in the Trijunction}\label{sec:Unitary}
\begin{figure}[h!]
	\centering
		\includegraphics[width=0.45\textwidth]{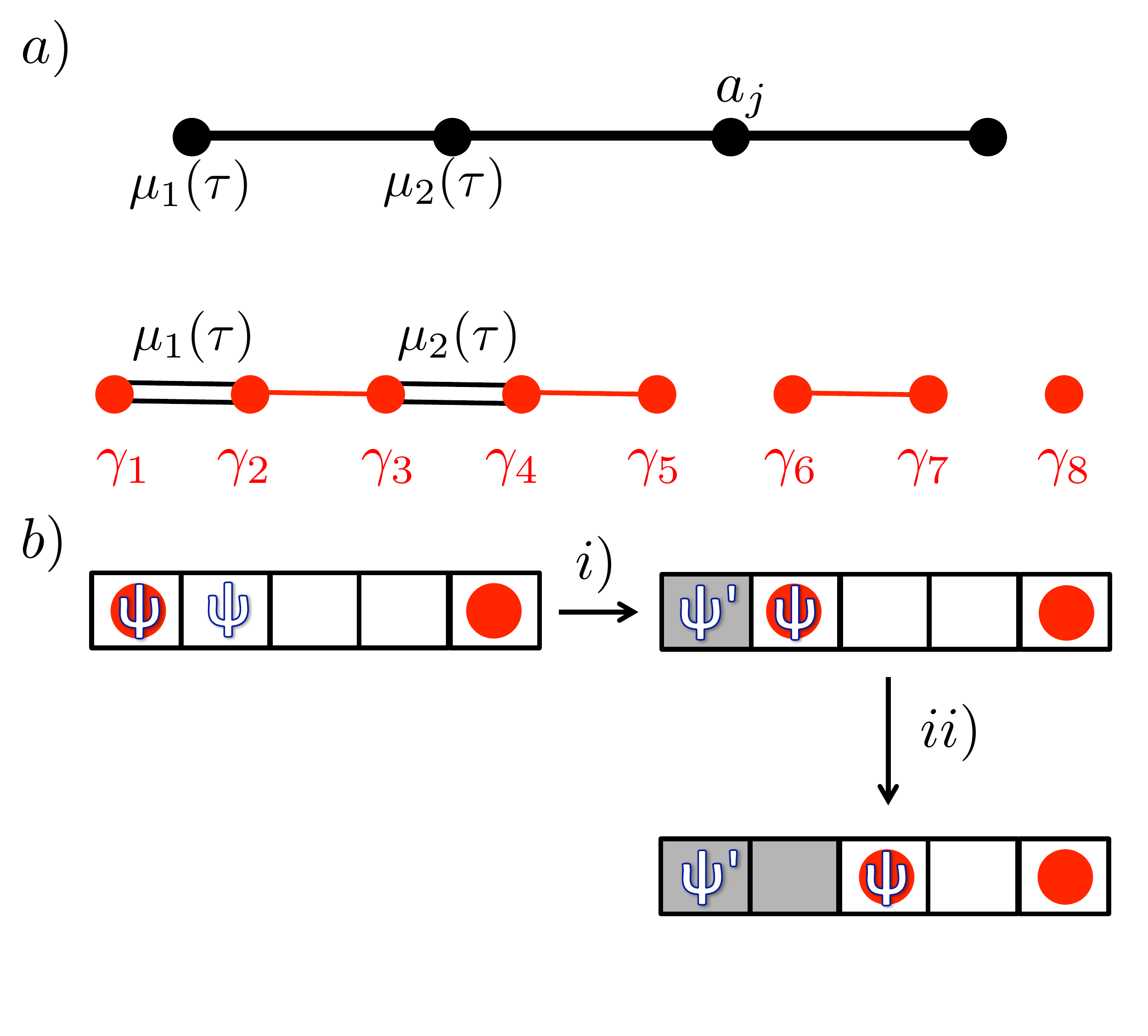}
	\caption{Four-site Kitaev wire model. $a)$ The black dots represent the sites of the Kitaev wire. The chemical potentials $\mu_{1}(\tau)$ and $\mu_{2}(\tau)$ on the first two sites are successively decreased according to $\mu_{1}(\tau)=-10^{-3}\vert\Delta\vert^{2} \tau$ and $\mu_{2}(\tau)=-10^{-3}\vert\Delta\vert \vartheta(\tau\vert\Delta\vert-10^{4})(\tau\vert\Delta\vert-10^{4})$.  The solid black lines describe hopping and superconducting pairing. Below you can find the Majorana representation of the same wire, where smaller (red) dots represent Majorana modes. The double lines describe the varying chemical potentials. $b)$ Box representation of the four-site model. The $\psi$- and $\psi^{\prime}$-excitations are explicitly shown and gray shaded boxes correspond to nontopological segments. The initial state supports two $\psi$-excitations that evolve during the unitary evolution $i)$ and $ii)$.}
	\label{fig:foursites}
\end{figure}
The motions of MBSs in a braiding sequence are performed unitarily. Therefore it is worth spending time to describe the unitary evolution of excited states when MBSs are moved. Indeed, it is essential to understand how moving MBSs interact with $\psi$-particles if one wants to simulate the dynamics of the system. The rules governing the interactions between moving MBSs  and excitations were reported in Ref.~\onlinecite{short}, here we present a detailed analysis leading to these rules. 

\subsection{Adiabaticity}\label{sec:ad}
For any braiding protocol to be valid, the MBSs  must be moved sufficiently slowly with respect to the gap separating the ground states and the rest of the spectrum; here this is the superconducting gap $\vert\Delta\vert$. In other words, the chemical potential $\mu_{j}(\tau)$ at site $j$ must be varied slowly enough.  This was a central assumption to our previous work Ref.~\onlinecite{short}; here we give an explicit formula for the time-dependent chemical potentials. We  implement adiabaticity by choosing
\begin{equation}
\mu_{j}(\tau)=-10^{-3} \vert\Delta\vert^{2}\, \vartheta(\tau\vert\Delta\vert-\tau_{j}\vert\Delta\vert)\,(\tau-\tau_{j})\,,
\end{equation}
where $\vartheta(\tau)$ is the Heaviside theta function and $\tau_{j}$ is fixed by the details of the braiding motions and determines when the chemical potential at site $j$ starts to change.  We have tested numerically whether the above functional form of the chemical potential is good enough to remain within the adiabatic regime. We have diagonalize a time-dependent four-site model; starting from the ground-state we have calculated its time evolution up to time $\tau$ and its overlap with the instantaneous ground state at time $\tau$. The overlap was very close to $1$ at any time; for example the value of the non-adiabatic matrix elements at time $\tau\vert\Delta\vert=10^{4}$ was not larger than $1.6\times 10^{-8}$. 

We find that a good rule of thumb is that an MBS   has moved from site $j$ to site $j+1$ when $\mu_{j}=10\vert\Delta\vert$; in other terms it takes a time $\tau\vert\Delta\vert\approx 10^{4}$ to move an MBS   from one site to a nearest-neighbor site.

\subsection{Linear Motion}
 For the linear motion along the horizontal or vertical wires,  our analysis is based on numerical diagonalization of a small Kitaev wire composed of four sites where we successively decrease the chemical potentials on the different sites, see Fig.~\ref{fig:foursites}a. This is done very slowly such that the system remains in an instantaneous energy eigenstate.  We have computed numerically the time-ordered exponential
\begin{equation}
U(0,\tau)=\mathcal{T}\exp\left(-i\int_{0}^{\tau}d\tau^{\prime} H_{S}(\tau^{\prime})\right)\,,
\end{equation}
that describes the unitary evolution under $H_{S}(\tau)$. Here $\mathcal{T}$ is the time-ordering operator. 

Consider the initial configuration of $\psi$ shown in Fig.~\ref{fig:foursites}b, i.e. one $\psi$ inside the leftmost MBS  and another one in the second box, such that $d_{1}^{\dagger}d_{1}=1$. The initial state is thus an eigenvector with eigenvalues $d_{0}^{\dagger}d_{0}=1,d_{1}^{\dagger}d_{1}=1,d_{2}^{\dagger}d_{2}=0,d_{3}^{\dagger}d_{3}=0$. At time $\tau=0$, the parity of the logical qubit is given by $2d_{0}^{\dagger}d_{0}-1=i\gamma_{1}\gamma_{8}=+1$. After decreasing the chemical potential according to $\mu_{1}(\tau)=-10^{-3}\vert\Delta\vert^{2}\,\tau$ until it reaches the value $\mu_{1}(10^{4}/\vert\Delta\vert)=-10\vert\Delta\vert$, the MBS  has moved to the right and the parity of the logical qubit is changed to $i\gamma_{3}\gamma_{8}$. The amplitude of the chemical potential on site $1$ being large, the operator $i\gamma_{1}\gamma_{2}$ becomes close to an eigenoperator of the Hamiltonian.

\begin{figure}[h!]
	\centering
		\includegraphics[width=0.40\textwidth]{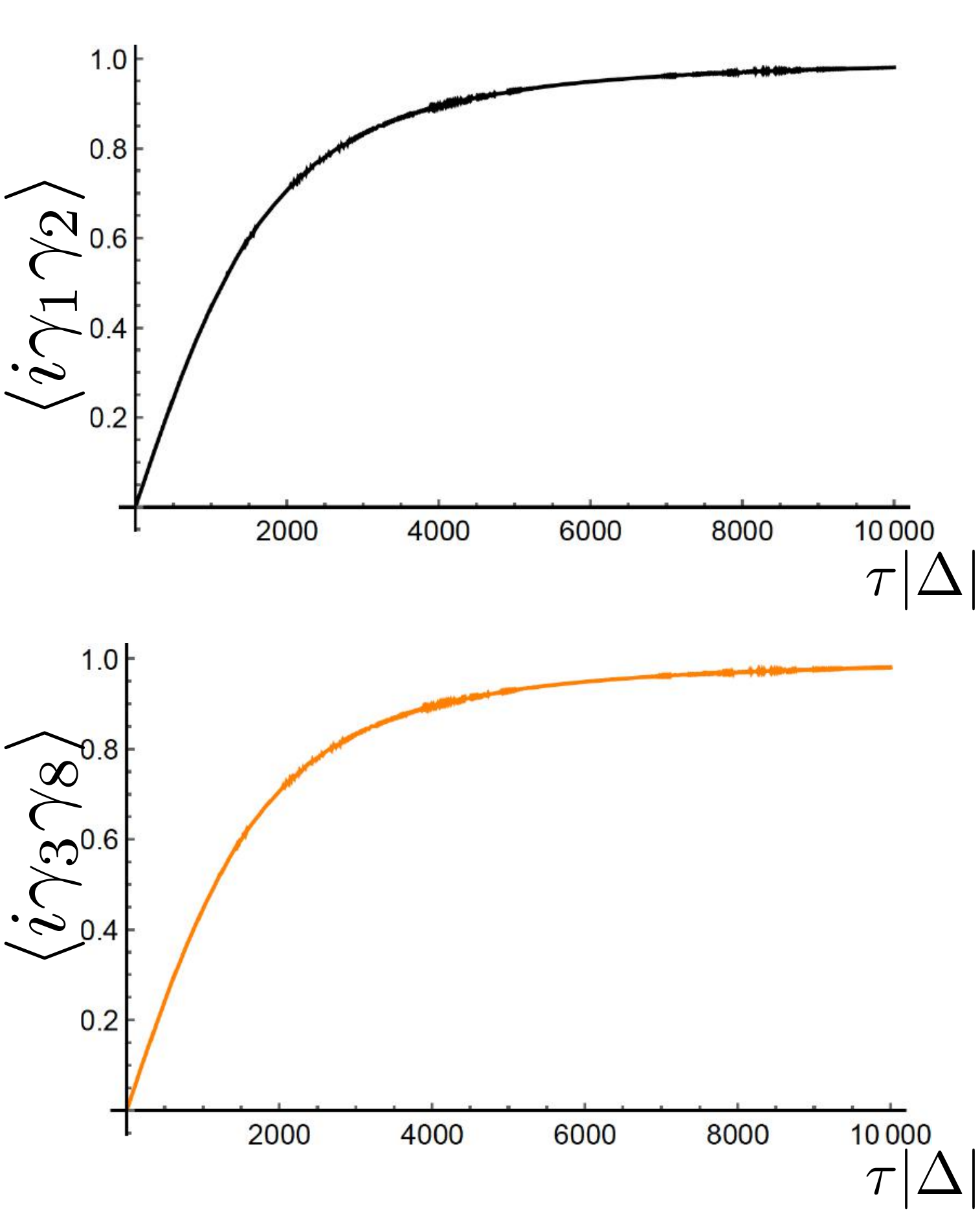}
	\caption{Expectations of $i\gamma_{1}\gamma_{2}$ and $i\gamma_{3}\gamma_{8}$ as a function of time for the four-site model of Fig.~\ref{fig:foursites}. The plots correspond to transition $i)$ of  Fig.~\ref{fig:foursites}b. We see that both quantities goes to $1$ as $\mu_{1}(\tau)$ is decreased from $0$ to $-10\vert\Delta\vert$. When the chemical potential on the first site is as low as $-10\vert\Delta\vert$, then the MBS  has moved to the right and the parity of the Majorana qubit becomes $i\gamma_{3}\gamma_{8}$. The fact that $\langle i\gamma_{1}\gamma_{2}\rangle\rightarrow 1$ and $\langle i\gamma_{3}\gamma_{8}\rangle\rightarrow 1$ show that the $\psi$ bound to the MBS  stays bound to the MBS  while the other $\psi$-excitation is transferred to the nontopological segment and thus becomes a $\psi^{\prime}$-particle.}
	\label{fig:MajoPsi1}
\end{figure}

In Fig.~\ref{fig:MajoPsi1}, we plot the expectation values of $i\gamma_{1}\gamma_{2}$ and $i\gamma_{3}\gamma_{8}$ as function of time. As both go to $+1$ at time $\tau\vert\Delta\vert=10^{4}$, we interpret the results as follows: the $\psi$ carried by the MBS  stays bound to the MBS, while the other $\psi$ is transferred from the topological segment into the nontopological one. An excitation in a nontopological segment is called a $\psi^{\prime}$ to notify that it has different attributes, e.g. a higher energy.  In order to localize the $\psi^{\prime}$-excitations, the chemical potentials in the nontopological segments have a gradient as shown in Fig.~\ref{fig:TopNonTop}b. 

\begin{figure}[h!]
	\centering
		\includegraphics[width=0.45\textwidth]{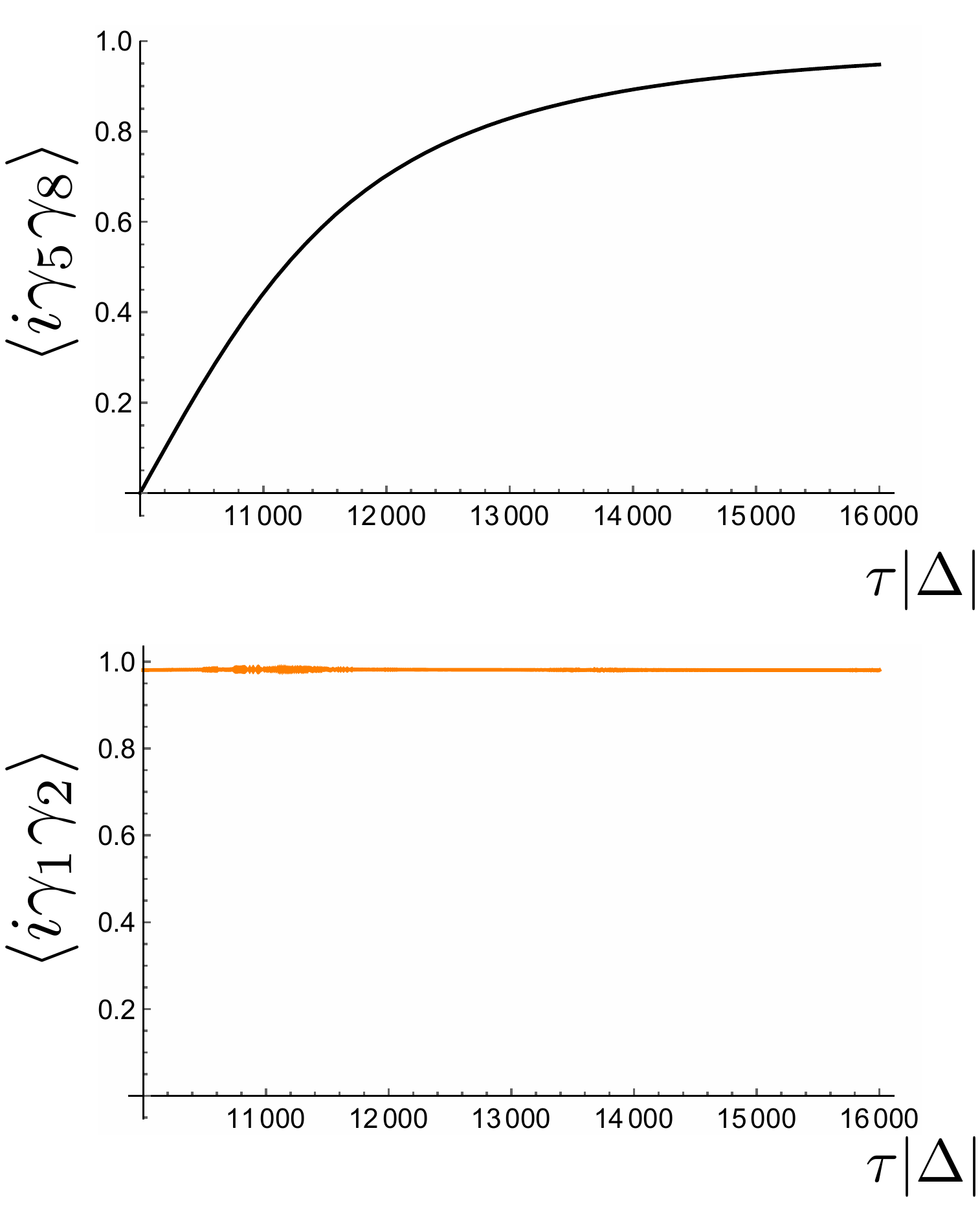}
	\caption{Expectations of $i\gamma_{1}\gamma_{2}$ and $i\gamma_{5}\gamma_{8}$ as a function of time for the four-site model of Fig.~\ref{fig:foursites}. The plots correspond to transition $ii)$ of Fig.~\ref{fig:foursites}b. Here the second chemical potential, $\mu_{2}(\tau)$, starts to decrease while the first chemical potential, $\mu_{1}(\tau)$, continues to decrease. When $\mu_{2}(\tau)$ is negative enough, then  the MBS  has moved to the nearest-neighbor site and the parity of the logical qubit becomes $i\gamma_{5}\gamma_{8}$. The fact that $\langle i\gamma_{1}\gamma_{2}\rangle$=1 confirms that $\psi^{\prime}$-particle remains bound at its position in the nontopological segment. Since $\langle i\gamma_{5}\gamma_{8}\rangle\rightarrow 1$ as the chemical potential is decreased, we conclude again that the $\psi$-particle inside the MBS  remains inside the MBS  during the unitary evolution.}
	\label{fig:MajoPsi2_2}
\end{figure}

Let us now decrease the second chemical potential according to $\mu_{2}(\tau)=-10^{-3}\,\vert\Delta\vert^{2}\,\tau$ while, at the same time, we continue to decrease $\mu_{1}(\tau)$ until time $\tau\vert\Delta\vert=2\times 10^{4}$, see Fig.~\ref{fig:foursites}b. We thus have $\mu_{1}(2\times 10^{4}/\vert\Delta\vert)=-20\,\vert\Delta\vert$ and $\mu_{2}(10^{4}/\vert\Delta\vert)=-10\,\vert\Delta\vert$. The parity of the logical qubit becomes $i\gamma_{5}\gamma_{8}$. We expect to see the $\psi^{\prime}$ excitation immobile in the nontopological segment, while the $\psi$ bound to the MBS  moves together with the MBS  further to the right. This is exactly what we observe in Fig.~\ref{fig:MajoPsi2_2} where we have plotted $\langle i\gamma_{1}\gamma_{2}\rangle$ and $\langle i\gamma_{5}\gamma_{8}\rangle$ as function of time.  We point out that it is necessary to maintain different chemical potentials on the different sites of the nontopological segment in order to localize the $\psi^{\prime}$-particles, see Fig.~\ref{fig:TopNonTop}b. 

\begin{figure}[h!]
	\centering
		\includegraphics[width=0.45\textwidth]{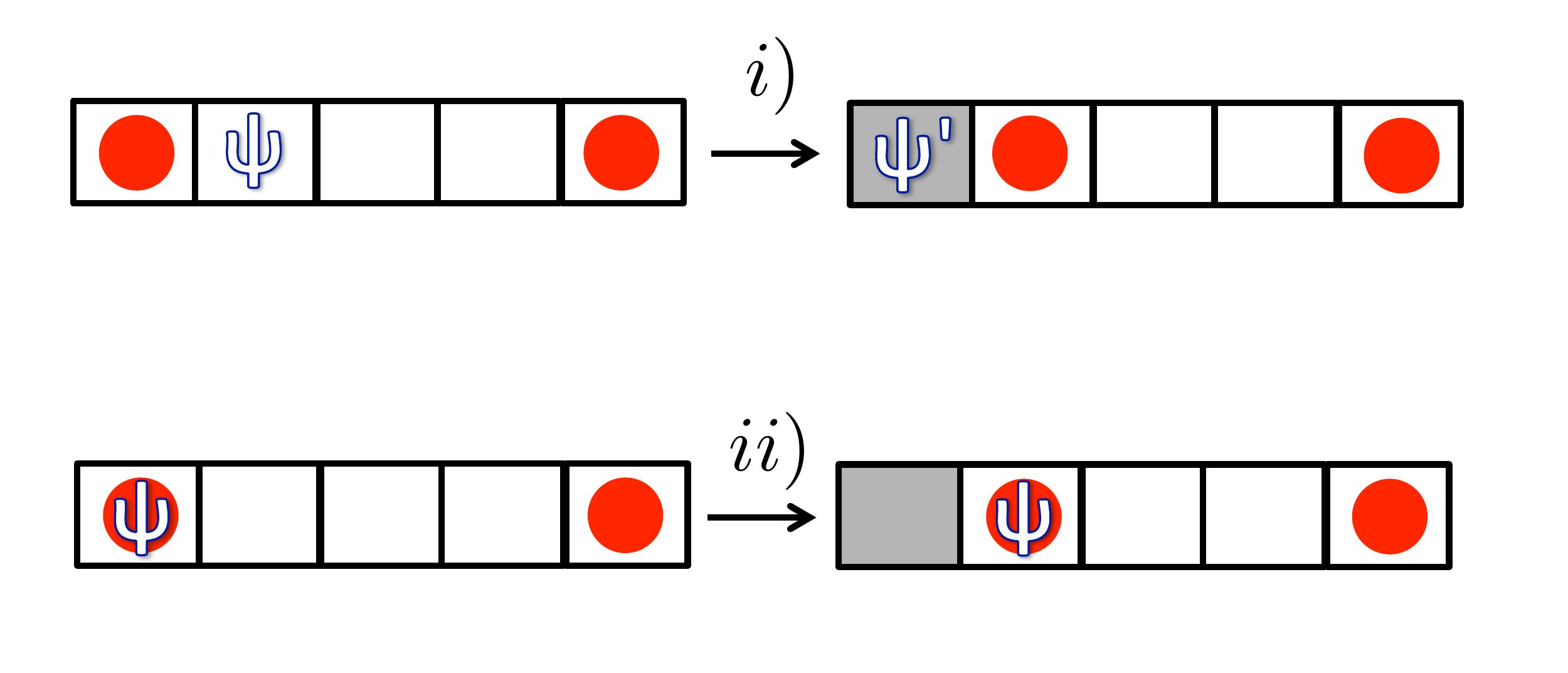}
	\caption{Box representation of the unitary evolution of moving MBSs in the presence of excitations. i) When a MBS moves over an existing $\psi$-excitation, the $\psi$-excitation is transferred to the nontopological segment and becomes a $\psi^{\prime}$-excitation. ii) A $\psi$-excitation trapped inside a MBS remains trapped during the motion of the MBS.}
	\label{fig:foursites2}
\end{figure}

We have performed several similar tests, starting from different configurations of $\psi$-excitations. All the conclusions are the same and can be summarized in terms of the following rules. i) When an MBS  moves into a topological segment and crosses a $\psi$-excitation, then the $\psi$-excitation is transferred to the first site to the left of the MBS  into the nontopological segment and stays immobile. ii) A $\psi$ inside an MBS  moves together with the MBS.
In case of reverse motion, i.e. when the MBS  moves into the nontopological segment, then a $\psi^{\prime}$ from the nontopological segment will be transferred back into the topological segment. We have represented these two rules pictorially in Fig.~\ref{fig:foursites2}.  It is important to recognize that while the overall parity of the system is conserved, the parity of individual topological segments is not preserved. This is a crucial difference as compared to the previous case with two immobile MBSs, as now logical $X$ and $Y$ errors are possible. 

In Ref.~\onlinecite{Supplement} we summarize all the unitary evolutions necessary to simulate the system;  the unitary motions of an MBS  over the trijunction point need to be obtained by simulating a minimal six-site trijunction model. 

\section{Adiabatic Davies Equation}\label{sec:TrijDavies}
We must take the time-dependence of the trijunction Hamiltonian into account when writing down the master equation.\cite{short} 
In the adiabatic limit,  it is correct to generalize Eq.~(\ref{eq:MasEq}) to
\begin{eqnarray}\label{eq:MasterEqAdiabatic}
\mathcal{D}(\rho_{S}(\tau))&=&\sum_{i,j}\sum_{\omega(t)}\gamma^{ij}(\omega(\tau))\left(A^{i}(\omega(\tau))\rho_{S}(\tau)(A^{j}(\omega(\tau)))^{\dagger}\right.\nonumber\\
&&\hspace{1cm}\left.-\frac{1}{2}\{(A^{j}(\omega(\tau)))^{\dagger}A^{i}(\omega(\tau)),\rho_{S}(\tau)\}\right)\,,\nonumber\\
\end{eqnarray}
where $\omega(\tau)$ are the time-dependent energy differences in the spectrum of $H_{\text{trij}}(\tau)$.

The populations follow an adiabatic Pauli master equation
\begin{eqnarray}\label{eq:popt}
&&\frac{d P(n(\tau),\tau)}{d\tau}=\sum_{m(\tau)}\left[W(n(\tau)\vert m(\tau))P(m(\tau),\tau)\right.\nonumber\\
&&\hspace{2.5cm}\left.-W(m(\tau)\vert n(\tau))P(n(\tau),\tau)\right]\,,
\end{eqnarray}
with
\begin{equation}\label{eq:Wmnt}
W(n(\tau)\vert m(\tau))=\gamma(\omega_{mn}(\tau))\left\vert\langle m(\tau)\vert A^{i_{mn}}(\omega_{mn}(\tau))\vert n(\tau)\rangle\right\vert^{2}\,.
\end{equation}

The bulk error processes and the associated rates remain the same as in the time-independent scenario, see Sec.~\ref{sec:Davies}. To be more precise, the error processes away from the moving MBSs, including at other MBSs  that are for the time being stationary, are the ones presented in Sec.~\ref{sec:Davies}. However, more complicated boundary processes appear because of the motion of the MBSs. In Appendix~\ref{sec:DaviesTime} we present some examples. An exhaustive table of the more than 200 distinct allowed processes can be found in Ref.~\onlinecite{Supplement}. 

It is worth pointing out that the system-bath interaction of Eq.~(\ref{eq:SB}) does not support the creation of excitations in the nontopological segments of the trijunction. Indeed, when the chemical potential is very negative, the eigenstates of a nontopological segment approaches the eigenstates of $H_{\text{W}}^{\text{nontop}}$ and the coincidence becomes better as the chemical potential becomes more negative. Since $[H_{\text{W}}^{\text{nontop}},H_{SB}]=0$, creation of excitations in the nontopological segment is suppressed as the chemical potential decreases.  However, this does not mean that no excitations will ever be present in the nontopological segments, as we discussed in Sec.~\ref{sec:Unitary}.

The time dependence of $H_{S}(\tau)$ must also be taken into account to calculate the rates of all the error processes. For example, when changing the chemical potential from time $0$ to time $\tau$ with $\tau\vert\Delta\vert\approx10^{4}$,  such that an MBS   has moved by one site, one obtains the rates associated with the possible error processes by integrating Eq.~(\ref{eq:Wmnt}); we defer a detailed discussion to Appendix~\ref{sec:DaviesTime}. 

\section{ Error Correction and Dangerous processes in the trijunction}\label{sec:ErrorCorr}
The error correcting procedure applied here is an adaptation of the ÔgreedyÕ algorithm proposed by Wootton in Ref.~\onlinecite{Wootton}. We point out that this is \emph{not} a MWPM algorithm in the sense described in Sec.~\ref{sec:EC}.  We have chosen this decoder because of its great simplicity. Also, choosing any other decoding scheme would not change the main message of our paper, as we will see.  More generally, many matching procedures might be applied to the same model, each of which potentially having a different threshold. \cite{BarbaraReview}  

Our error-correcting scheme here is passive, meaning that we apply it at the end of the quantum computing protocol (that below will consist in a single braid only). This is in contrast with active error correction where error correcting steps are performed in the midst of a braid sequence of a full quantum computing protocol. We believe that it is useful to make the distinction between the following active scenarios: 
\begin{enumerate}
\item Error correction is executed at the end of each braid.
\item Error correction is performed repeatedly \emph{during each} braid.
\end{enumerate}
As we will see, our results imply that passive error correction and even active error correction that is performed at the completion of braids, do not lead to a lifetime that increases with the size of the trijunction; a scheme where error correction is executed during each braid is required to cure the dangerous errors.

\begin{figure}[h!]
	\centering
		\includegraphics[width=0.4\textwidth]{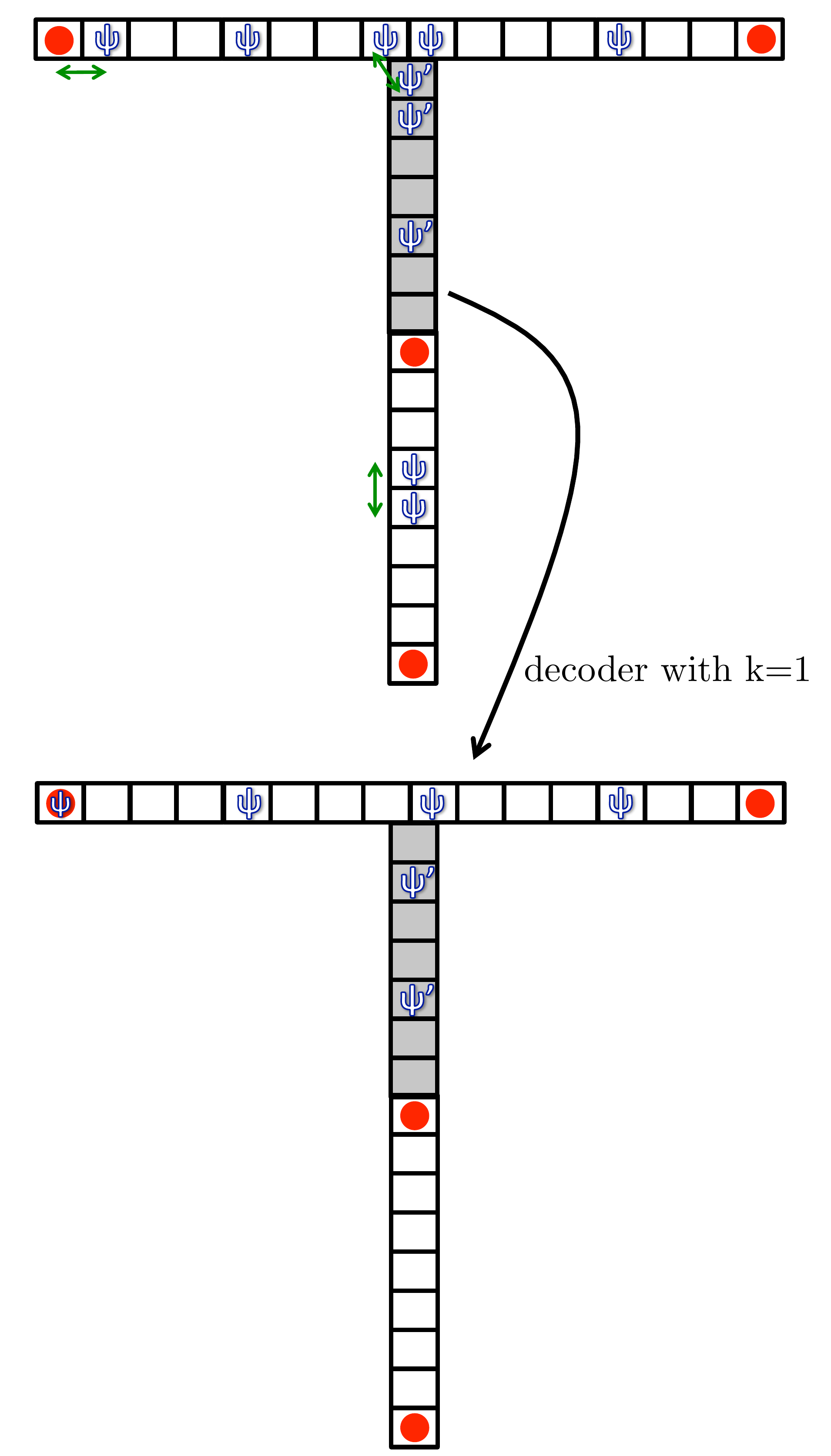}
	\caption{First iteration ($k=1$) of the ÒgreedyÓ decoder used in our trijunction simulations. After having identified the pairs of excitations at distance $k$, we annihilate them according to the protocol described in Sec.~\ref{sec:ErrorCorr}. Our algorithm is a straightforward adaptation of the one presented in Ref.~\onlinecite{Wootton}.}
	\label{fig:ErrorCorrection}
\end{figure}

Below we summarize the main steps of the error correction algorithm for the trijunction:

\begin{enumerate}
\item Loop through all sites of the trijunction to find the pairs of quasi-particles ($\psi$, $\psi^{\prime}$, or MBS ) that are at minimal distance $k$. Start with $k=1$.
\item If there a no multiple possibilities, annihilate the corresponding pairs. If there are multiple possibilities, e.g. in the case that three quasi-particles are positioned such that one of them is at distance $k$ from the two others,   we apply the following rules:
\begin{itemize}
\item If all the excitations are on the horizontal wire, we pair the particles from left to right.
\item If all the excitations are on the vertical wire, we pair the particles from top to bottom.
\item If two excitations are on the horizontal wire and a third is on the vertical wire, we annihilate the pair composed of the leftmost quasi-particle on the horizontal wire and the uppermost quasi-particle on the vertical wire. 
\item If two excitations are on the vertical wire and a third is on the horizontal wire, we annihilate the pair composed of the uppermost quasi-particles on the vertical wire and the quasi-particle on the horizontal wire. 
\end{itemize}
\item If there are still some excitations in the bulk of the trijunction, repeat the procedure with $k+1$.
\end{enumerate}
To illustrate this procedure we present a pictorial representation of one error correction step in Fig.~\ref{fig:ErrorCorrection}.

\subsection{Dangerous Errors}\label{sec:Dangerous}
This section contains the central result of our work: the identification of so-called dangerous errors that prevent the lifetime of the Majorana trijunction from increasing with the system size.

\begin{figure}[h!]
	\centering
		\includegraphics[width=0.5\textwidth]{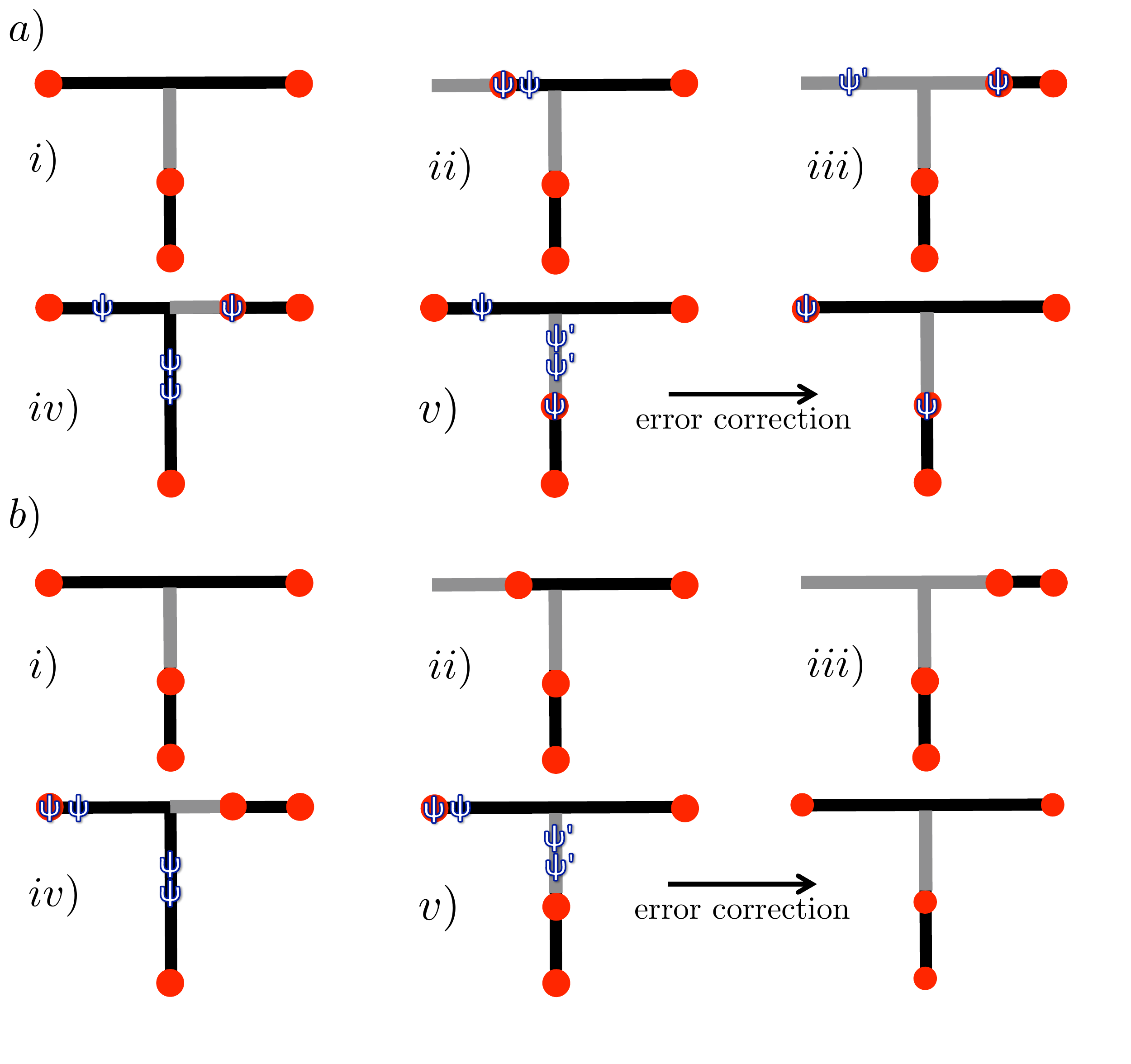}
	\caption{Pictorial representation of two error sequences leading to the same error syndrome. The difference between $a)$ and $b)$ is the occurrence of $\psi$ inside MBSs. Since $\psi$ trapped inside an MBS  is invisible to an external observer, both situations have the same error syndrome. Independently of the details of the error-correcting algorithm, if one situation is successfully corrected, the other one will lead to failure.}
	\label{fig:Figure1_2}
\end{figure}

Consider the situation in which an MBS  is moving and a pair of excitation is created, one inside the MBS  and one inside the bulk of the trijunction, see Fig.~\ref{fig:Figure1_2}. While the two $\psi$'s are originally created as a pair in neighboring boxes, the motion of the MBS  drags along one of the $\psi$ and separates it from its partner. In other terms, the braiding renders an originally local error source completely nonlocal. We thus call an error process that creates a $\psi$ inside a \emph{mobile} MBS  \emph{dangerous}.
The effective non-locality of the noise prevents our algorithm from successfully recovering the stored quantum information. For example in Fig.~\ref{fig:Figure1_2}a,  a single error event will not be cured by our algorithm and will lead to a $X$ error.  Note that in Fig.~\ref{fig:Figure1_2}, we have drawn two error processes: a dangerous error process, where a $\psi$ is created inside an MBS,  and an inoffensive error process where a pair of excitations is generated in the bulk of the vertical wire.

A natural question that arises is whether a better algorithm could take into account the nonlocality of the noise in a clever manner. Unfortunately this is impossible if error correction is not performed during braiding. The reason is that different error processes can lead to exactly the same error syndrome. In Fig.~\ref{fig:Figure1_2}a and b, we depict two error processes that generate the same syndrome. The main difference between them is the occurrence of $\psi$-particles inside MBSs. This can be traced back to the moment where a dangerous error happens. In Fig.~\ref{fig:Figure1_2}a it happens at the beginning of the braid, while it happens at the end of the braid in Fig.~\ref{fig:Figure1_2}b. If one syndrome is successfully cured by an algorithm, the other one will lead to failure. A central ingredient for the emergence of such ambiguity is that different MBSs travel over the same segments of the trijunction during braiding, making it impossible to identify which bulk $\psi$ should be paired with which MBS.  Since the probability of dangerous events is finite and does not depend on the size of the trijunction, we expect the lifetime of coherence of the trijunction qubit to be independent of $L$.

We point out that the physics of dangerous error processes is the same at high ($\beta\ll\vert1/\Delta\vert$) and low ($\beta\gg1/\vert\Delta\vert$) temperatures. Therefore, we expect the restriction due to dangerous error processes to be qualitatively identical in both regimes.

\section{Monte Carlo Simulation of the Trijunction}\label{sec:MonteCarloTrij}
To confirm our predictions, we perform a standard Monte Carlo simulation  for the trijunction and determine the evolution of the stored quantum information under the adiabatic master equation (\ref{eq:popt}). We focus here on the low-temperature regime with $\beta=4/\vert\Delta\vert$.

The Monte Carlo simulation consists essentially of the same five steps as in the simulation described in Sec.~\ref{sec:MonteCarlo} for the single Majorana wire. However, some care has to be taken in the low-temperature regime. Indeed, in that case the total probability $W_{\text{tot}}$ that an error event occurs, see Eq.~(\ref{eq:Wtot}), is strongly \emph{state-dependent} and we cannot always approximate it by a constant. This is the case because the spectral function $\gamma(\omega)$ depends very much on the value of $\omega$ at low temperatures. 

One possible way to solve this issue would be to apply an alternative set of five Monte Carlo steps:  i) Register all the relevant parameters of the system, in particular the actual configuration of excitations. ii) Calculate the time $\delta\tau$ for the next error process to occur, drawing $\delta\tau$ from an exponential distribution $\propto \exp(1/W_{\text{tot}})$. iii) Update the time to $\tau+\delta\tau$. If $\tau+\delta\tau\leqslant\tau_{\text{sim}}$, go to step iv). Otherwise go directly to step v). iv) Apply an error event randomly according to their relative rates. Go back to step i). v) Perform the error corecting algorithm described in Sec.~\ref{sec:ErrorCorr} and finally record whether the error correction was successful or not.

Such a procedure is perfectly valid in the low-temperature regime, but only when the MBSs are immobile. Indeed, when MBSs are in the process of being moved this method cannot be applied. The main problem resides in the fluctuating $\delta\tau$ drawn from the exponential distribution. When MBSs  are moved, one needs to define a time-step $\delta\tau_{\text{M}}$ for an MBS   to be carried to the nearest-neighbor site. For example, here we have chosen $\delta\tau_{\text{M}}\vert\Delta\vert=10^{4}$. It is however clear that, most of the time, $\delta\tau$ drawn from the exponential distribution would never be an exact multiple of $\delta\tau_{\text{M}}$ and thus we cannot decide by how many steps the MBS  must be moved during the time interval $\delta\tau$. Therefore, this method is applicable only when the MBS  are kept immobile; so, this method is applicable during the time in between braids.

During the braiding motions, we apply the sequence of steps i)-v) from Sec.~\ref{sec:MonteCarlo} but by taking care that $W_{\text{tot}}$ is state dependent and by ensuring that the number of errors $n$ drawn from the Poisson distribution is either $0$ or $1$; we assure this by choosing a small enough coupling constant $\kappa=2\times 10^{-4}\vert\Delta\vert$.  We point out that this choice of $\kappa$ is only relevant for our numerical procedure to be valid, but it does not hide any important physical issue. In fact,  allowing $n>1$ would be problematic (in the low-temperature regime) since each time an error is applied, the change in total probability $W_{\text{tot}}$ is very drastic and must be taken into account.

\begin{figure}[h!]
	\centering
		\includegraphics[width=0.45\textwidth]{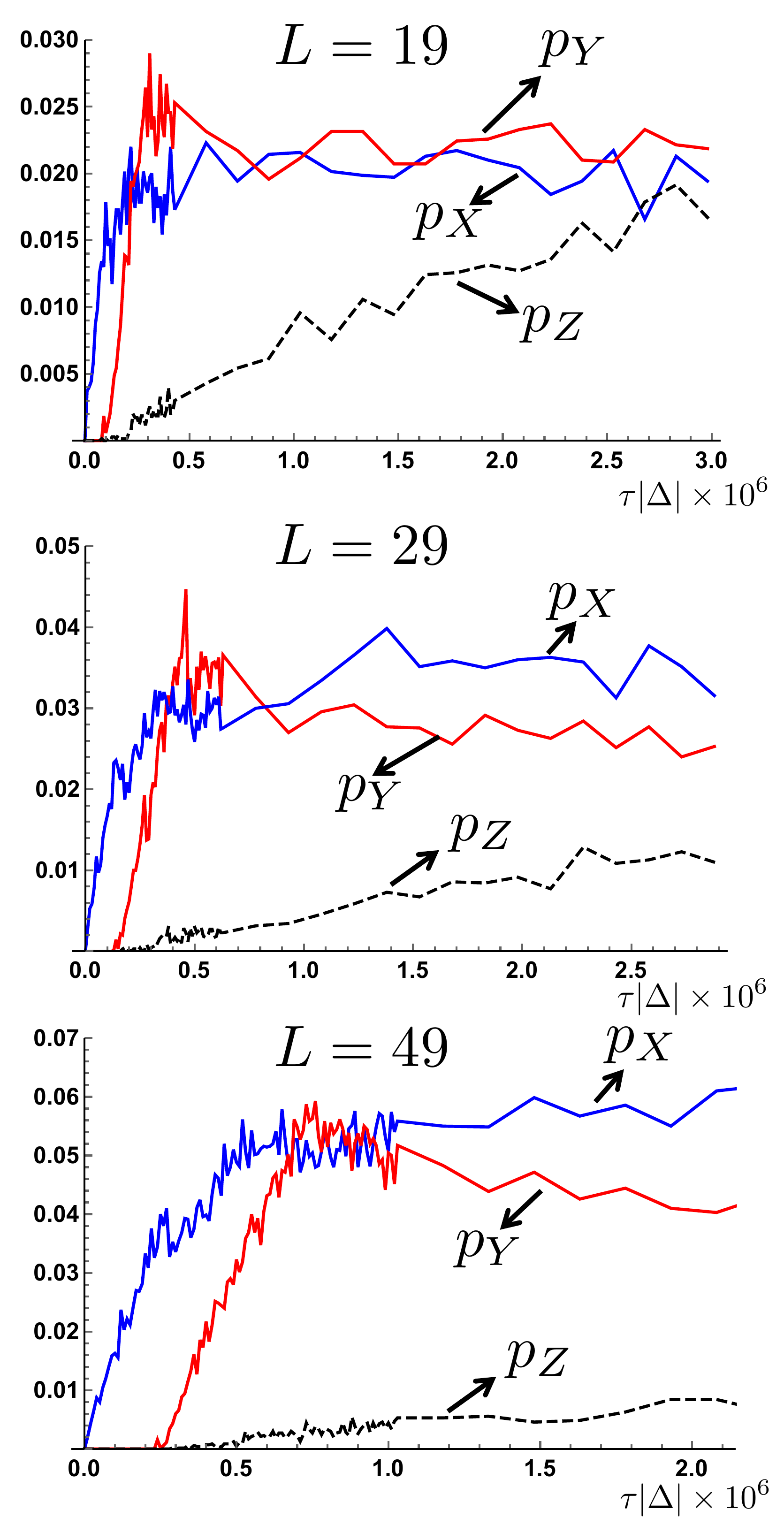}
	\caption{Plots of $p_{X}$, $p_{Y}$, and $p_{Z}$ as function of time. MBSs  $1$ and $3$ are exchanged while MBSs  $2$ and $4$ remain immobile, see Fig.~\ref{fig:Figure1_Long}b. The time to braid is proportional to the size of the trijunction and we have $\tau_{\text{braid}}=(2L+2)10^{4}/\vert\Delta\vert$. For $\tau<\tau_{\text{braid}}$, i.e. when the exchange is not finished, we exchange the MBSs  until time $\tau$ with the coupling to the thermal bath being on. Then we unitarily (no coupling to the thermal bath) finish the braid and perform error correction at the end. Here we use $\kappa=2\times 10^{-4}\vert\Delta\vert$ and $\beta=4/\vert\Delta\vert$.}
	\label{fig:Result}
\end{figure}

In Fig.~\ref{fig:Result}, we present  the probabilities $p_{X}$, $p_{Y}$, and $p_{Z}$ that a logical $X-$, $Y-$, or $Z-$error occurs for trijunctions of various lengths, see also Ref.~\onlinecite{short}. When we assume perfect error correction, the noise acting on the logical qubit is unital, i.e. the fixed point is a completely mixed state, and takes the form of a generalized depolarizing channel \cite{NielsenChuang}
\begin{equation}
\rho_{\text{trij}}\rightarrow \sum_{O=I,X,Y,Z}p_{O}\,O\,\rho_{\text{trij}}\,O\,,
\end{equation}
where $I$ is the identity operator and $p_{I}+p_{X}+p_{Y}+p_{Z}=1$. This is in contrast with the underlying physical noise that is not unital here.
 
We observe that $p_{X}$ and $p_{Y}$ significantly increase during the braiding time, while $p_{Z}$ remains small. This is due to the presence of dangerous error processes during braiding as explained in Sec.~\ref{sec:Dangerous}. It is also interesting to distinguish between $X$ and $Y$ errors. It is clear from the plots that $p_{X}$ increases faster than $p_{Y}$ and there is a period of time where $p_{Y}$ does not increase. In fact, considering only single-error events (a very accurate approximation at short times) the environment cannot produce a $Y$ error during the first $L/2+1$ time steps. Indeed, the only possible configuration of $\psi$-excitations corresponding to a $Y$-error after error correction, resulting from a single error event during braiding, is the one in the bottom left of  Fig.~\ref{fig:Errors_2}. Such a situation arises when a dangerous error happens during the braiding stage $i)$ of Fig.~\ref{fig:Figure1_Long}. However, the $\psi$-particle created in the bulk of the horizontal wire must lie closer to the right boundary than to the left boundary. Therefore, such a $Y$-error indeed cannot happen during the first $L/2+1$ time steps of braiding.  On the contrary, an $X$ error can be produced by a single error events at any time during the braid. We also point out that dangerous error processes are more probable than creation of a pair somewhere in the bulk, since the energy cost is lower.  A transition energy of $2\vert\Delta\vert$ vs. $4\vert\Delta\vert$ makes a considerable difference, and the strong dependence of $\gamma(\omega)$ on $\omega$ also plays a role.

\begin{figure}[h!]
	\centering
		\includegraphics[width=0.45\textwidth]{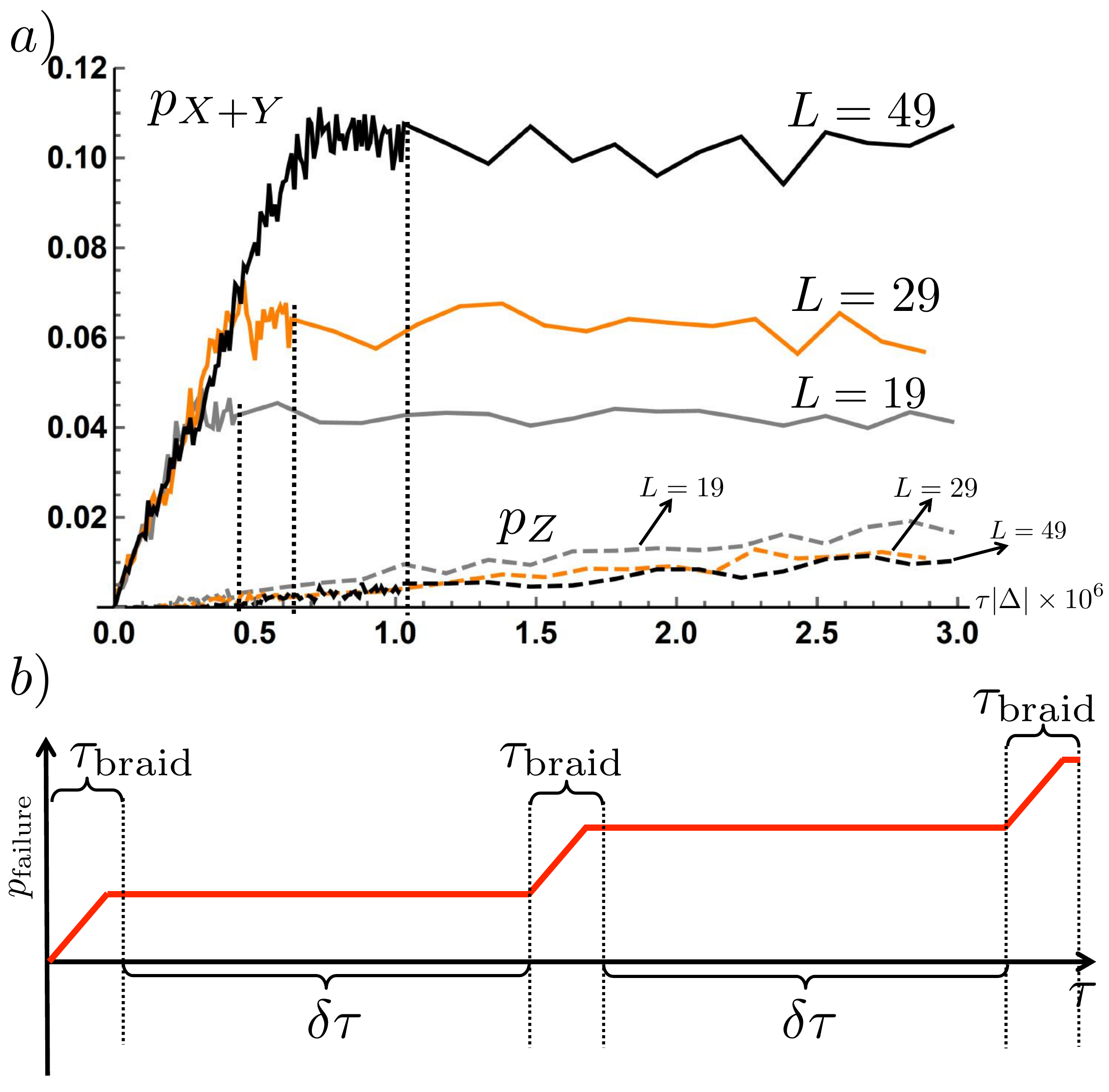}
	\caption{$a)$ Probability $p_{X+Y}$ (solid) and $p_{Z}$ as function of time for trijunctions of length $L=19,29,49$. The details of the plot are the same as in Fig.~\ref{fig:Result}. $b)$ Artistic representation of the probability of failure as function of time when three consecutive braids are executed. The time interval between two braids is $\delta\tau$ and the braiding time is $\tau_{\text{braid}}$. }
	\label{fig:Figure3}
\end{figure}

In Fig.~\ref{fig:Figure3} we plot $p_{X+Y}=p_{X}+p_{Y}$  for different $L$. After the end of the braid $p_{X+Y}$ remains constant since no dangerous errors are possible anymore when MBSs stay immobile, although $X$ and $Y$ errors can be interconverted during this period. Indeed, $X$ and $Y$ errors are possible only when the parity of topological segments is broken and this is solely possible when MBSs move.
The most important feature of the growth of $p_{X+Y}$ is that it is completely independent of $L$ at small times; the lifetime of the memory does not grow with $L$. This is in complete agreement with our discussion of dangerous errors in Sec.~\ref{sec:Dangerous}. It is worth pointing out that the braiding time grows linearly with the size of the trijunction. Therefore, the probability that a dangerous error occurs during braiding is higher for a larger trijunction. This is observed in Fig.~\ref{fig:Figure3}, where $p_{X+Y}$ at the end of the braid is bigger for larger trijunctions. 

At small temperature, the origin of a non vanishing $p_{Z}$ probability is the creation of a pair of $\psi$'s that diffuse across the trijunction. However, it takes more time for a pair to reach the MBSs when the trijunction is longer, therefore we expect $p_{Z}$ to decrease with increasing $L$, and this is the case in Fig.~\ref{fig:Figure3}a.

In the context of a full quantum computing protocol, where several braids are executed, our results show that performing error correction either at the end of all the braids or at the end of each braid is not enough to cure the failure induced by dangerous errors.
\section{Conclusions}\label{sec:Conclusions}
In this work we have investigated the self-correcting properties of Majorana 1D quantum computing architectures. In particular, starting from a microscopic model, we focused on the situation where MBSs are braided in a trijunction setup coupled to a parity-preserving bosonic environment.

While a single wire with immobile MBSs represents a truly self-correcting quantum memory with a lifetime that increases with the size of the wire, this is not true anymore when MBSs are exchanged in a trijunction architecture. The main reason is the occurrence of so-called dangerous errors that are solely due to the motion of MBSs; an MBS  can trap an excitation and drag it along during braiding, thus rendering a local source of noise highly nonlocal in its effect. In this case, error correction at the end of the braid is insufficient to recover the stored quantum information.

In the context of a full quantum computing protocol, where several braids are executed, our results imply that passive error correction (at the end of all braids) and even active error correction, in which correction is performed at the end of each braid, is too weak to counteract the negative effects of dangerous errors. The only possibility we envision to preserve the stored quantum information is to perform repeatedly error correction \emph{during} each braid: in the very simple example of Fig.~\ref{fig:Figure1_2}, performing error correction both at stages $iii)$ and $v)$ would allow successful recovery of the logical qubit. Our results are in agreement with the more and more popular view that active error correction is necessary even in non-abelian topological systems. \cite{WoottonLoss, BrellPRX, Hutter, WoottonHutter} In light of this discussion, we point out that increasing $L$, far from improving the lifetime of the trijunction logical qubit, actually makes the situation worse because the time to braid MBSs adiabatically is proportional to $L$. \cite{Beverland}

We also comment that while our results put restrictions on the self-correcting properties of this specific quantum computing architecture, one can expect other schemes, such as interaction-based braiding of MBSs, \cite{Hassler2012, Fulga} to behave in a more favorable way. Finally, it is a priori not clear whether braiding MBSs in 2D systems suffers from the same restrictions as the 1D case studied here. In 2D setups, the paths followed by the braided MBSs must cross at least once, but do not need to overlap over a large region. Therefore, dangerous error processes should occur less frequently than in 1D implementations; but we keep in mind that a single uncorrectable error is enough to prevent the lifetime of the topological qubit from increasing with the system size.

\section{Acknowledgements}
We are happy to thank valuable discussions with Stefano Chesi, Fabian Hassler, Adrian Hutter, Olivier Landon-Cardinal, and Daniel Loss. We are grateful for support from the Alexander von Humboldt foundation and from QALGO. NEB was supported in part by US DOE Grant No. FG02-
97ER45639.

\newpage
\appendix

\section{Davies Prescription: Time-Independent Hamiltonian}\label{app:Davies}
In this Appendix we aim to derive explicit expressions for the jump operators $A^{i}(\omega)$ appearing in the master equation (\ref{eq:MasterEq}) for the time-independent  problem; for simplicity we focus here on a single wire. The time-dependent case is treated in an exactly similar fashion because we consider only adiabatic time variation. 

Let us start from the system bath Hamiltonian
\begin{equation}
H_{SB}=-2\sum_{j}B_{j}\otimes a_{j}^{\dagger}a_{j}\,,
\end{equation}
where the constant in Eq.~(\ref{eq:SB}) has been ignored because it only leads to a renormalization of the bath Hamiltonian $H_{B}$.  Rewrite $H_{SB}$ in terms of the eigenoperators $d_{j}$ and $d_{j}^{\dagger}$ that diagonalize $H_{\text{W}}^{\text{top}}$, see Eq.~(\ref{eq:HSdiag}). We have
\begin{eqnarray}
H_{SB}&=&-i 2B_{1}\otimes (d_{0}+d_{0}^{\dagger})(d_{1}+d_{1}^{\dagger})\nonumber\\
&&-2\sum_{j=1}^{\lceil L/2\rceil-1}B_{2j}\otimes(d_{2j-1}-d_{2j-1}^{\dagger})(d_{2j}+d_{2j}^{\dagger})\nonumber\\
&&-2\sum_{j=1}^{\lceil L/2\rceil-1}B_{2j+1}\otimes(d_{2j}-d_{2j}^{\dagger})(d_{2j+1}+d_{2j+1}^{\dagger})\nonumber\\
&&+i 2B_{L}\otimes (d_{L-1}-d_{L-1}^{\dagger})(d_{0}-d_{0}^{\dagger})\,.
\end{eqnarray}
It is now useful to decompose $H_{SB}$ into three physically relevant terms:

\noindent
\textbf{Hopping}:
\begin{eqnarray}
&&A_{\text{hopping}}:=-i2B_{1}\otimes d_{0}d_{1}^{\dagger}-2\sum_{j=1}^{\lceil L/2\rceil-1}B_{2j}\otimes d_{2j-1}d_{2j}^{\dagger}\nonumber\\
&&-2\sum_{j=1}^{\lceil L/2\rceil-1}B_{2j+1}\otimes d_{2j}d_{2j+1}^{\dagger}-i2B_{L}\otimes d_{L-1}d_{0}^{\dagger}+\text{h.c.}\nonumber\\
\end{eqnarray}
\textbf{Pair creation:}
\begin{eqnarray}
&&A_{\text{creation}}:=\nonumber\\
&&-i2B_{1}\otimes d_{0}^{\dagger}d_{1}^{\dagger}+2\sum_{j=1}^{\lceil L/2\rceil-1}(B_{2j}\otimes d_{2j-1}^{\dagger}d_{2j}^{\dagger}+B_{2j+1}\otimes d_{2j}^{\dagger}d_{2j+1}^{\dagger})\nonumber\\
&&+i2B_{L}\otimes d_{L-1}^{\dagger}d_{0}^{\dagger}\nonumber\\
\end{eqnarray}
\textbf{Pair annihilation}
\begin{eqnarray}
&&A_{\text{annihilation}}:=\nonumber\\
&&i2B_{1}\otimes d_{1}d_{0}+2\sum_{j=1}^{\lceil L/2\rceil-1}(B_{2j}\otimes d_{2j}d_{2j-1}+B_{2j+1}\otimes d_{2j+1}d_{2j})\nonumber\\
&&-i2B_{L}\otimes d_{0}d_{L-1}
\end{eqnarray}
Following the Davies prescription, we calculate the Fourier transforms of the above operators by first writing their time evolution with respect to $H_{S}$,
\begin{eqnarray}
&&e^{i H_{S} t }\,A_{\zeta}\,e^{-i H_{S}t}\nonumber\\
&=&\sum_{m,n,k,\ell}\vert m\rangle\langle m\vert e^{i H_{S} t } \vert k\rangle\langle k\vert A_{\zeta}\vert \ell\rangle\langle \ell \vert e^{-i H_{S}t}\vert n\rangle\langle n \vert\nonumber\\
&=&\sum_{m,n}e^{it(\epsilon_{m}-\epsilon_{n})}\vert m\rangle\langle m\vert A_{\zeta}\vert n\rangle\langle n \vert\,,
\end{eqnarray}
where $\vert i\rangle$ is an eigenbasis of $H_{S}$ with eigenenergies $\epsilon_{i}$ and $\zeta\in\{\text{hopping, creation, annihilation}\}$. We have used the decomposition of unity $1=\sum_{m}\vert m\rangle\langle m\vert$. The Fourier components of $A_{\zeta}$ are then simply given by
\begin{equation}
A_{\zeta}(\omega)=\sum_{\epsilon_{m}-\epsilon_{n}=\omega}\vert m\rangle\langle m\vert A_{\zeta}\vert n\rangle\langle n \vert\,.
\end{equation}
We can then easily identify the following relevant system operators:

\noindent
\textbf{Hopping terms:}
\begin{eqnarray}
&&A_{\text{hopping}}^{1}(-2\vert\Delta\vert)=\sum_{\epsilon_{m}-\epsilon_{n}=-2\vert\Delta\vert}\vert m\rangle\langle m\vert 2 d_{0}d_{1}^{\dagger}\vert n\rangle\langle n \vert\,,\nonumber\\
&&A_{\text{hopping}}^{L}(-2\vert\Delta\vert)=\sum_{\epsilon_{m}-\epsilon_{n}=-2\vert\Delta\vert}\vert m\rangle\langle m\vert 2 d_{0}d_{L-1}^{\dagger}\vert n\rangle\langle n\vert\,,\nonumber\\
&&A_{\text{hopping}}^{1}(2\vert\Delta\vert)=\sum_{\epsilon_{m}-\epsilon_{n}=2\vert\Delta\vert}\vert m\rangle\langle m\vert 2 d_{1}d_{0}^{\dagger}\vert n\rangle\langle n\vert\,,\nonumber\\
&&A_{\text{hopping}}^{L}(2\vert\Delta\vert)=\sum_{\epsilon_{m}-\epsilon_{n}=2\vert\Delta\vert}\vert m\rangle\langle m\vert 2 d_{L-1}d_{0}^{\dagger}\vert n\rangle\langle n\vert\,,\nonumber\\
&&A_{\text{hopping}}^{2j}(0)=\sum_{\epsilon_{m}-\epsilon_{n}=0}\vert m\rangle\langle m\vert 2 d_{2j-1}d_{2j}^{\dagger}+2 d_{2j}d_{2j-1}^{\dagger}\vert n\rangle\langle n\vert\,,\nonumber\\
&&A_{\text{hopping}}^{2j+1}(0)=\sum_{\epsilon_{m}-\epsilon_{n}=0}\vert m\rangle\langle m\vert 2 d_{2j}d_{2j+1}^{\dagger}+ 2 d_{2j+1}d_{2j}^{\dagger}\vert n\rangle\langle n\vert\,,\nonumber\\
\end{eqnarray}
with $j=1,\dots,\lceil L/2\rceil-1$.

\noindent
\textbf{Creation terms:}
\begin{eqnarray}
&&A_{\text{creation}}^{1}(-2\vert\Delta\vert)=\sum_{\epsilon_{m}-\epsilon_{n}=-2\vert\Delta\vert}\vert m\rangle\langle m\vert 2d_{0}^{\dagger}d_{1}^{\dagger}\vert n\rangle\langle n\vert\,,\nonumber\\
&&A_{\text{creation}}^{L}(-2\vert\Delta\vert)=\sum_{\epsilon_{m}-\epsilon_{n}=-2\vert\Delta\vert}\vert m\rangle\langle m\vert 2d_{0}^{\dagger}d_{L-1}^{\dagger}\vert n\rangle\langle n\vert\,,\nonumber\\
&&A_{\text{creation}}^{2j}(-4\vert\Delta\vert)=\sum_{\epsilon_{m}-\epsilon_{n}=-4\vert\Delta\vert}\vert m\rangle\langle m\vert 2d_{2j-1}^{\dagger}d_{2j}^{\dagger}\vert n\rangle\langle n\vert\,,\nonumber\\
&&A_{\text{creation}}^{2j+1}(-4\vert\Delta\vert)=\sum_{\epsilon_{m}-\epsilon_{n}=-4\vert\Delta\vert}\vert m\rangle\langle m\vert 2d_{2j}^{\dagger}d_{2j+1}^{\dagger}\vert n\rangle\langle n\vert\,,\nonumber\\
\end{eqnarray}

\noindent
\textbf{Annihilation terms:}
\begin{eqnarray}
&&A_{\text{annihilation}}^{1}(2\vert\Delta\vert)=\sum_{\epsilon_{m}-\epsilon_{n}=2\vert\Delta\vert}\vert m\rangle\langle m\vert 2d_{1}d_{0} \vert n\rangle\langle n\vert\,,\nonumber\\
&&A_{\text{annihlation}}^{L}(2\vert\Delta\vert)=\sum_{\epsilon_{m}-\epsilon_{n}=2\vert\Delta\vert}\vert m\rangle\langle m\vert 2d_{L-1}d_{0}\vert n\rangle\langle n\vert\,,\nonumber\\
&&A_{\text{annihilation}}^{2j}(4\vert\Delta\vert)=\sum_{\epsilon_{m}-\epsilon_{n}=4\vert\Delta\vert}\vert m\rangle\langle m\vert 2d_{2j}d_{2j-1}\vert n\rangle\langle n\vert\,,\nonumber\\
&&A_{\text{annihilation}}^{2j+1}(4\vert\Delta\vert)=\sum_{\epsilon_{m}-\epsilon_{n}=4\vert\Delta\vert}\vert m\rangle\langle m\vert 2d_{2j+1}d_{2j} \vert n\rangle\langle n\vert\,,\nonumber\\
\end{eqnarray}
with $j=1,\dots,\lceil L/2\rceil-1$.

In the box representation, see Fig.~\ref{fig:Processes_Time_Ind}, the terms with energy argument $\pm2\vert\Delta\vert$ correspond to hopping at the ends of the wire, where the $\psi$ hops out from the MBS  to inside the neighboring box or vice versa, or to a process at the boundaries where a $\psi$ is created or annihilated inside the MBS  in conjunction with a second $\psi$ at the neighboring box inside the bulk of the wire. The terms with energy argument $0$ are associated with hopping processes inside the bulk of the wire, where a $\psi$ hops from one box into another one without energy cost. Finally, the terms with energy argument $\pm 4\vert\Delta\vert$ correspond to processes where a pair of excitations is created or annihilated in the bulk of the wire.

\subsection{Pauli Master Equation}\label{app:Pauli}
We present a proof that the diagonal terms decouple from the off-diagonal terms in the master equation (\ref{eq:MasEq}), leading to the Pauli master equation (\ref{eq:Populations}) for populations.

Consider the eigenstates $\vert n_{i}\rangle$ of $H_{S}$ such that $H_{S}\vert n_{i}\rangle=\epsilon_{n}\vert n_{i}\rangle$. Here $i=1,2,\ldots$ indexes the degeneracy of the $\epsilon_{n}$ energy level.  

\textbf{Assumption}:
If $\langle m_{\alpha}\vert A_{\eta}^{i}(\omega)\vert n_{k}\rangle\neq0$, (here $\epsilon_{m}-\epsilon_{n}=\omega$) then there are \emph{no other system operators} $A_{\eta^{\prime}}^{i^{\prime}}$ that cause a transition between $\vert n_{k}\rangle$ and any of the degenerate states $\vert m_{\beta}\rangle$ with energy $\epsilon_{m}$. 

Note that it is straightforward to see that the above assumption is satisfied in our case. We thus have,
\begin{widetext}
\begin{eqnarray}
&&\langle n_{k}\vert \mathcal{D}(\rho_{S})\vert n_{k}\rangle=
\sum_{i,j,\eta,\eta^{\prime}}\sum_{\omega}\gamma^{ij}(\omega)\left(\langle n_{k}\vert A_{\eta}^{i}(\omega)\rho_{S}A_{\eta^{\prime}}^{j}(\omega)^{\dagger}\vert n_{k}\rangle-\frac{1}{2}\langle n_{k}\vert A_{\eta^{\prime}}^{j}(\omega)^{\dagger}A_{\eta}^{i}(\omega)\rho_{S}\vert n_{k}\rangle-\frac{1}{2}\langle n_{k}\vert \rho_{S}A_{\eta^{\prime}}^{j}(\omega)^{\dagger}A_{\eta}^{i}(\omega)\vert n_{k}\rangle\right)\nonumber\\
&=&\sum_{\omega,i,j,\eta,\eta^{\prime}}\gamma^{ij}(\omega)\left(\sum_{m,\kappa,\alpha,\beta}\langle n_{k}\vert A_{\eta}^{i}(\omega)\vert m_{\alpha}\rangle\langle m_{\alpha}\vert\rho_{S}\vert \kappa_{\beta}\rangle\langle \kappa_{\beta}\vert A_{\eta^{\prime}}^{j}(\omega)^{\dagger}\vert n_{k}\rangle-\sum_{\ell,u,\gamma,\xi}\frac{1}{2}\langle n_{k}\vert A_{\eta^{\prime}}^{j}(\omega)^{\dagger}\vert\ell_{\gamma}\rangle\langle\ell_{\gamma}\vert A_{\eta}^{i}(\omega)\vert u_{\xi}\rangle\langle u_{\xi}\vert\rho_{S}\vert n_{k}\rangle\right)\nonumber\\
&&-\frac{1}{2}\sum_{\omega,i,j,\eta,\eta^{\prime}}\sum_{r,s}\sum_{\delta,\xi}\gamma^{ij}(\omega)\langle n_{k}\vert \rho_{S}\vert s_{\delta}\rangle\langle s_{\delta}\vert A_{\eta^{\prime}}^{j}(\omega)^{\dagger}\vert r_{\xi}\rangle\langle r_{\xi}\vert A_{\eta}^{i}(\omega)\vert n_{k}\rangle\label{eq:jlp}\\
&=&\sum_{\omega,i,\eta}\gamma^{ii}(\omega)\langle n_{k}\vert A_{\eta}^{i}(\omega)\vert m_{\alpha}(i,\eta)\rangle\langle m_{\alpha}(i,\eta)\vert\rho_{S}\vert m_{\alpha}(i,\eta)\rangle\langle m_{\alpha}(i,\eta)\vert A_{\eta}^{i}(\omega)^{\dagger}\vert n_{k}\rangle\nonumber\\
&&-\frac{1}{2}\sum_{\omega,i,\eta}\gamma^{ii}(\omega)\langle n_{k}\vert(A_{\eta}^{i}(\omega)^{\dagger}\vert m_{\alpha}(i,\eta)\rangle\langle m_{\alpha}(i,\eta)\vert A_{\eta}^{i}(\omega)\vert n_{k}\rangle\langle n_{k}\vert\rho_{S}\vert n_{k}\rangle\nonumber\\
&&-\frac{1}{2}\sum_{\omega,i,\eta}\gamma^{ii}(\omega)\langle n_{k}\vert \rho_{S}\vert n_{k}\rangle\langle n_{k}\vert A_{\eta}^{i}(\omega)^{\dagger}\vert m_{\alpha}(i,\eta)\rangle\langle m_{\alpha}(i,\eta)\vert A_{\eta}^{i}(\omega)\vert n_{k}\rangle\,.
\end{eqnarray}
\end{widetext}
The Pauli master equation (\ref{eq:Populations}) follows then directly.

Here we have used the orthonormality relation $\langle n_{\alpha}\vert m_{\beta}\rangle=\delta_{mn}\delta_{\alpha\beta}$. The state $\vert m_{\alpha}\rangle$ in the sums (\ref{eq:jlp}) that has non vanishing matrix element $\langle m_{\alpha}\vert A_{\eta}^{i}(\omega)\vert n_{k}\rangle$ depends on the system operator $A_{\eta}^{i}(\omega)$. For the sake of clarity we have thus introduced the notation $\vert m_{\alpha}\rangle\equiv\vert m_{\alpha}(i,\eta)\rangle$.

\section{Error Processes: Time-dependent Case}\label{sec:DaviesTime}
The calculation of the rates of the error processes happening during the motion of MBSs  must be performed with great care. In the following we present a detailed analysis of two cases. All the remaining cases shown in Ref.~\onlinecite{Supplement} are treated similarly.
\subsection{Linear case}
\begin{figure}[h!]
	\centering
		\includegraphics[width=0.45\textwidth]{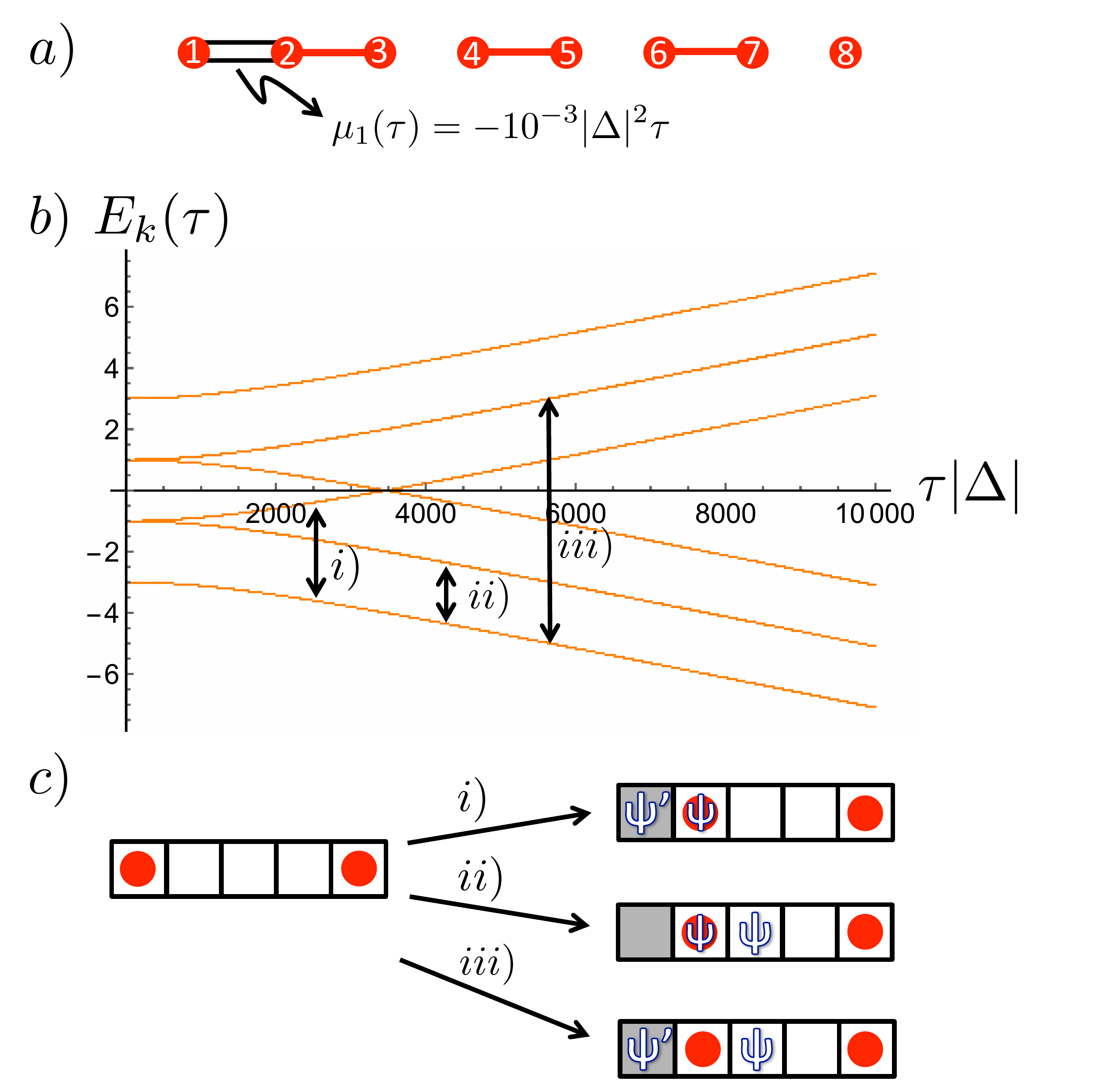}
	\caption{$a)$ Pictorial representation of the four-site model; the dots represent the Majorana modes and the single lines in between represent the pairings. The double lines depict the time-dependent chemical potential on the first site. It varies according to $\mu_{1}(\tau)=-10^{-3}\vert\Delta\vert^{2}\tau$. $b)$ Energy levels $E_{k}(\tau)$ as function of time. The arrows describe the transitions above the vacuum produced by the system operator $\gamma_{1}\gamma_{2}$, inducing transition $i)$, and $\gamma_{3}\gamma_{4}$ inducing transitions $ii)$ and $iii)$. $c)$ Box representation of transitions $i)$, $ii)$, and $iii)$. When the chemical potential is sufficiently negative, then the MBS  has moved to its nearest-neighbor site and a pair of excitations has been produced.}
	\label{fig:Linear_Rates}
\end{figure}

The linear case is easily understood in terms of a four-site model where the chemical potential on the first site is varied according to $\mu_{1}(\tau)=-10^{-3}\vert\Delta\vert^{2} \tau$. The red dots in Fig.~\ref{fig:Linear_Rates}a represent the Majorana modes, while the single solid lines represent the coupling between them in the Hamiltonian. The double solid lines represent the time-dependent chemical potential. We assume that a region is nontopological when the chemical potential satisfies $\vert\mu\vert\gtrsim 10\vert\Delta\vert$. Therefore, we say that the MBS  has moved to a nearest-neighbor site after a time $\tau\vert\Delta\vert\gtrsim 10^{4}$.
The black arrows between different branches of the spectrum in Fig.~\ref{fig:Linear_Rates}b describe transitions caused by the system operators. These transitions are also shown in box representation in Fig.~\ref{fig:Linear_Rates}$c$. 

The time dependence of the problem requires that the rates are obtained by integrating Eq.~(\ref{eq:Wmnt}) over time. The upper integration bound $T$ must be big enough such that the MBS  has moved,  i.e. as mentioned above we choose $T\vert\Delta\vert = 10^{4}$.  We have 
\begin{eqnarray}\label{eq:cnso}
&&W(n\vert m)^{\text{int}}=\frac{1}{T}\int_{0}^{T}d\tau \left\vert\langle m(\tau)\vert A_{1,2}\vert n(\tau)\right\vert\rangle^{2}\gamma(\omega_{mn}(\tau))\,.\nonumber\\
\end{eqnarray}

We calculate numerically the integral (\ref{eq:cnso}) by discretizing the interval $\left[0,T\right]$ and by replacing the integral by a sum.


\subsection{Trijunction}
\begin{figure}[h!]
	\centering
		\includegraphics[width=0.45\textwidth]{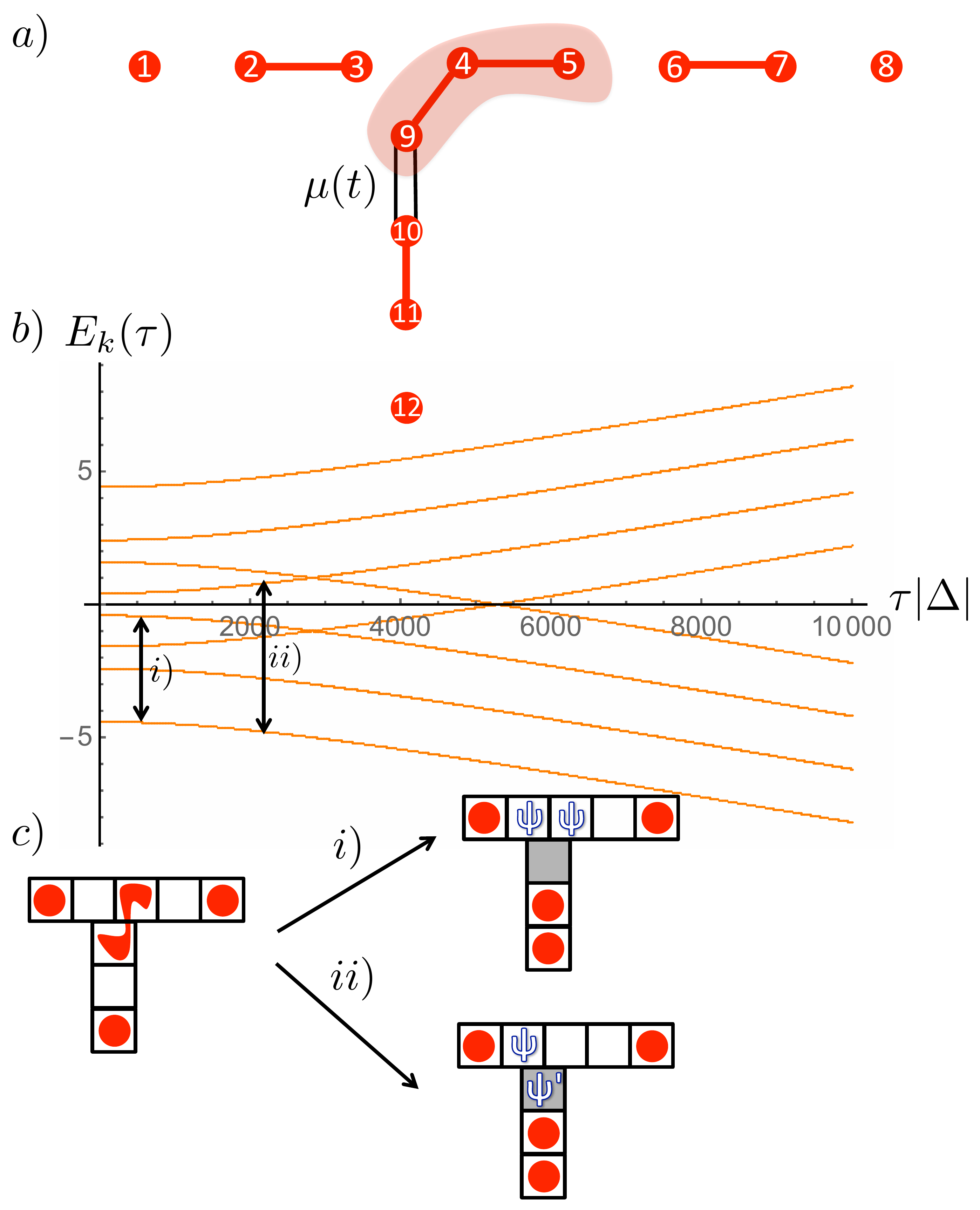}
	\caption{ $a)$ Six-site trijunction. The dots represent Majorana modes and the single lines the interactions between them. The shaded area shows that a Majorana mode is delocalized over the trjiunction. The double lines represent the time-dependent chemical potential varying according to $\mu(\tau)=-10^{-3}\vert\Delta\vert^{2} \tau$. $b)$ Spectrum as function of time. The black arrows represent the transitions above the vacuum induced by the system operator $\gamma_{3}\gamma_{4}$ . $c)$ Box representation of the two transitions $i)$ and $ii)$. The red shape on the left trijunction represents the delocalized MBS  at the trijunction.  Note that the two Majorana modes on the vertical wire are localized on neighboring boxes because we diagonalize here a small trijunction. In a longer trijunction, however, the third box on the vertical wire would correspond to a fermionic mode and the fourth MBSs  would lie at the bottom of the vertical wire. In order to calculate all the rates, it is enough to diagonalize such a small trijunction but one must keep in mind that creating a $\psi$ inside the MBS  at the bottom box of the vertical wire would correspond to creating a bulk $\psi$-excitation in the longer trijunction (and thus would imply some cost of energy).}
	\label{fig:Trijunction_Process}
\end{figure}

The processes that happen at the trijunction are analyzable with a six-site model, see Fig.~\ref{fig:Trijunction_Process}a). In Fig.~\ref{fig:Trijunction_Process}b we show the spectrum as function of time. The black arrows identify three transitions caused by the system operator $\gamma_{3}\gamma_{4}$, see the box representation in Fig.~\ref{fig:Trijunction_Process}c.
Similar to the linear case, the rates are obtained by integrating Eq.~(\ref{eq:Wmnt}).


%

\clearpage



\section*{Supplementary material}

## Supplementary Material to "Monte Carlo studies of the properties of the Majorana quantum error correction code: is self-correction possible during braiding?"

The goal of this supplementary material is to present an *exhaustive* list of the hundreds of processes that are relevant for the Monte Carlo simulation of the trijunction Hamiltonian in Eq. (30) of the main text. We use our box representation to describe all the relevant processes. We present both the unitary evolutions as well as the dissipative error processes that are caused by the system-bath coupling $H_{SB}$, see Eq. (12) of the main text. Of course, we present only the nontrivial unitary evolutions while more unitary evolutions, where a Majorana Bound State (MBS) is moving and excitations are inert, have to be taken into account in the simulations. Compared with the case where MBSs are immobile, many more processes must be taken into account when the MBSs are braided.

For the sake of clarity, we explain the language we will use below: when we say that an MBS is moving over the trijunction point, we have in mind the moment of braiding corresponding to Fig. 1.

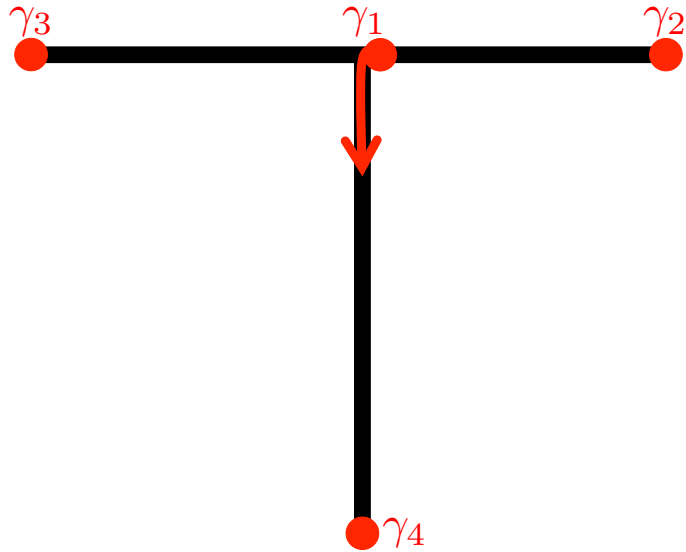

FIG. 1. Pictorial representation of the trijunction setup supporting four MBSs $\gamma_{1,...,4}$. When we say that an MBS is moving over the trijunction, we have in mind the precise moment of the braiding where $\gamma_1$ moves from the horizontal wire into the vertical wire.

We distinguish among the following cases where an error process can happen:

1. Inside a topological segment, away from MBSs.

2. Close to immobile MBSs.

3. Close to mobile MBSs that are moving in the direction of the topological segment. In other words, the considered topological segment becomes shorter.

4. Close to mobile MBSs that are moving in the direction of the nontopological segment. In other words, the considered nontopological segment becomes shorter.

5. Close to a mobile MBS during its motion over the trijunction point.

6. Close to a mobile MBS just before its motion over the trijunction point

In the following tables, the cases enumerated above i.e. 1,2,3,4,5, and 6, are written in the upper left corner. On the top of the table we also mention whether the depicted processes correspond to unitary evolution or to dissipative dynamics. For each case, the first figure, i.e. the one we call *small model*, is a model of the system at the considered time during braiding. We use this model to identify and describe the system operators $\gamma_i \gamma_j$ that are at the origin of the diffusive processes; we write on the left of each diffusive process the system operator causing the transition.

Also, we would like to comment on the meaning of Figs. 2a and b, which illustrate elements occurring in the tables. The red shapes at the trijunction point represent a delocalized Majorana mode. Since the trijunction region is composed of two sites (site $L/2$ on the horizontal wire and site $L+1$ on the vertical wire), it can support two

fermionic modes. Therefore, it is possible that the trijunction hosts a delocalized Majorana and a $\psi$-excitation at the same time, as depicted in Fig. 2b. It is worth pointing out that the Majorana modes contributing to a delocalized MBS cannot be nearest neighbors. For example, the delocalized MBS at the trijunction on page 15 has contributions from Majorana modes 5 and 9 but not from mode 4.

We note that some of the figures in the tables below contain a single excitation ($\psi$ or $\psi'$). However, since the bath is parity-preserving, the total number of excitations cannot be odd. Therefore, when a single excitation is present, one needs to keep in mind that at least another inert *undrawn* excitation must also be present somewhere in the trijunction.

Below the $\psi-$ and $\psi'-$particles are represented in a slightly different manner as compared to the main text. However, this difference is purely graphical and there is no difference at all between the quasi-particles represented below and the ones from the main text.

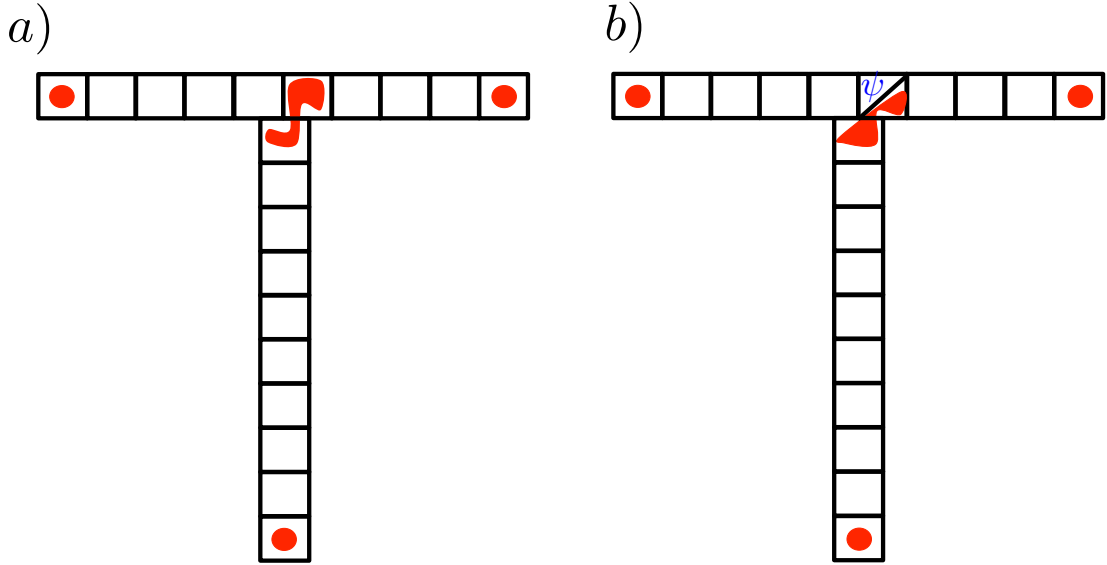

FIG. 2. Pictorial representation of the moment in braiding when the MBS is moving over the trijunction point. The red shapes at the trijunction represent the delocalized MBS. While in *a)* the trijunction point carries only an MBS mode, in *b)* an additional $\psi$-excitation is present.

| Case 1. | Error processes supported by the bath |
|---------|----------------------------------------|
| | 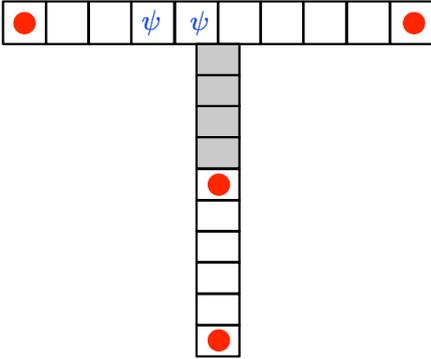 |
| | 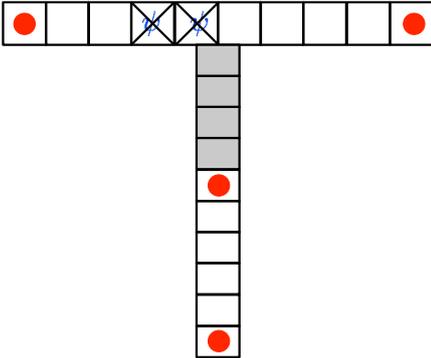 |
| | 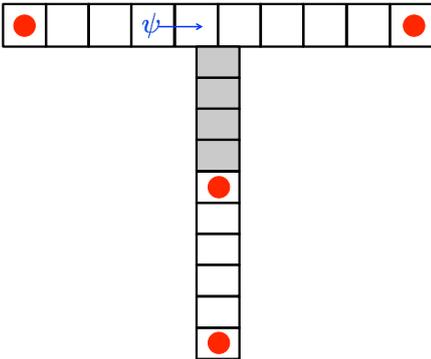 |

| Case 2. | Error processes supported by the bath |
|---|---|

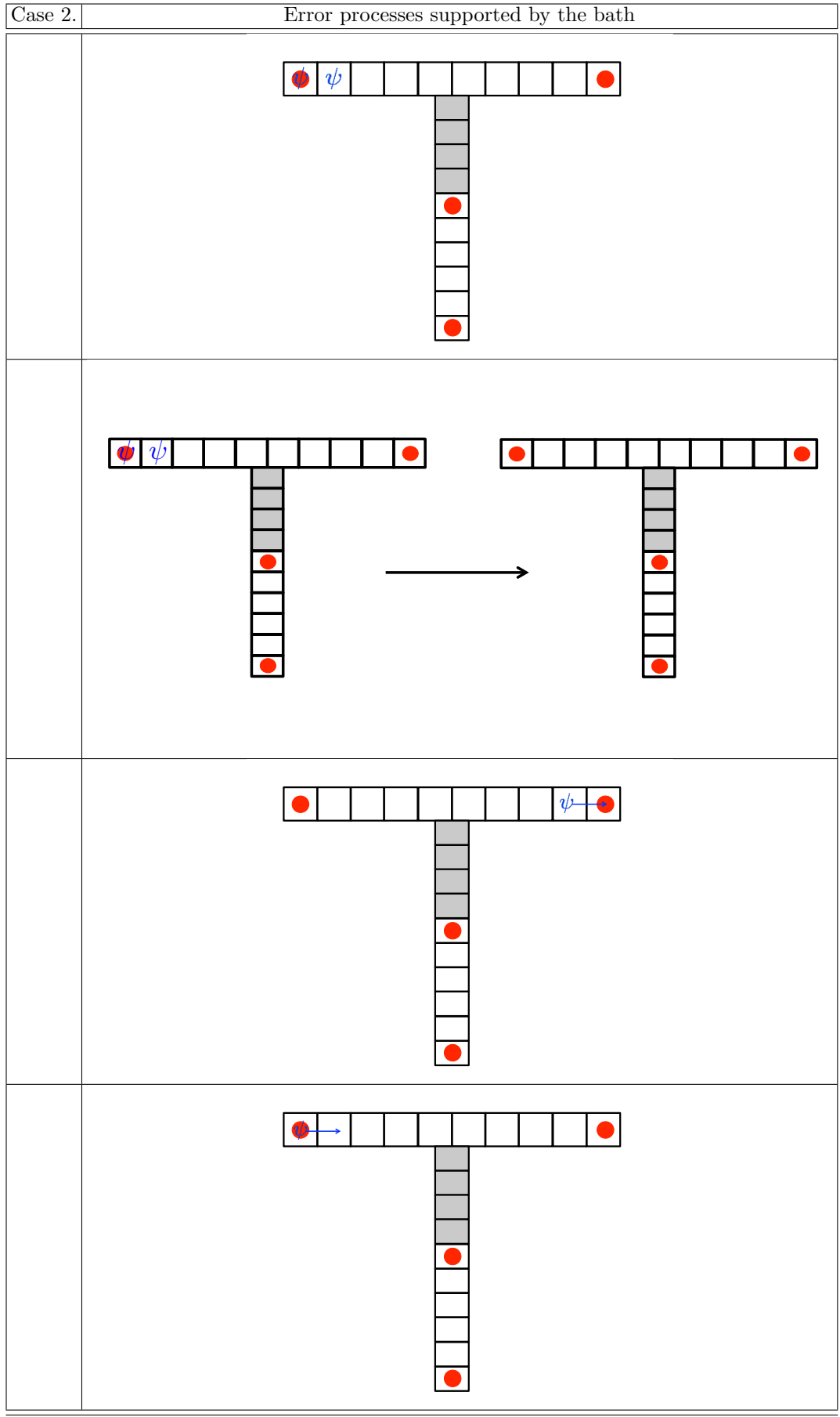

| Case 3. | Four-site model |
|---|---|
| | 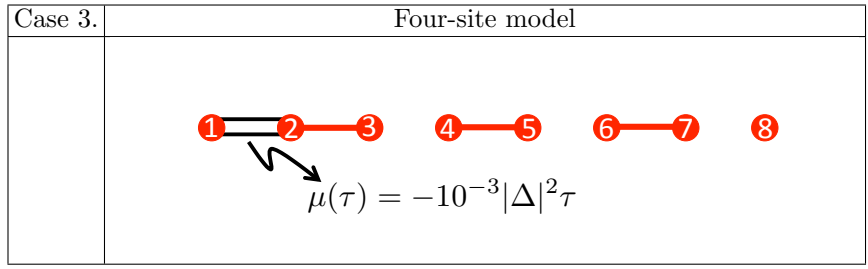 |

| Case 3. | Unitary Evolution |
|---|---|
| | 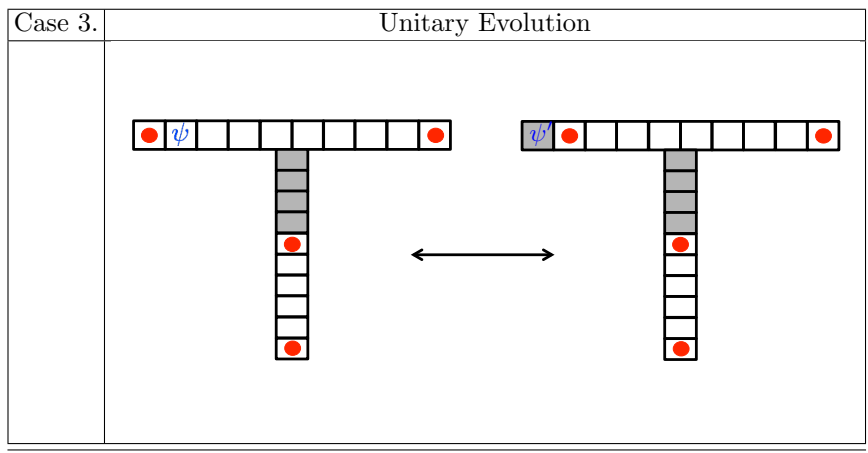 |

| Case 3. | Error processes supported by the bath |
|---------|---------------------------------------|
| $\gamma_1 \gamma_2$ | 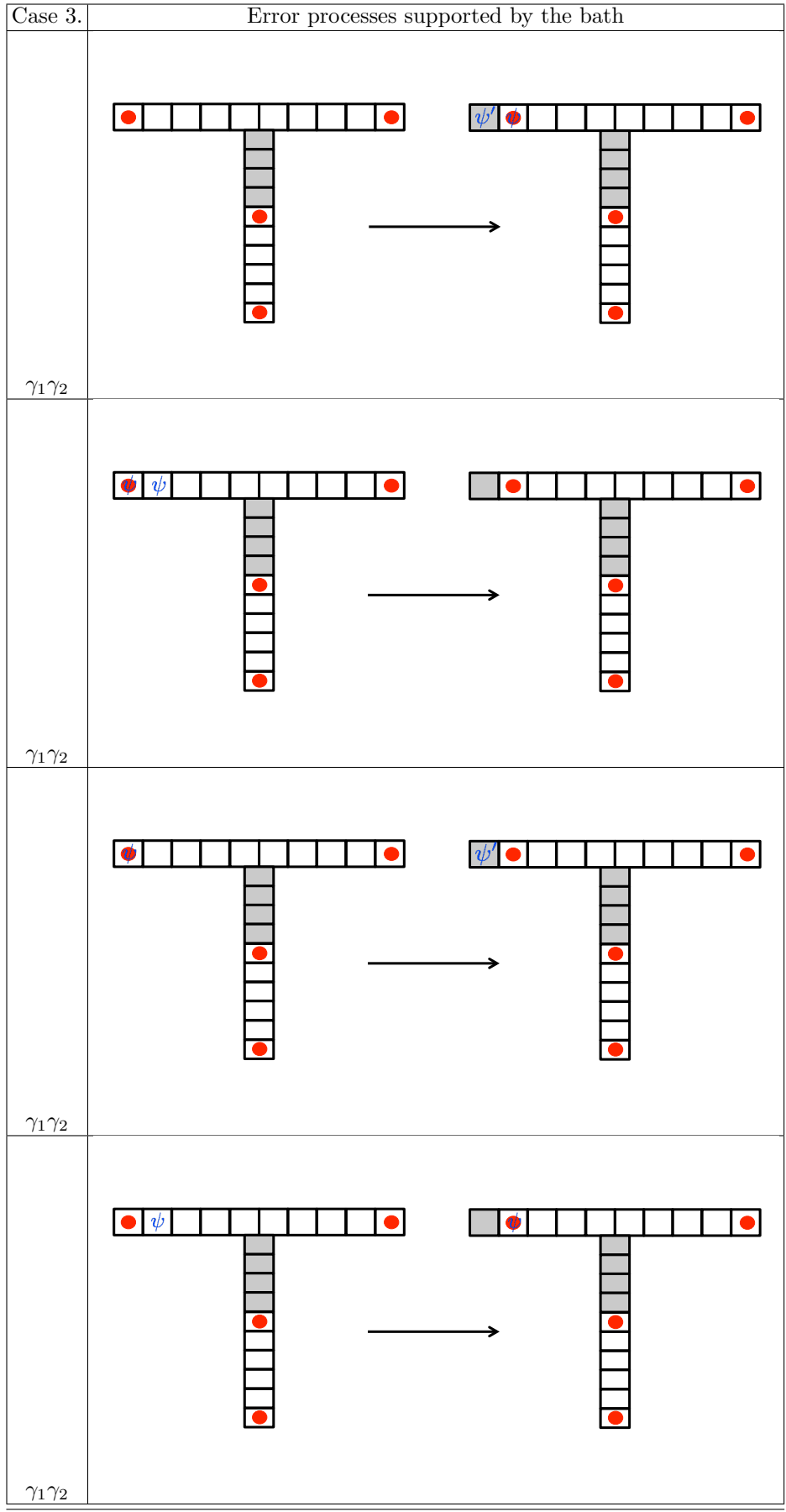 |
| $\gamma_1 \gamma_2$ | |
| $\gamma_1 \gamma_2$ | |
| $\gamma_1 \gamma_2$ | |

7



| Case 3. | Error processes supported by the bath |
|---|---|
| $\gamma_3\gamma_4$ | |
| $\gamma_3\gamma_4$ | |
| $\gamma_3\gamma_4$ | |
| $\gamma_3\gamma_4$ | |

| Case 3. | Error processes supported by the bath |
|---|---|

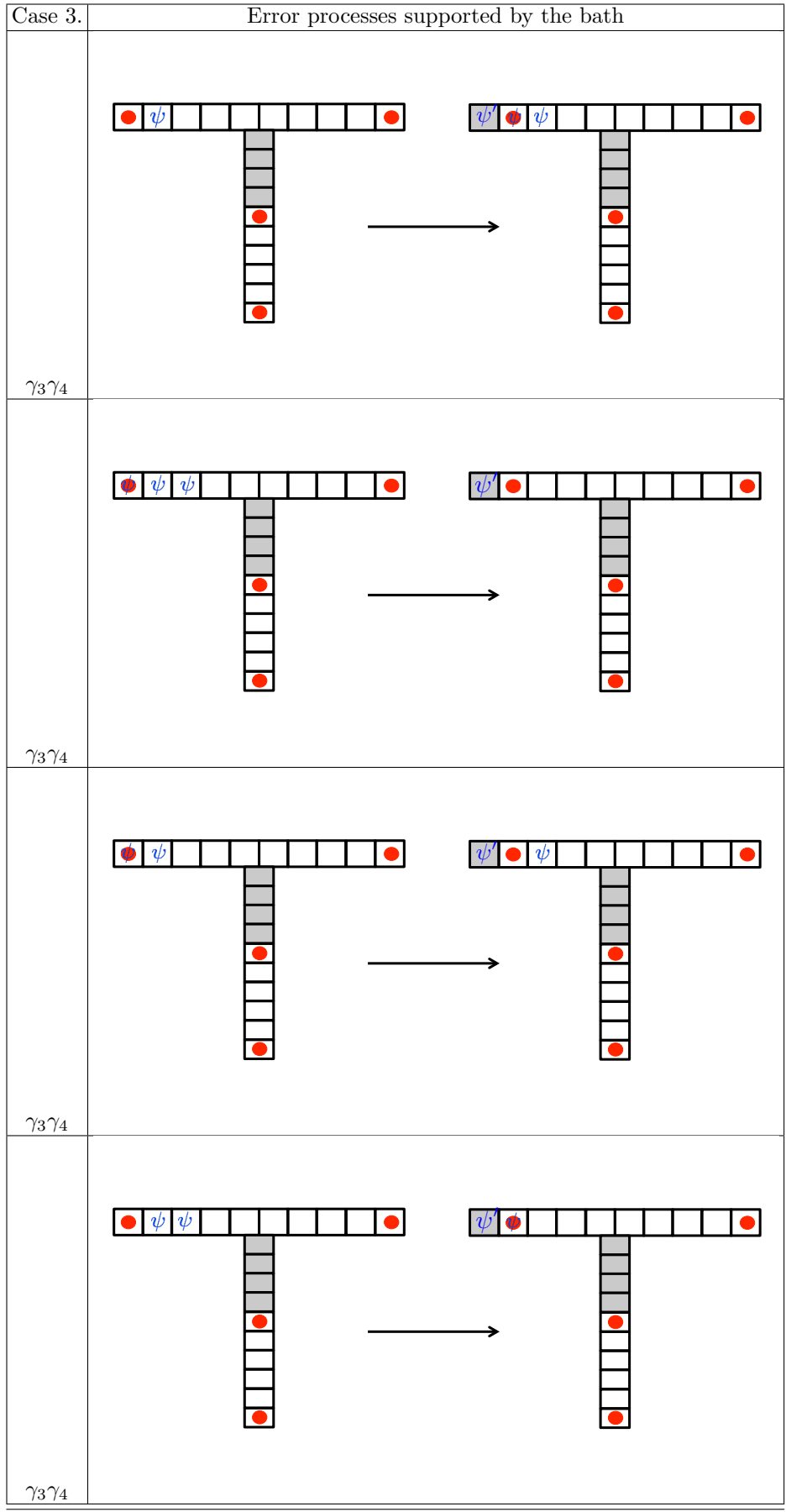

$\gamma_3 \gamma_4$

$\gamma_3 \gamma_4$

$\gamma_3 \gamma_4$

$\gamma_3 \gamma_4$

| Case 3. | Error processes supported by the bath |
|---------|----------------------------------------|
| $\gamma_3\gamma_4$ | 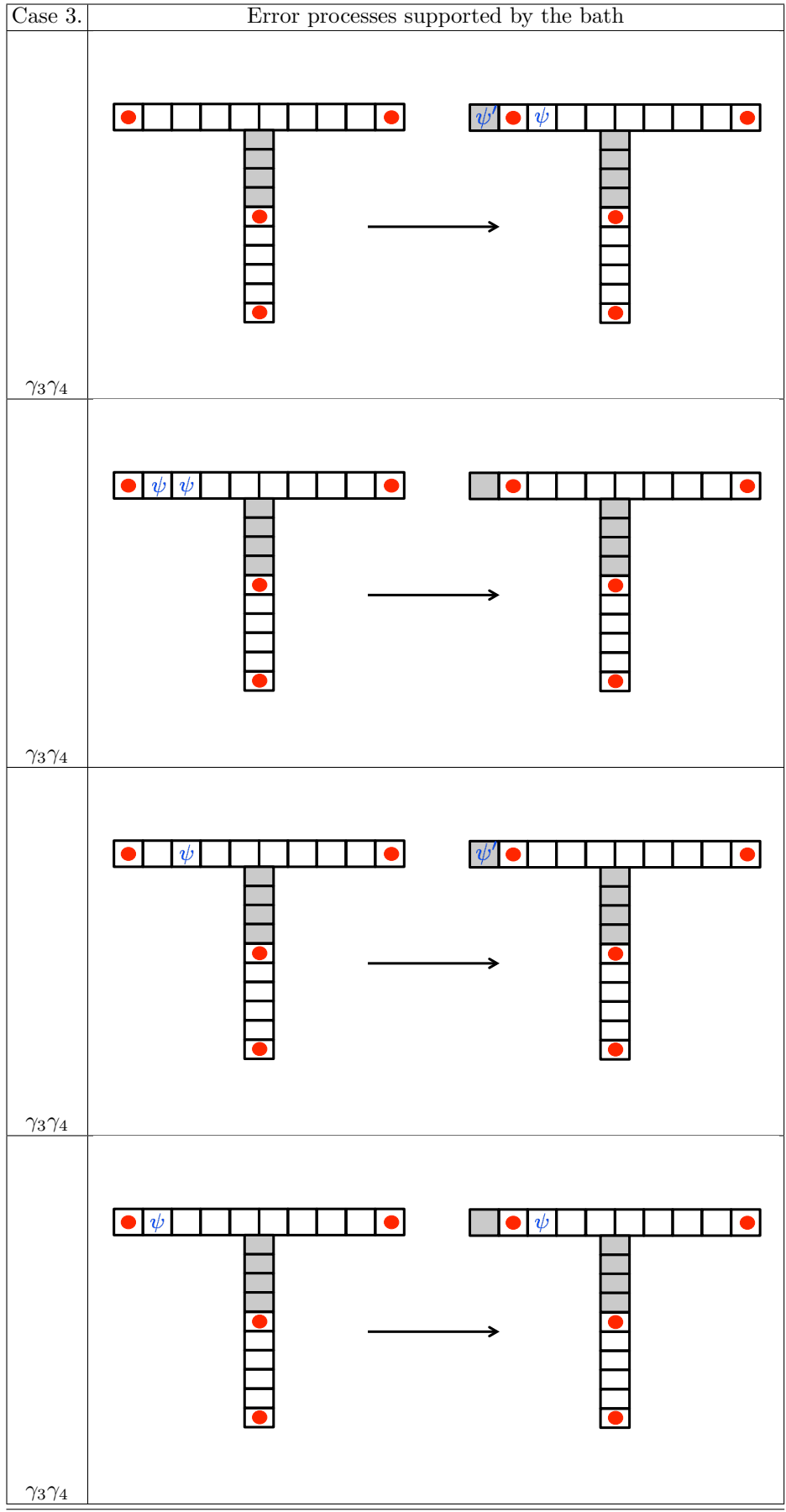 |
| $\gamma_3\gamma_4$ | |
| $\gamma_3\gamma_4$ | |
| $\gamma_3\gamma_4$ | |

| Case 4. | Five-site model |
|---------|-----------------|

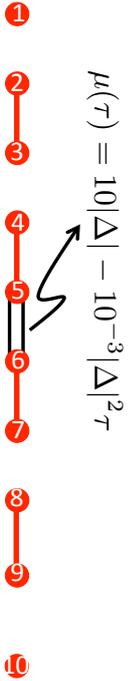

$$\mu(\tau) = 10|\Delta| - 10^{-3}|\Delta|^2\tau$$

| Case 4. | Unitary evolution |
|---------|-------------------|

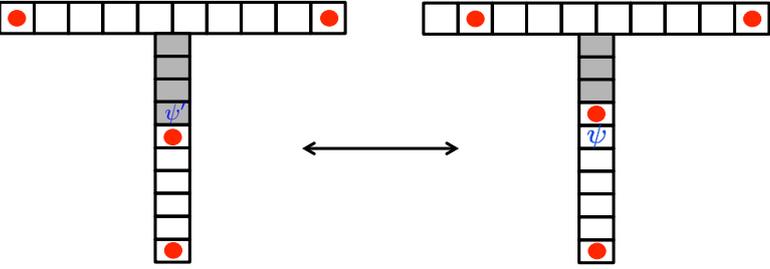

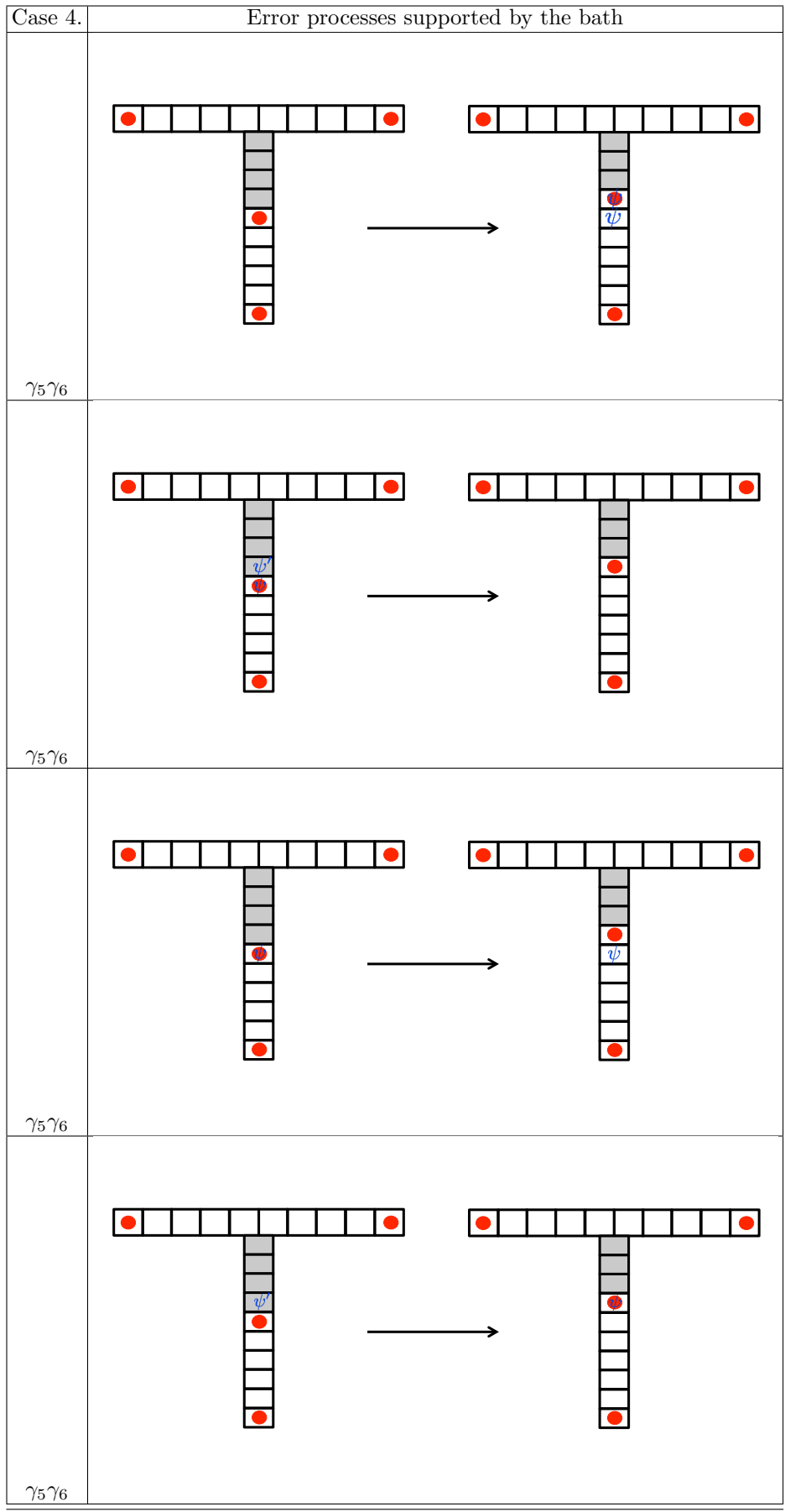

| Case 4. | Error processes supported by the bath |
|---------|---------------------------------------|

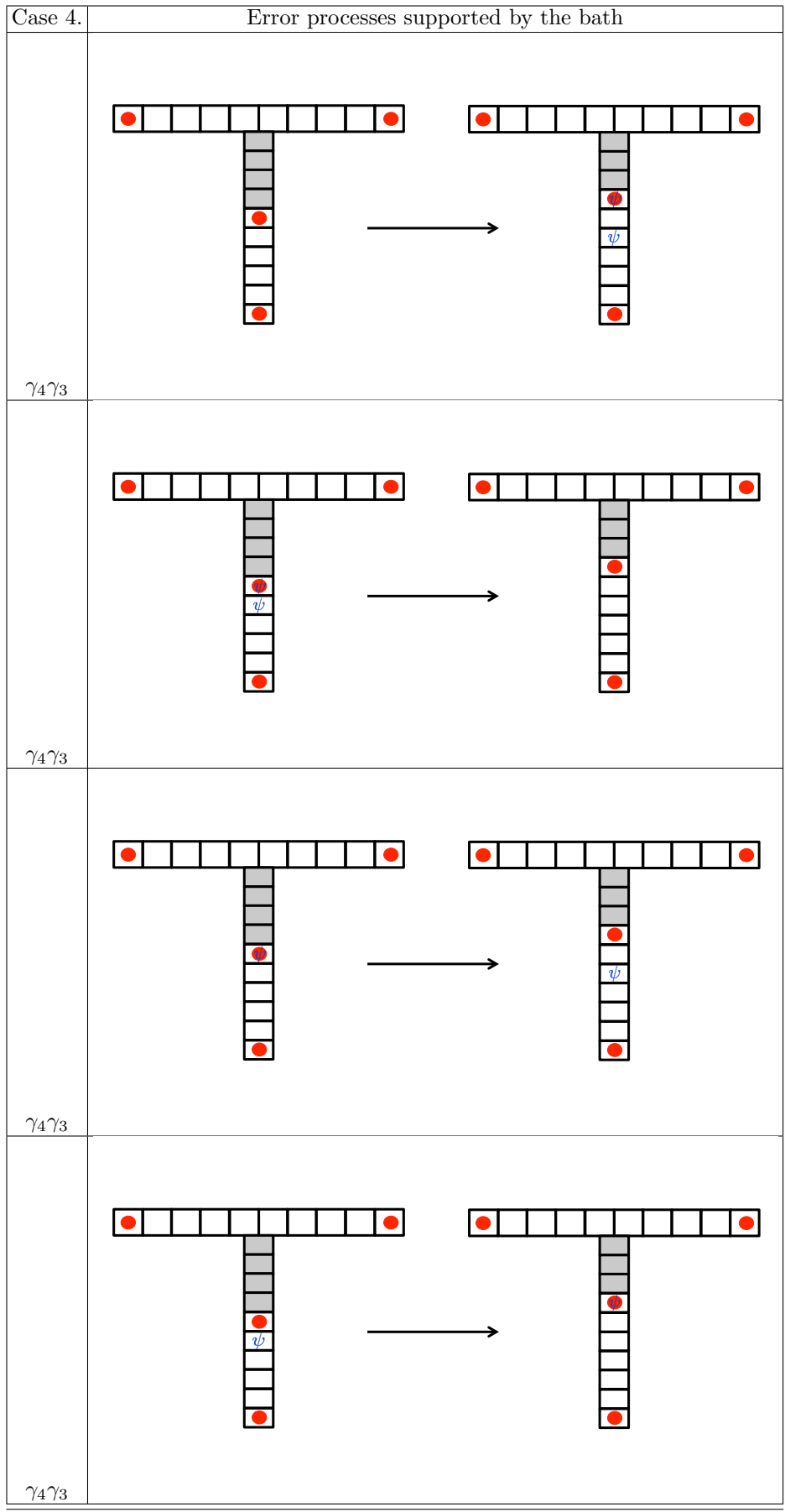

$\gamma_4\gamma_3$

$\gamma_4\gamma_3$

$\gamma_4\gamma_3$

$\gamma_4\gamma_3$

| Case 4. | Error processes supported by the bath |
|---|---|
| 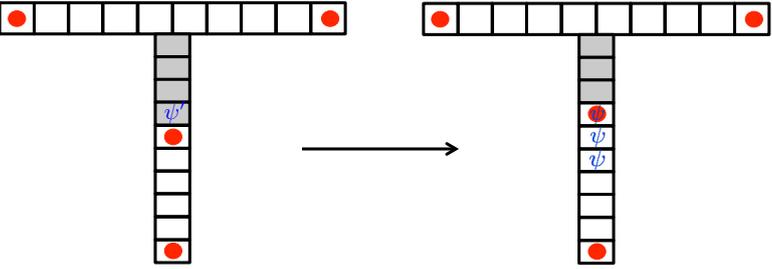 $\gamma_4\gamma_3$ | |
| 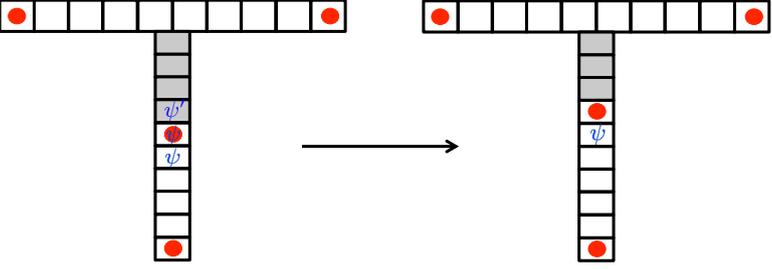 $\gamma_4\gamma_3$ | |
| 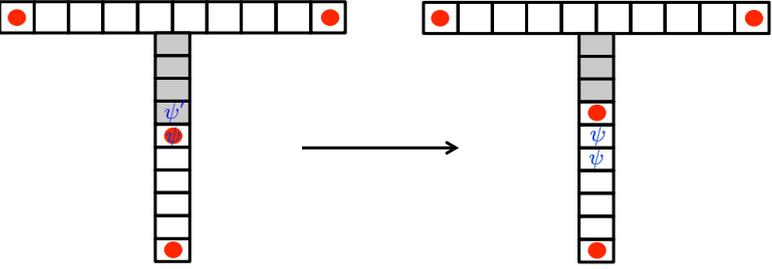 $\gamma_4\gamma_3$ | |
| 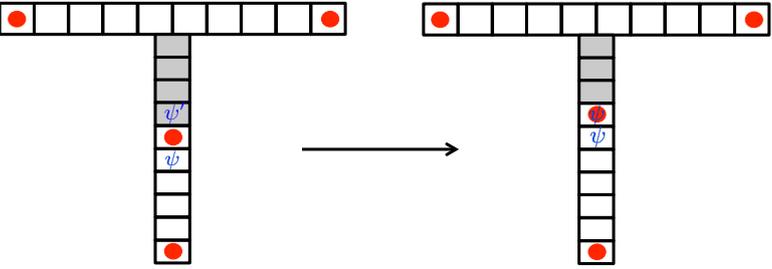 $\gamma_4\gamma_3$ | |

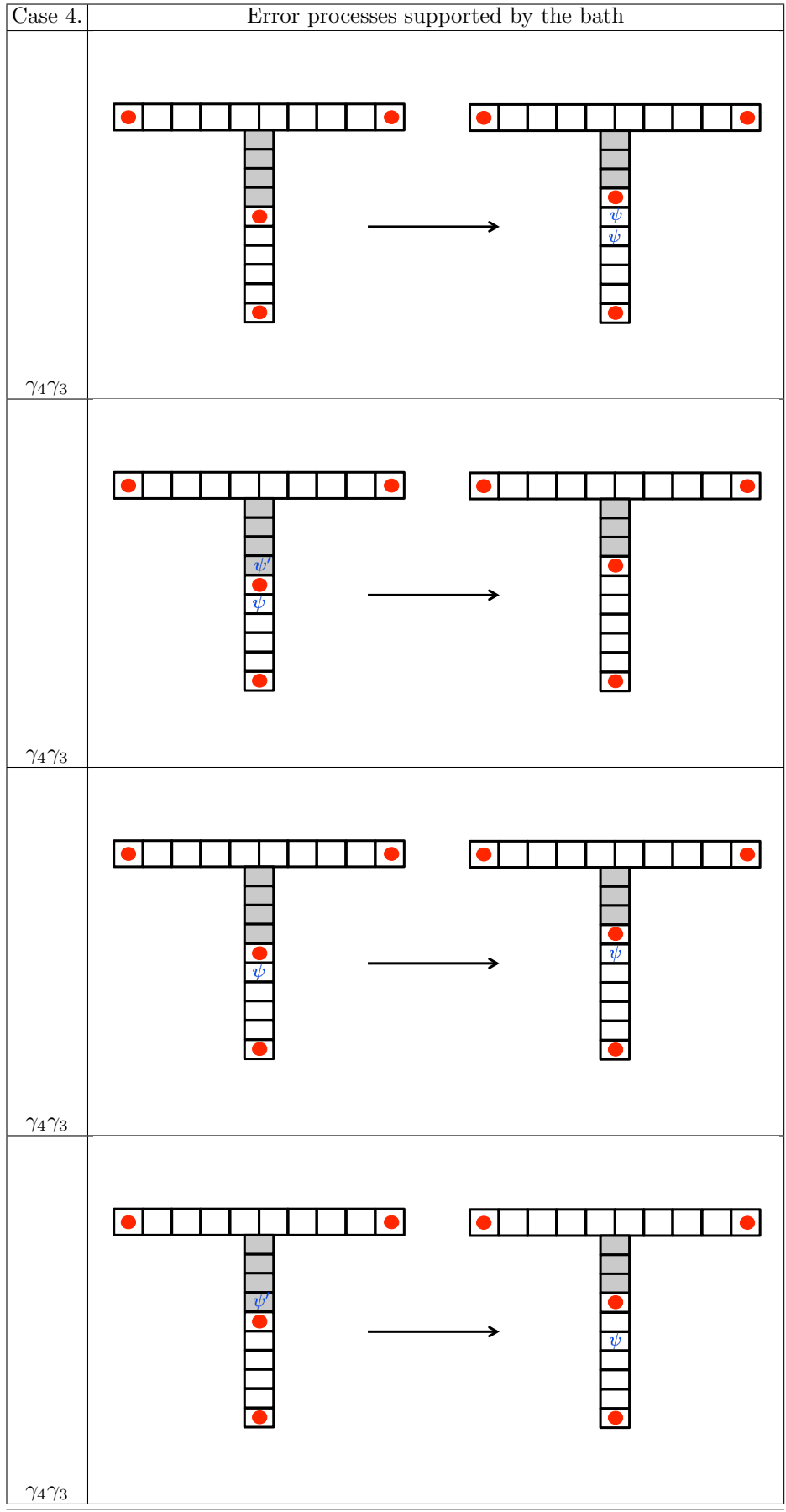

We point out that the following trijunction processes only occur during stage iii) of Fig. 7b in the main text (or Fig. 1 of this supplement). In fact, when the MBS is moving over the trijunction during stage ii) of Fig. 7b in the main text, the error processes are exactly the same as the one occurring in a single (bent) wire.

We recall that the delocalized MBS at the trijunction has contributions only from Majorana modes 5 and 9. Therefore, the system operator $\gamma_3\gamma_4$ that we will consider below cannot create or annihilate a $\psi$ inside the delocalized Majorana at the trijunction point; thus we only depict the case without a $\psi$ inside the Majorana at the trijunction when we study the effect of $\gamma_3\gamma_4$.

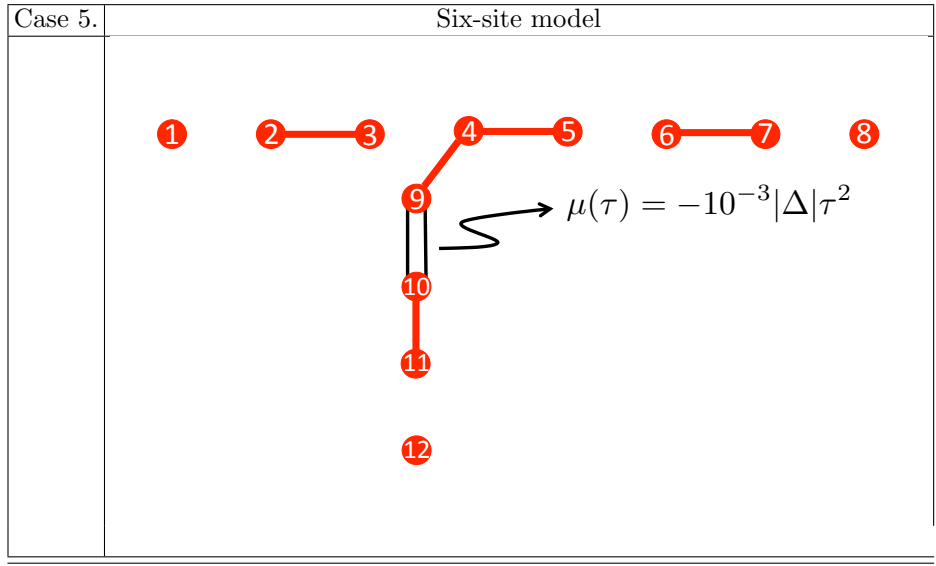

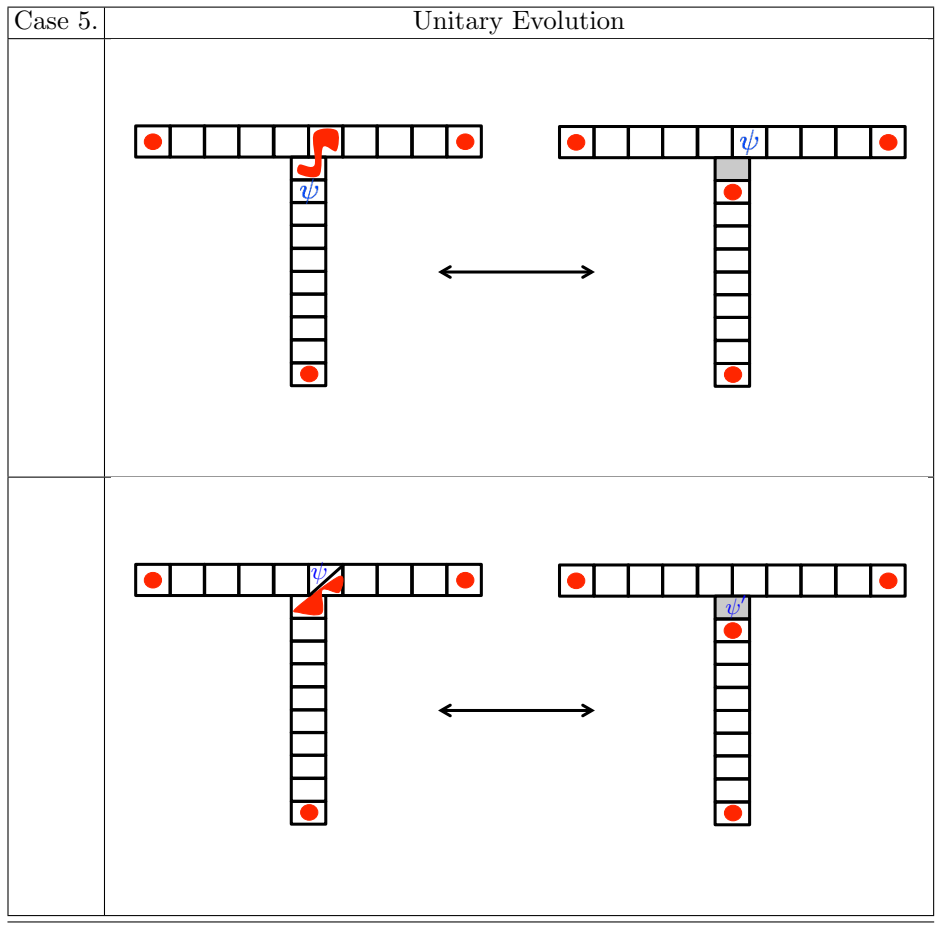

16

| Case 5. | Error processes supported by the bath |
|---|---|
|  $\gamma_3\gamma_4$ | |
|  $\gamma_3\gamma_4$ | |
|  $\gamma_3\gamma_4$ | |
|  $\gamma_3\gamma_4$ | |

| Case 5. | Error processes supported by the bath |
|---|---|
| $\gamma_3\gamma_4$ |  |
| $\gamma_3\gamma_4$ |  |
| $\gamma_3\gamma_4$ |  |
| $\gamma_3\gamma_4$ |  |

18

| Case 5. | Error processes supported by the bath |
|---|---|
| $\gamma_3\gamma_4$ | 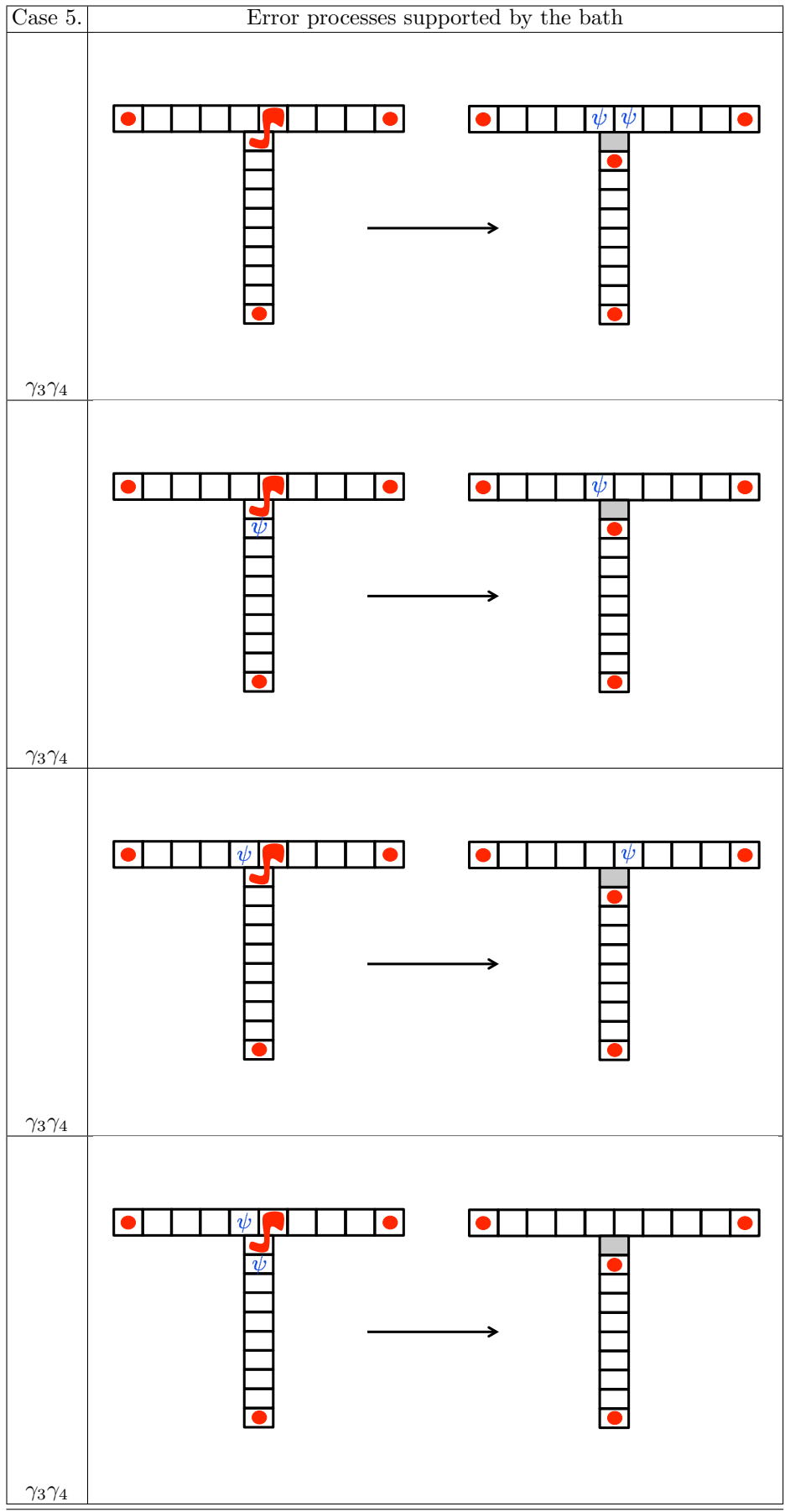 |
| $\gamma_3\gamma_4$ | |
| $\gamma_3\gamma_4$ | |
| $\gamma_3\gamma_4$ | |

| Case 5. | Error processes supported by the bath |
|---------|----------------------------------------|
| $\gamma_3\gamma_4$ |  |
| $\gamma_3\gamma_4$ |  |
| $\gamma_3\gamma_4$ |  |
| $\gamma_3\gamma_4$ |  |

2020Case 5.Error processes supported by the bath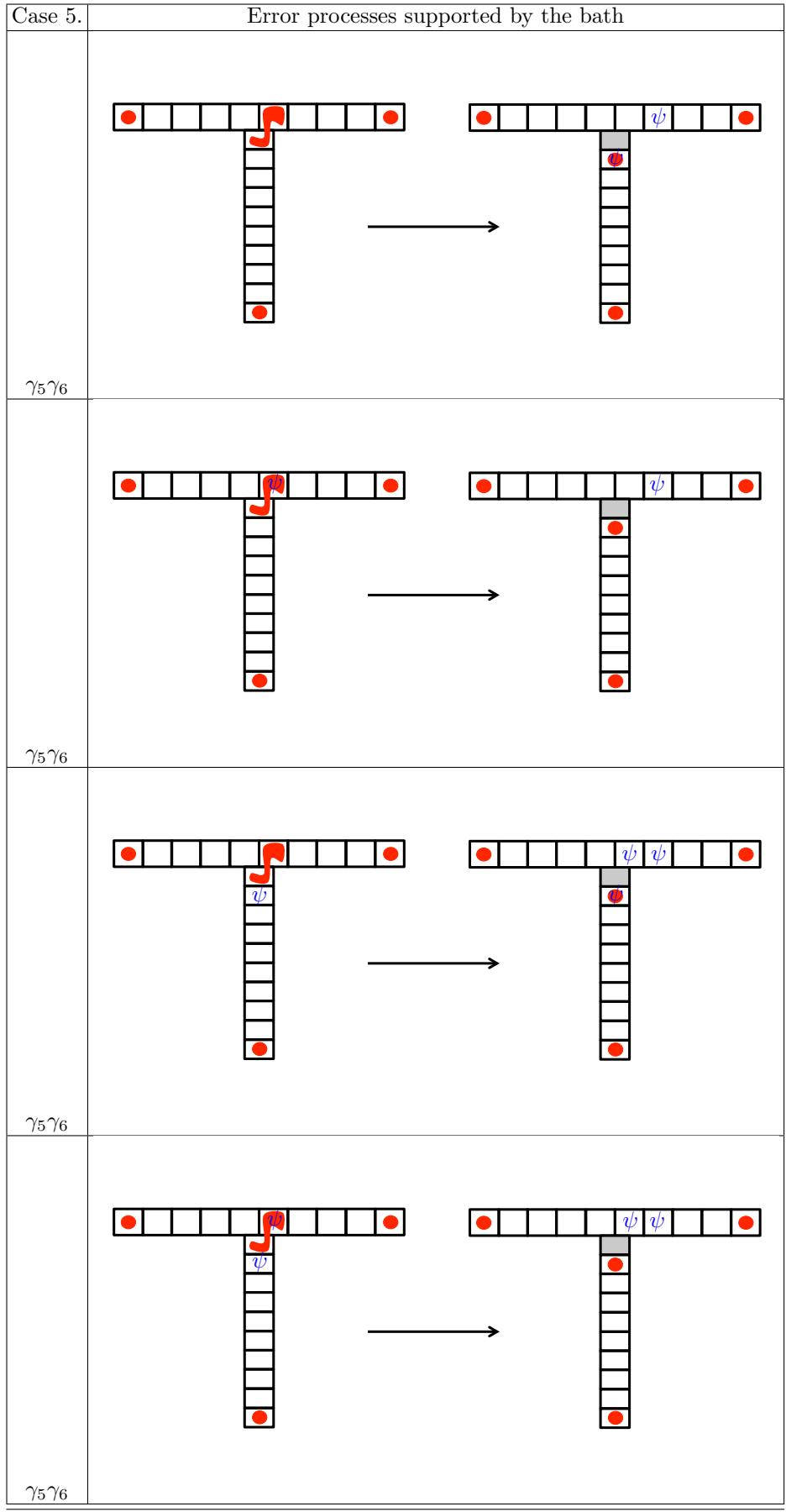

| Case 5. | Error processes supported by the bath |
|---|---|

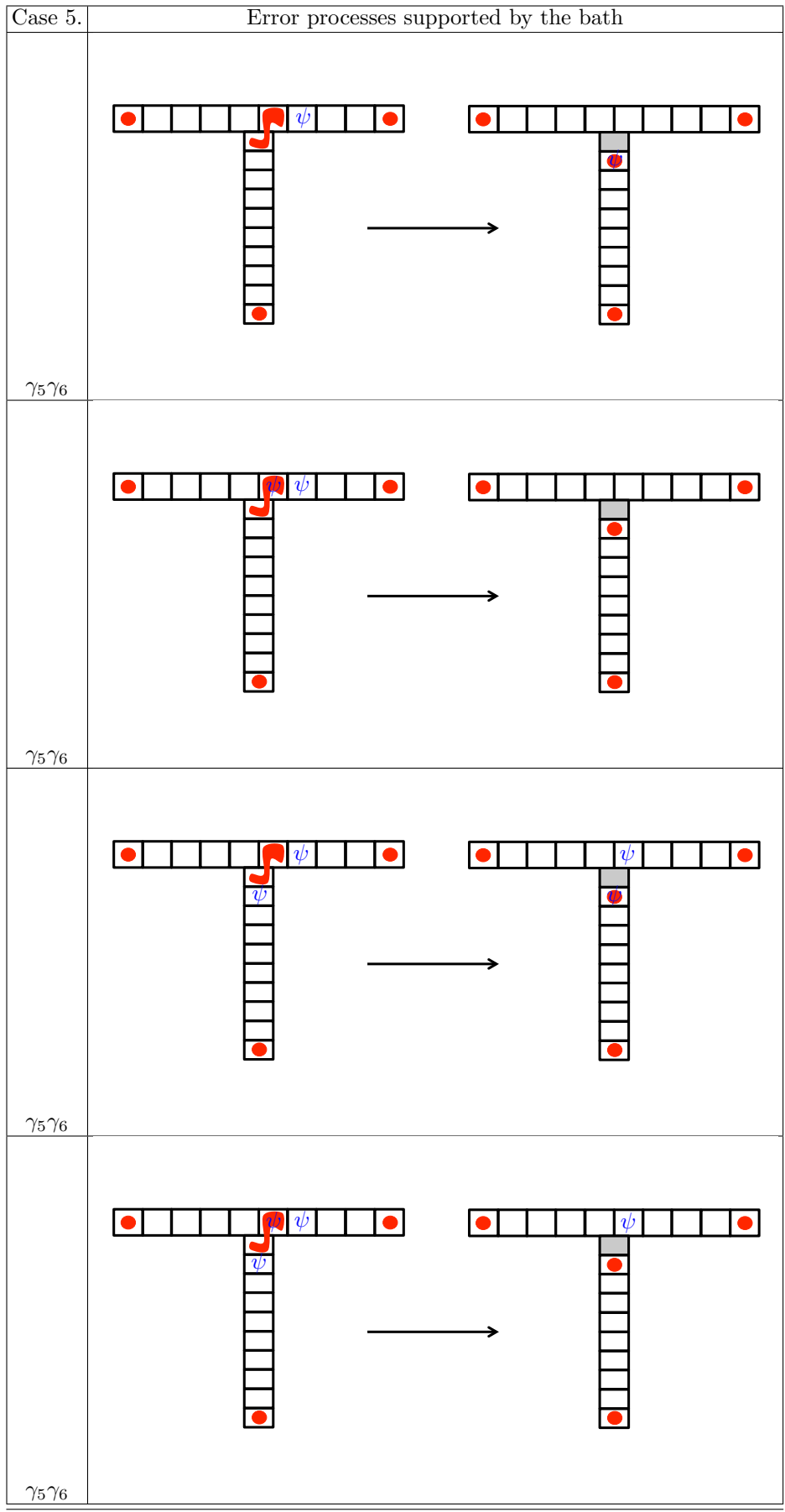

$\gamma_5\gamma_6$

$\gamma_5\gamma_6$

$\gamma_5\gamma_6$

$\gamma_5\gamma_6$

| Case 5. | Error processes supported by the bath |
|---|---|
| | 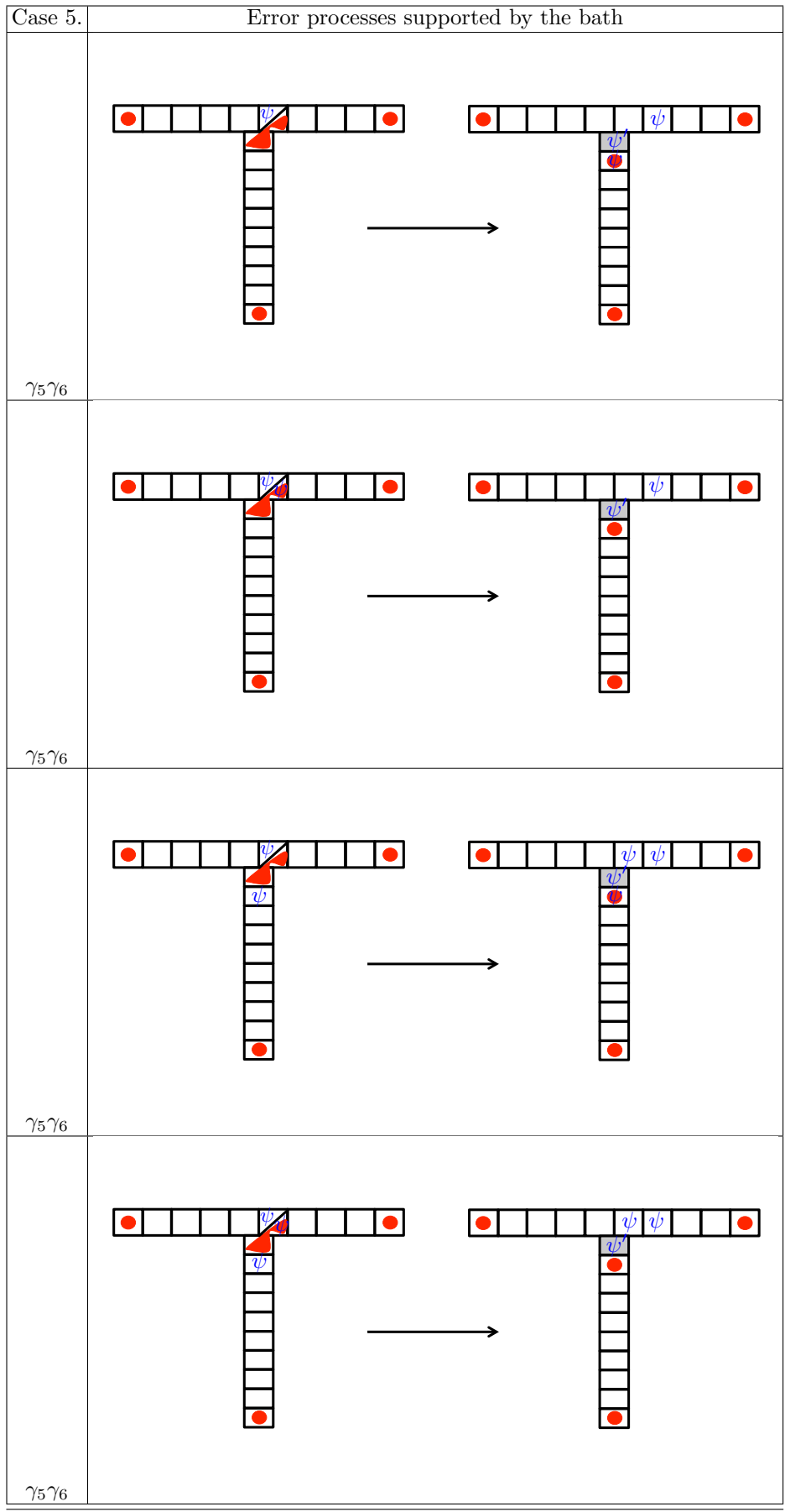 |
| $\gamma_5\gamma_6$ | |
| $\gamma_5\gamma_6$ | |
| $\gamma_5\gamma_6$ | |
| $\gamma_5\gamma_6$ | |

| Case 5. | Error processes supported by the bath |
|---|---|

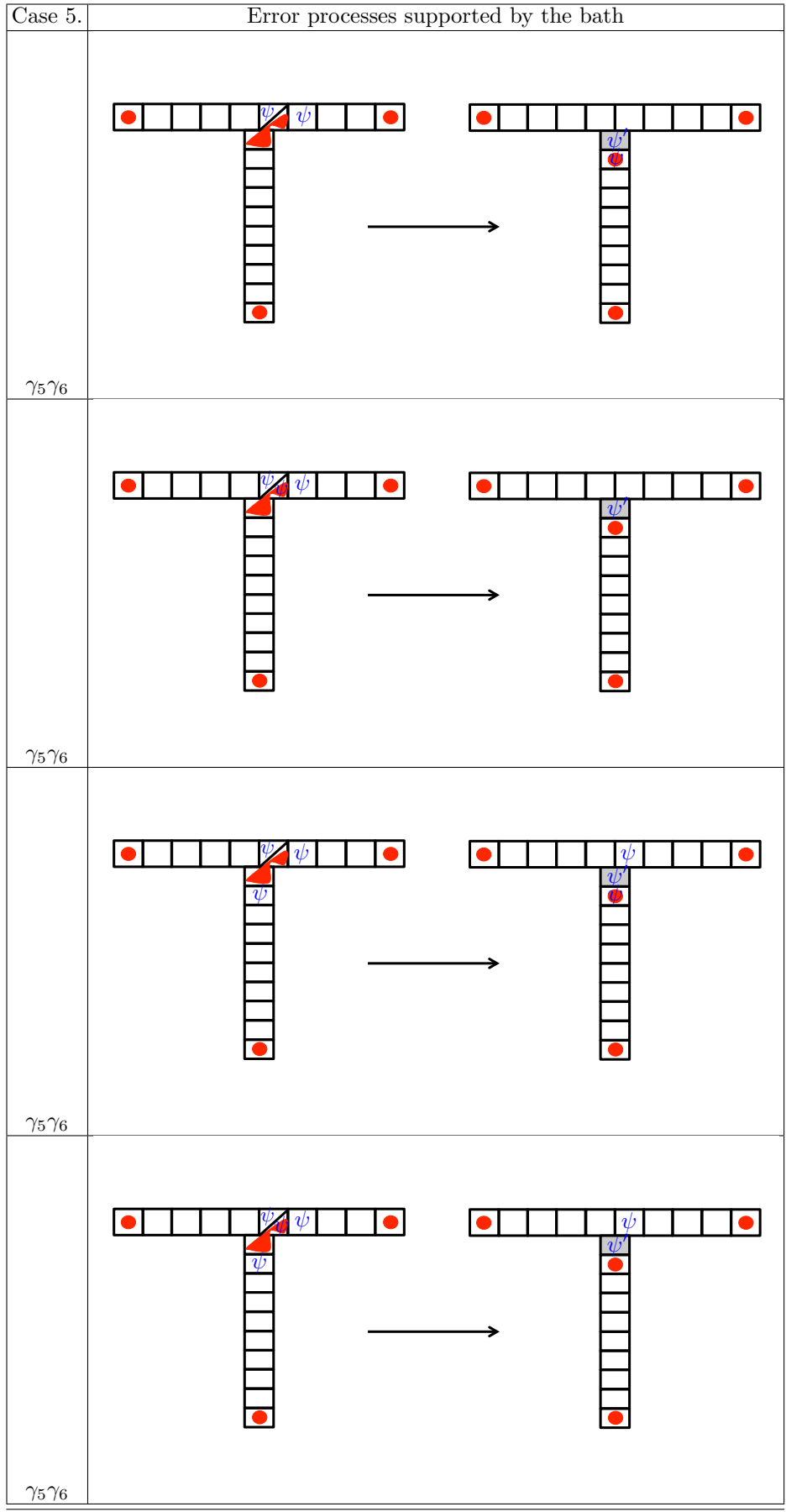

$\gamma_5\gamma_6$

$\gamma_5\gamma_6$

$\gamma_5\gamma_6$

$\gamma_5\gamma_6$

| Case 5. | Error processes supported by the bath |
|---------|----------------------------------------|
| | 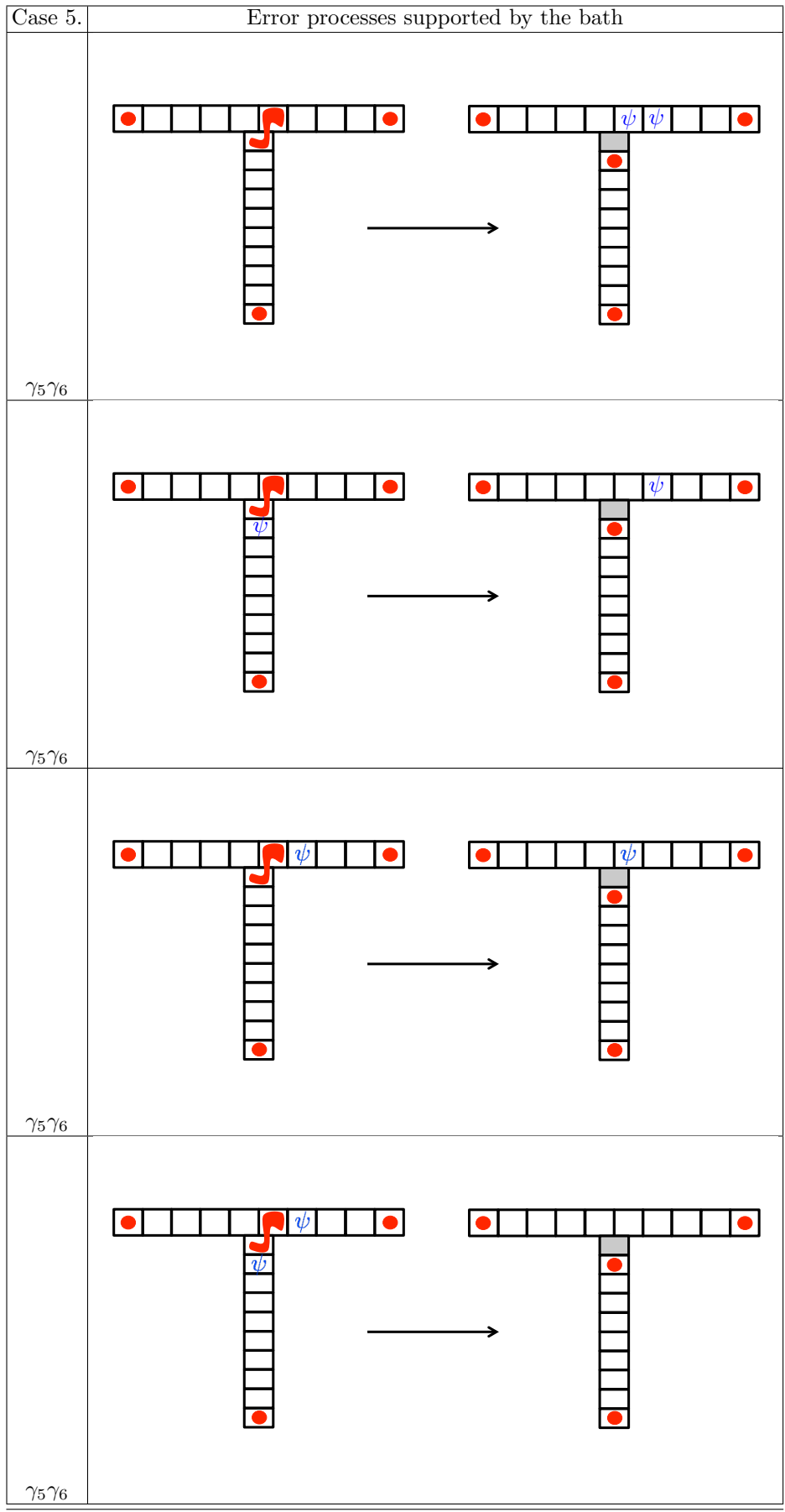 |

...

| Case 5. | Error processes supported by the bath |
|---|---|

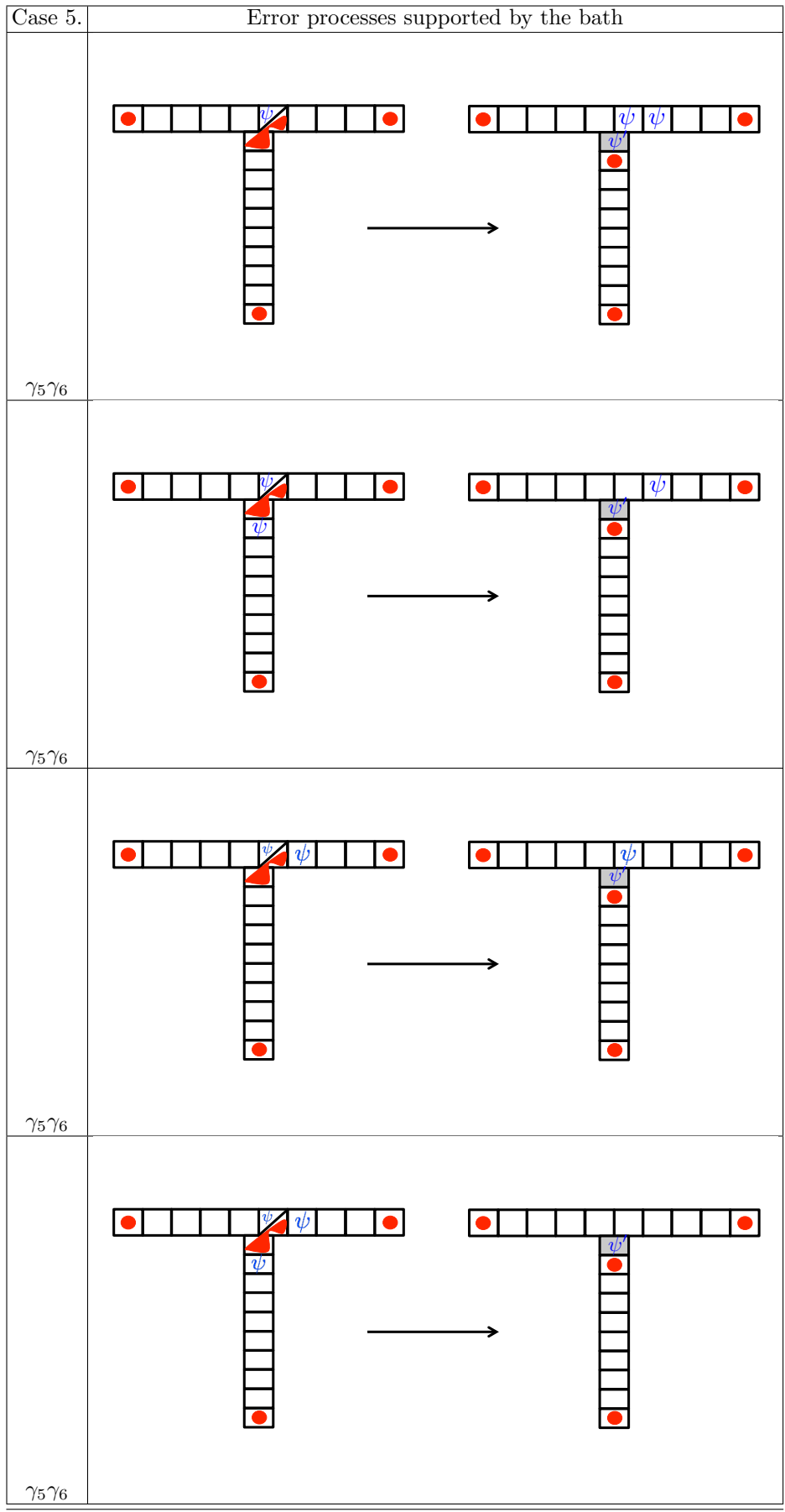

$\gamma_5\gamma_6$

$\gamma_5\gamma_6$

$\gamma_5\gamma_6$

$\gamma_5\gamma_6$

| Case 5. | Error processes supported by the bath |
|---------|----------------------------------------|
| | 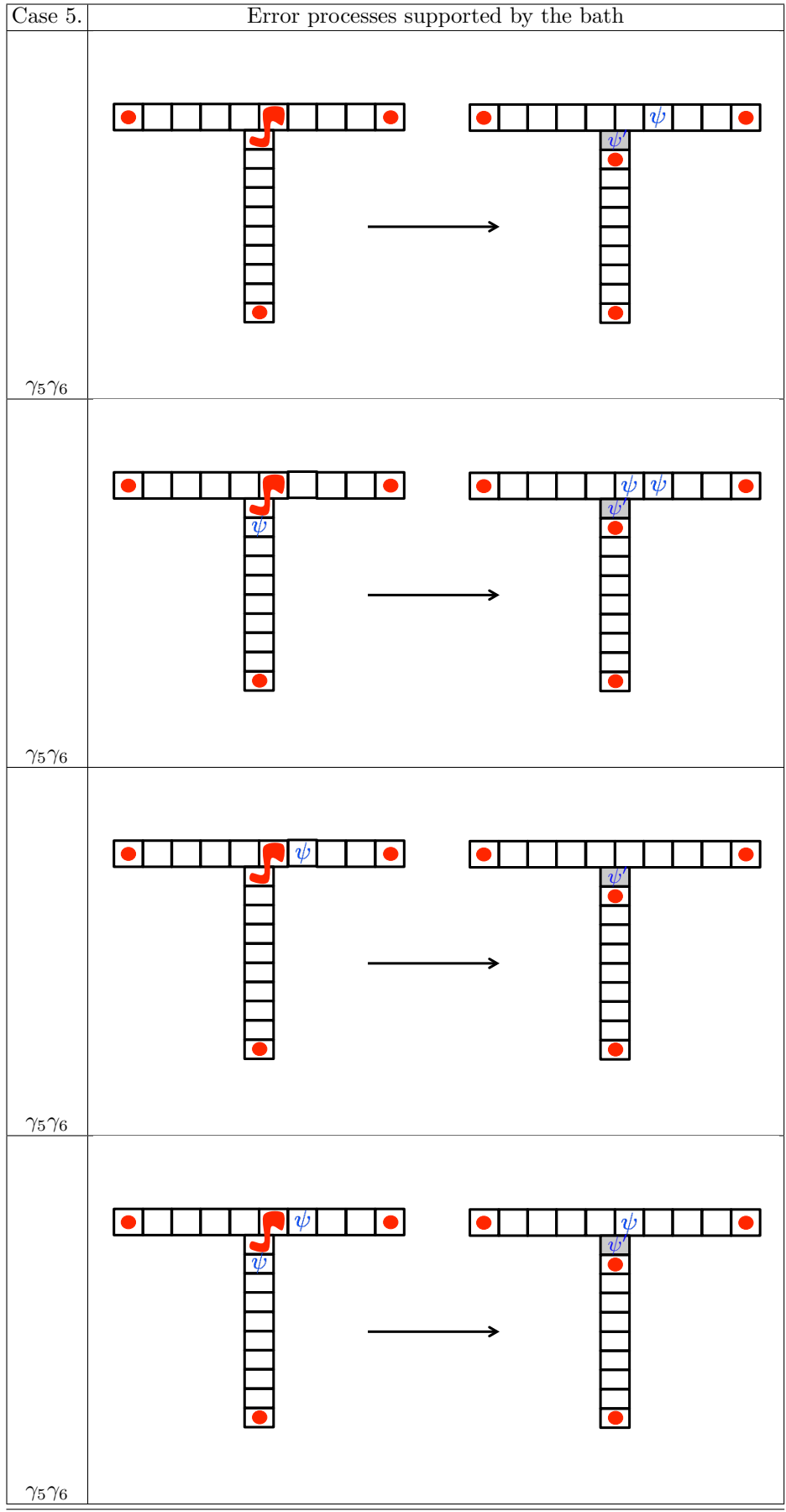 |

| Case 5. | Error processes supported by the bath |
|---|---|

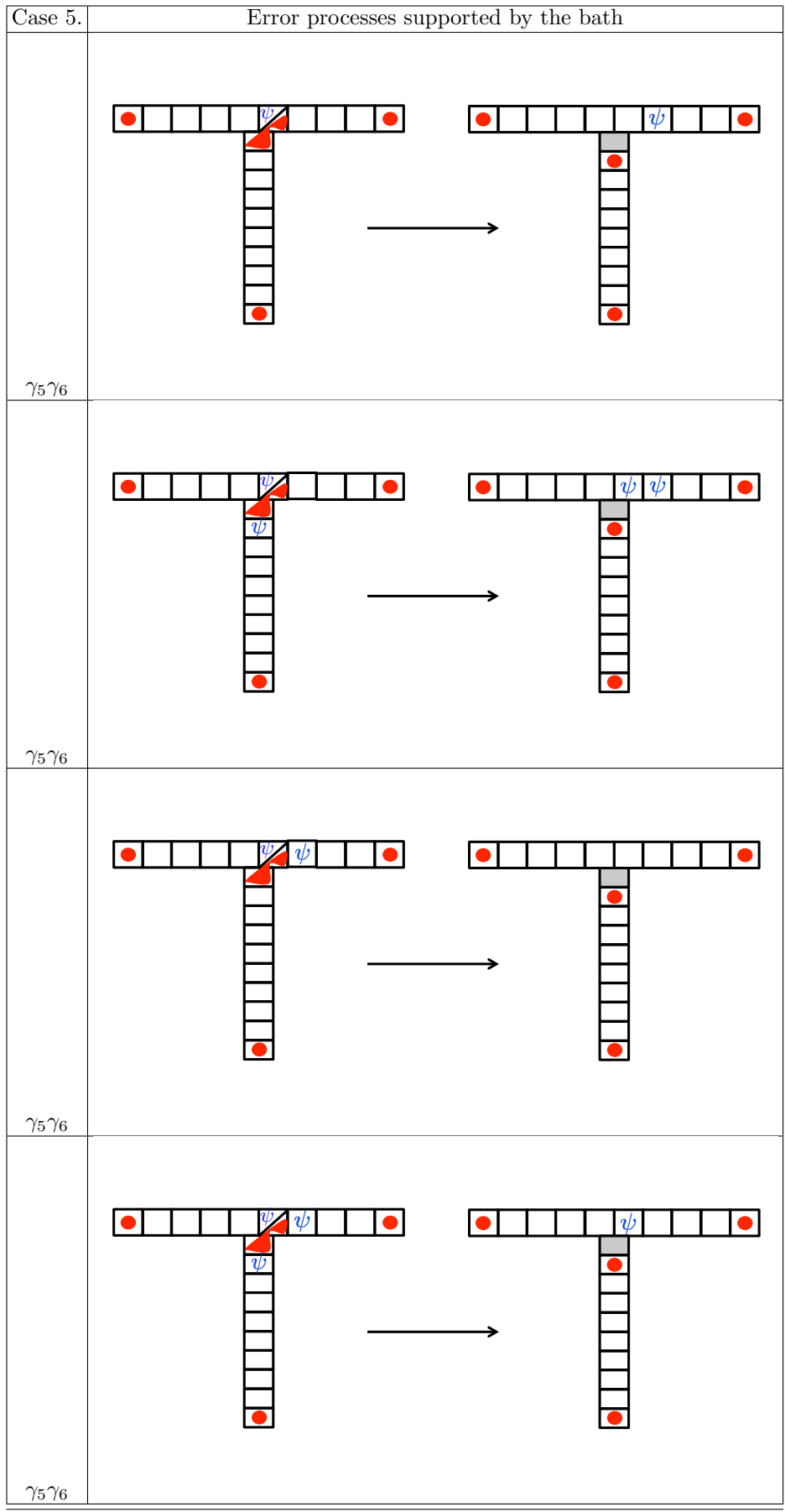

$\gamma_5\gamma_6$

$\gamma_5\gamma_6$

$\gamma_5\gamma_6$

$\gamma_5\gamma_6$

| Case 5. | Error processes supported by the bath |
|---|---|
| | 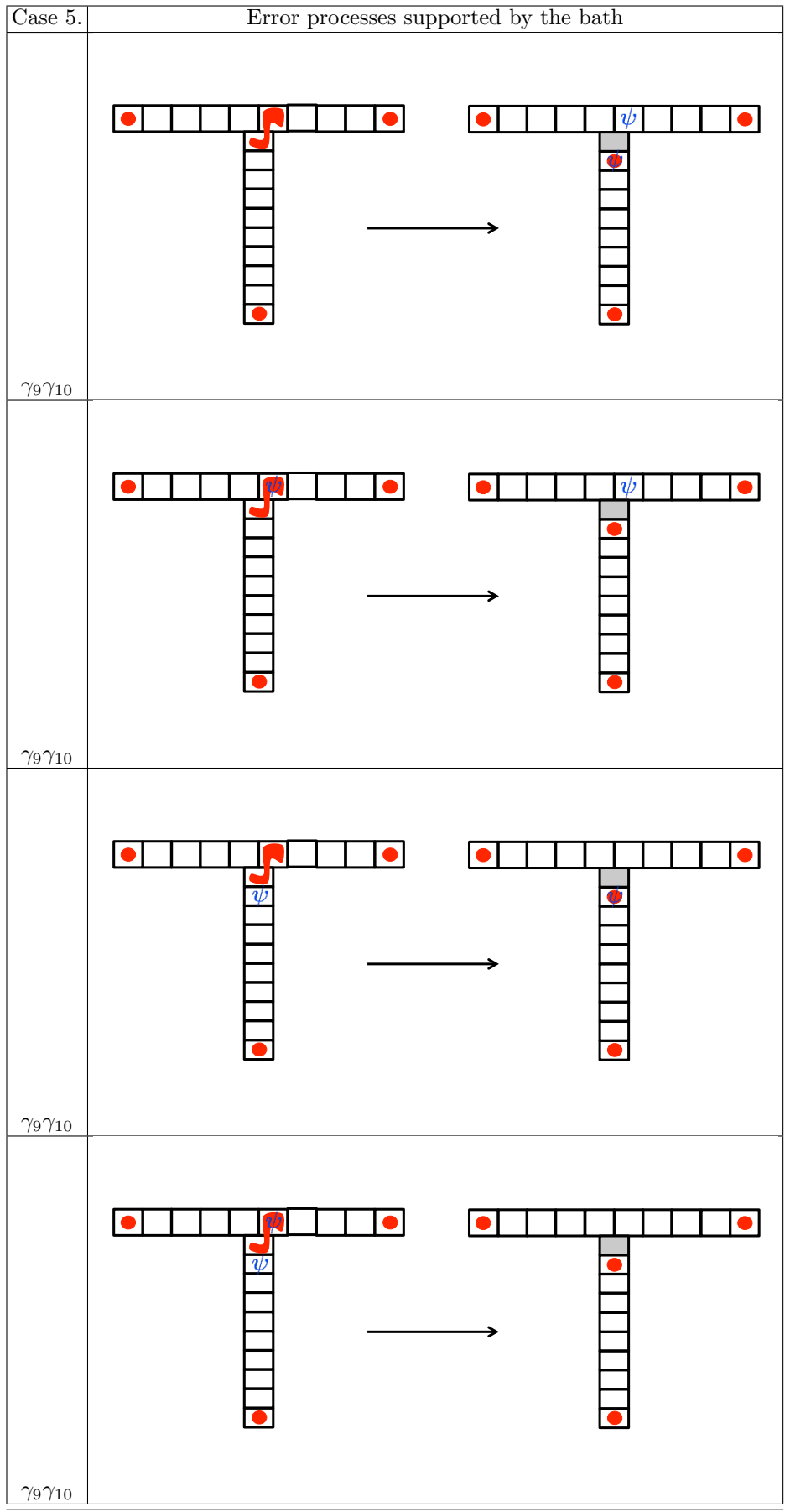 |

29

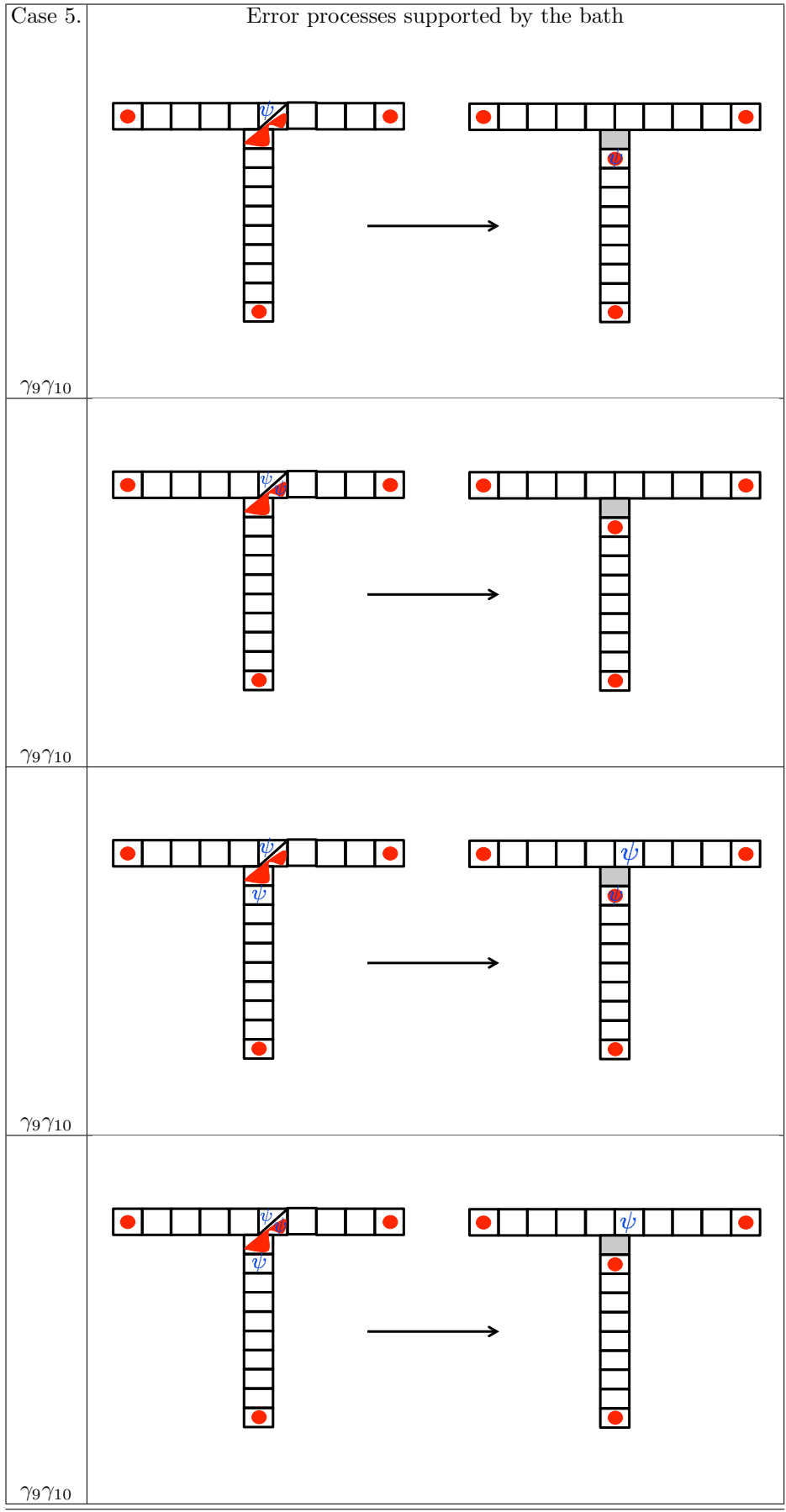

| Case 5. | Error processes supported by the bath |
|---|---|

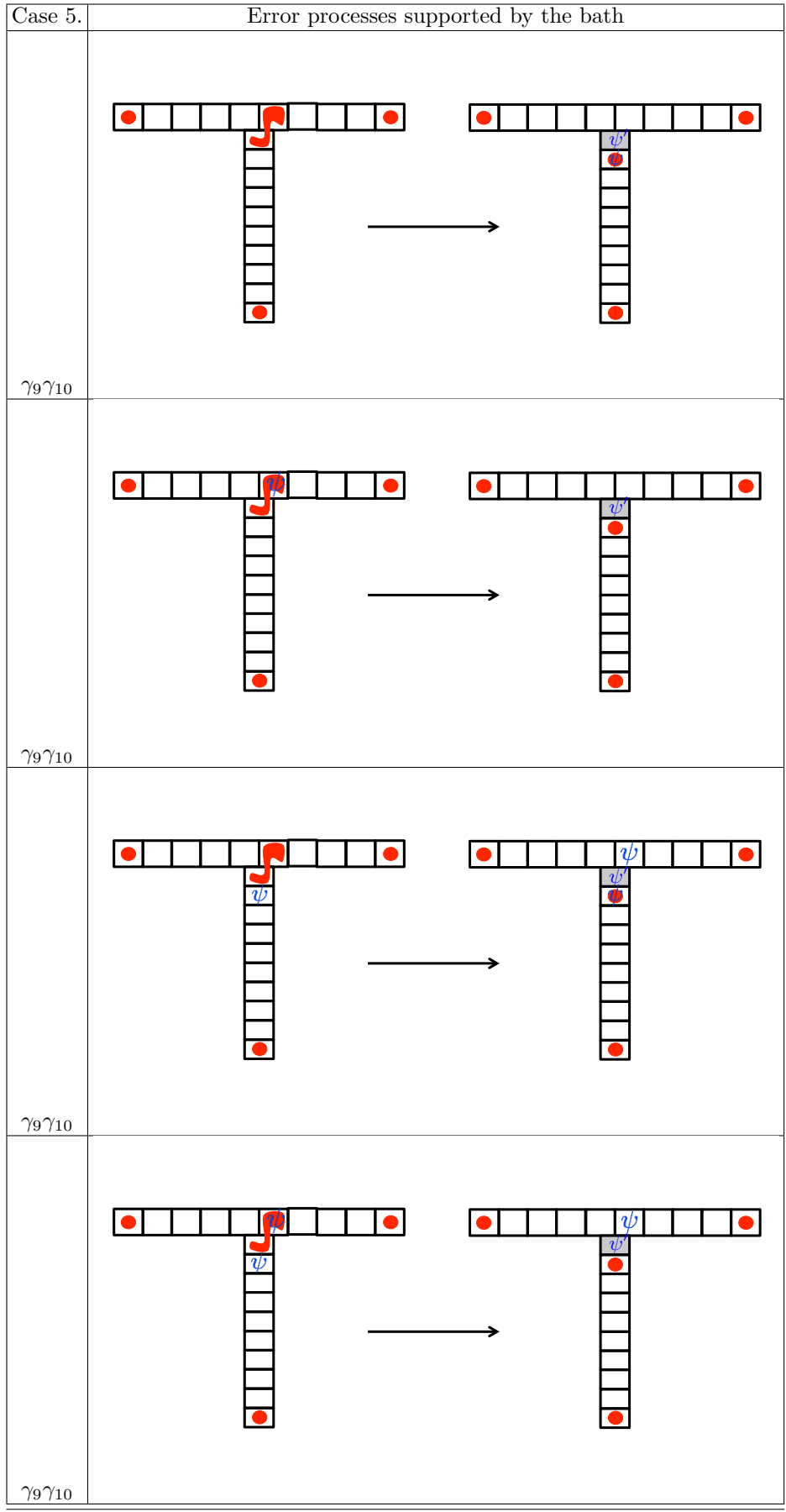

$\gamma_9\gamma_{10}$

$\gamma_9\gamma_{10}$

$\gamma_9\gamma_{10}$

$\gamma_9\gamma_{10}$

| Case 5. | Error processes supported by the bath |
|---------|----------------------------------------|

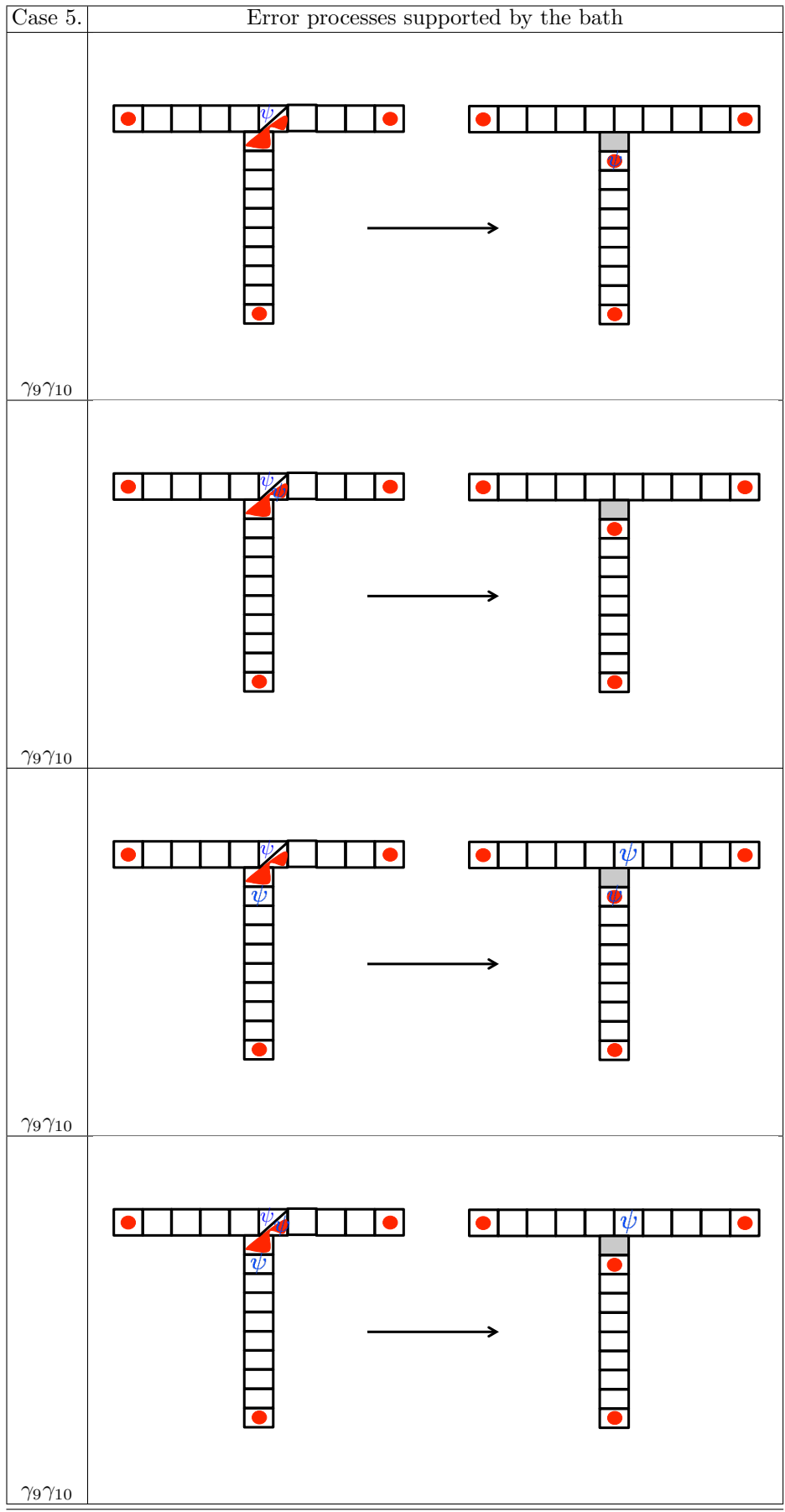

$\gamma_9\gamma_{10}$

$\gamma_9\gamma_{10}$

$\gamma_9\gamma_{10}$

$\gamma_9\gamma_{10}$

| Case 5. | Error processes supported by the bath |
|---|---|

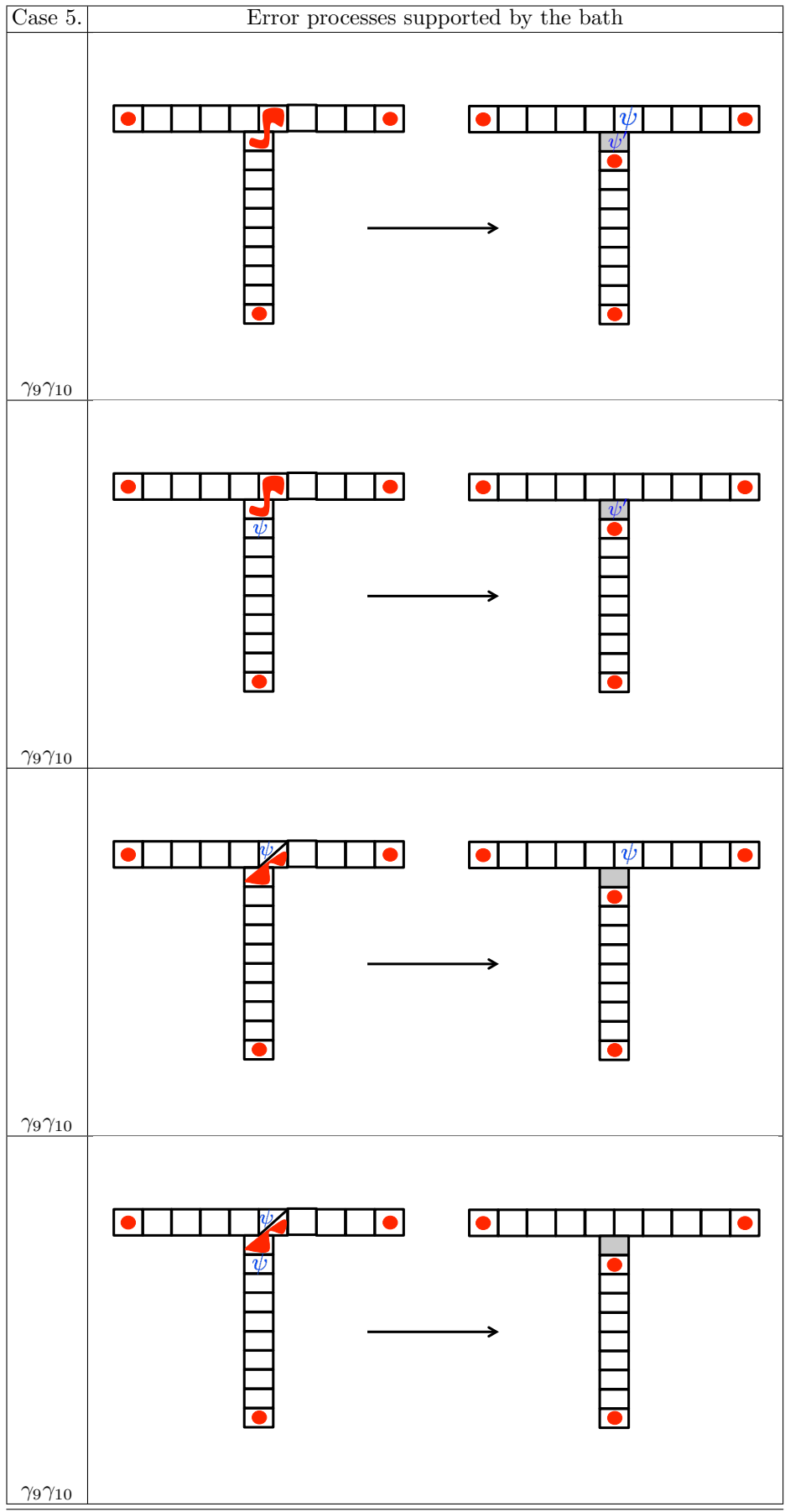

$\gamma_9\gamma_{10}$

$\gamma_9\gamma_{10}$

$\gamma_9\gamma_{10}$

$\gamma_9\gamma_{10}$

| Case 5. | Error processes supported by the bath |
|---------|----------------------------------------|

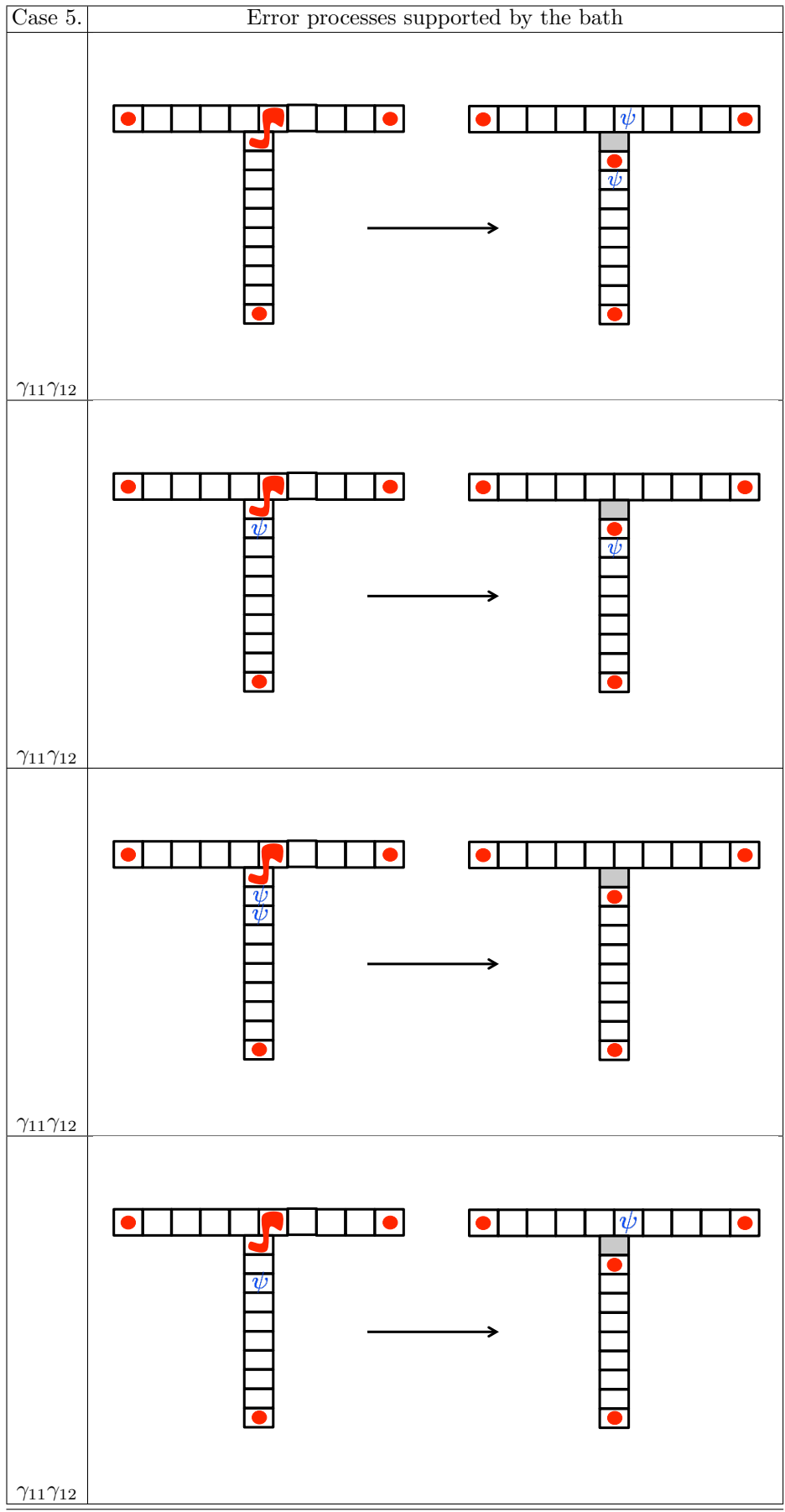

$\gamma_{11}\gamma_{12}$

$\gamma_{11}\gamma_{12}$

$\gamma_{11}\gamma_{12}$

$\gamma_{11}\gamma_{12}$

| Case 5. | Error processes supported by the bath |
|---------|----------------------------------------|
|  | |
| $\gamma_{11}\gamma_{12}$ | |
|  | |
| $\gamma_{11}\gamma_{12}$ | |
|  | |
| $\gamma_{11}\gamma_{12}$ | |
|  | |
| $\gamma_{11}\gamma_{12}$ | |

| Case 5. | Error processes supported by the bath |
|---------|----------------------------------------|
| | 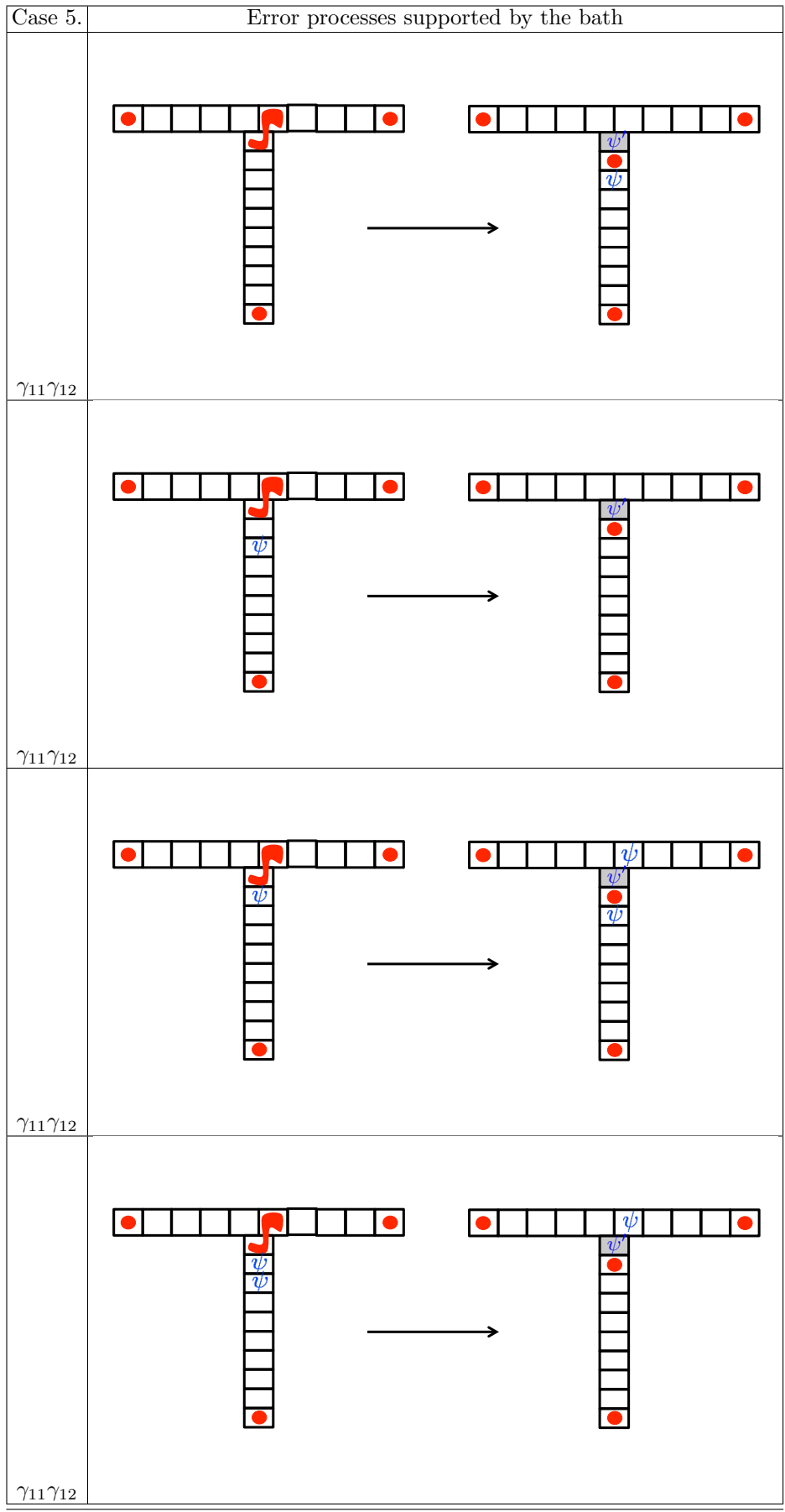 |

The row labels along the left column read, from top to bottom: $\gamma_{11}\gamma_{12}$, $\gamma_{11}\gamma_{12}$, $\gamma_{11}\gamma_{12}$, $\gamma_{11}\gamma_{12}$.

| Case 5. | Error processes supported by the bath |
|---------|----------------------------------------|
| $\gamma_{11}\gamma_{12}$ |  |
| $\gamma_{11}\gamma_{12}$ |  |
| $\gamma_{11}\gamma_{12}$ |  |
| $\gamma_{11}\gamma_{12}$ |  |

| Case 5. | Error processes supported by the bath |
|---------|---------------------------------------|
| | 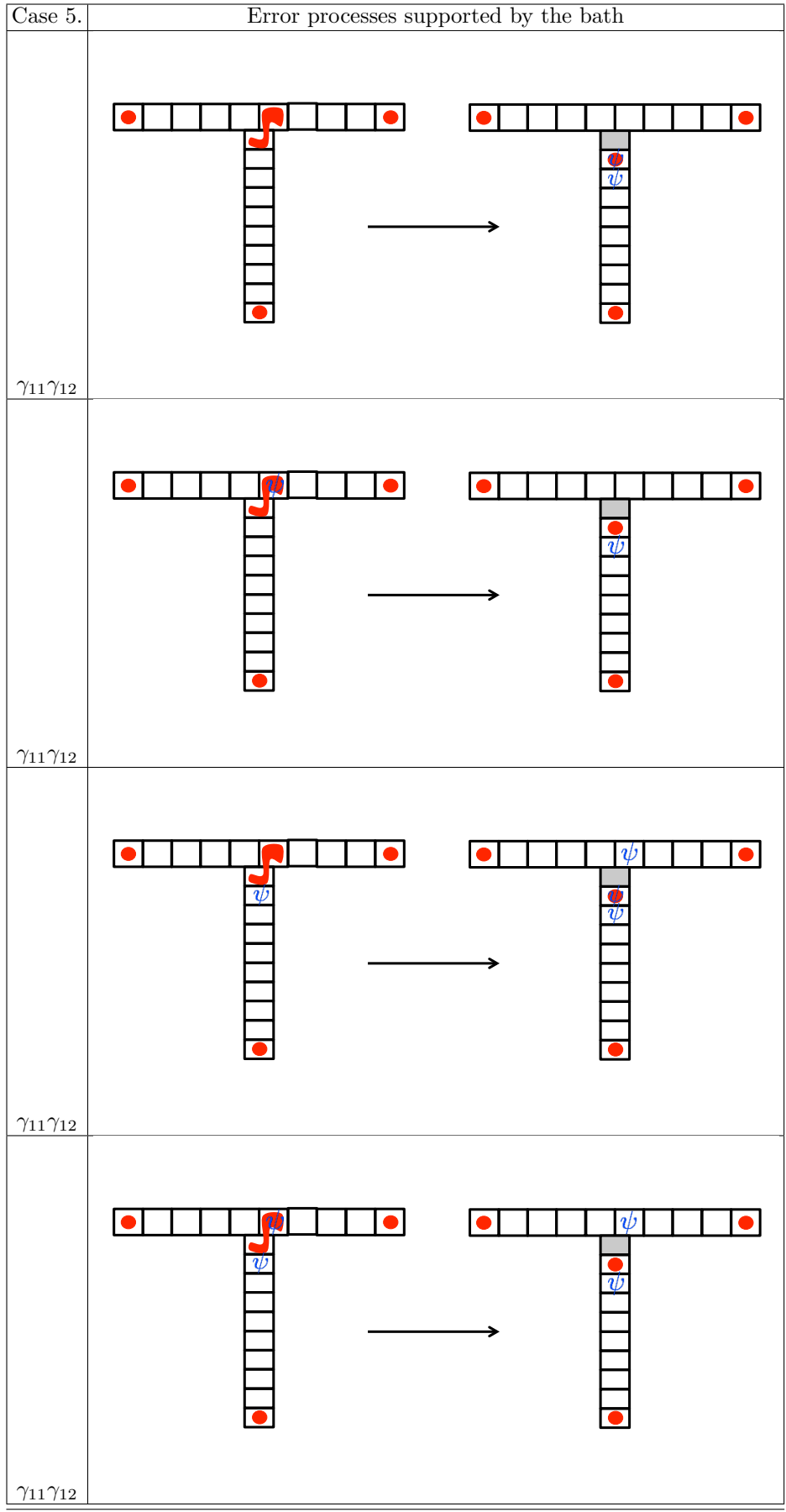 |

$\gamma_{11}\gamma_{12}$

$\gamma_{11}\gamma_{12}$

$\gamma_{11}\gamma_{12}$

$\gamma_{11}\gamma_{12}$

| Case 5. | Error processes supported by the bath |
|---------|---------------------------------------|

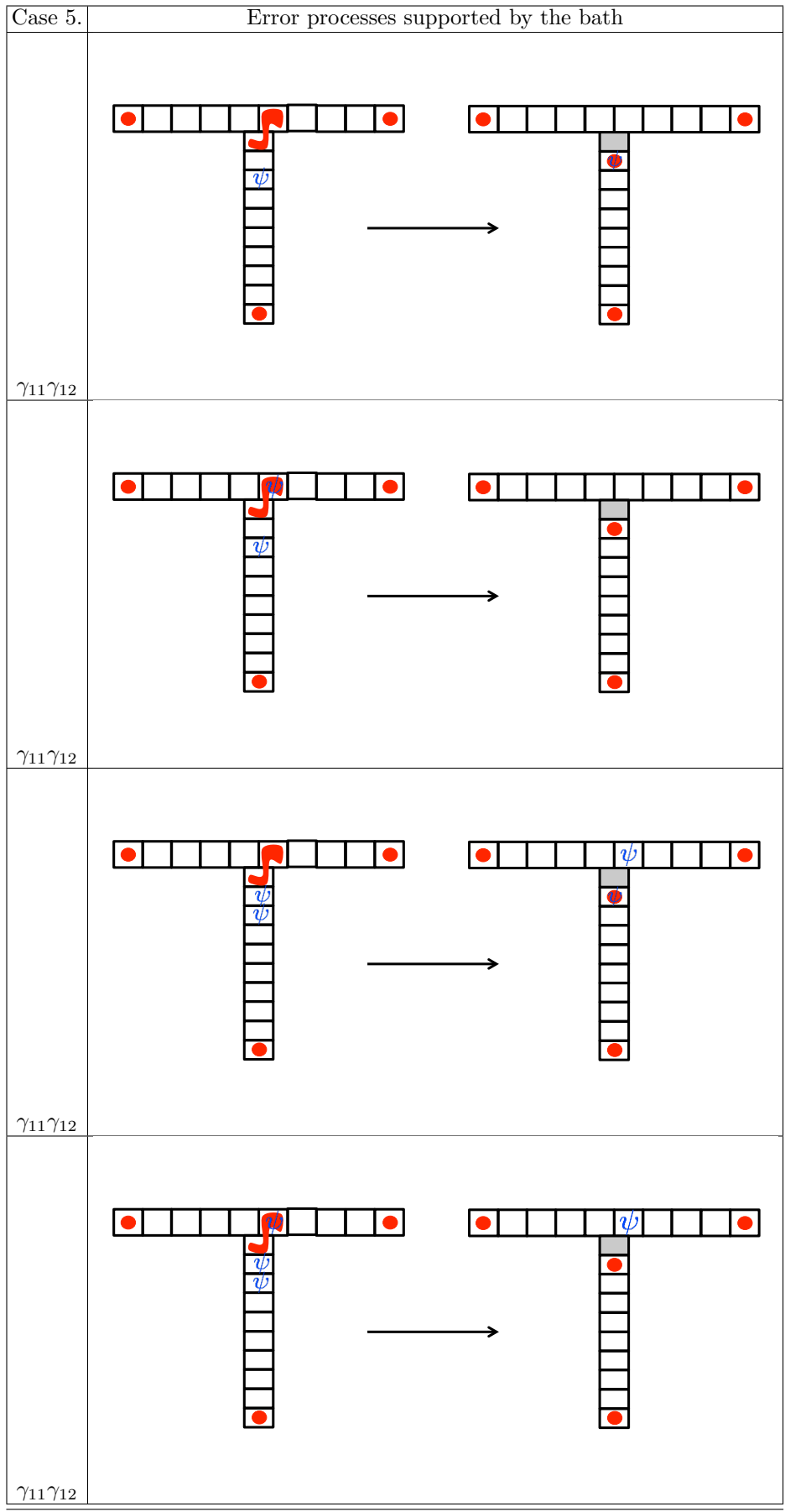

$\gamma_{11}\gamma_{12}$

$\gamma_{11}\gamma_{12}$

$\gamma_{11}\gamma_{12}$

$\gamma_{11}\gamma_{12}$

39

| Case 5. | Error processes supported by the bath |
|---|---|
| $\gamma_{11}\gamma_{12}$ |  |
| $\gamma_{11}\gamma_{12}$ |  |
| $\gamma_{11}\gamma_{12}$ |  |
| $\gamma_{11}\gamma_{12}$ |  |

| Case 5. | Error processes supported by the bath |
|---------|----------------------------------------|
| $\gamma_{11}\gamma_{12}$ |  |
| $\gamma_{11}\gamma_{12}$ |  |
| $\gamma_{11}\gamma_{12}$ |  |
| $\gamma_{11}\gamma_{12}$ |  |

| Case 6. | Six-site model |
|---|---|

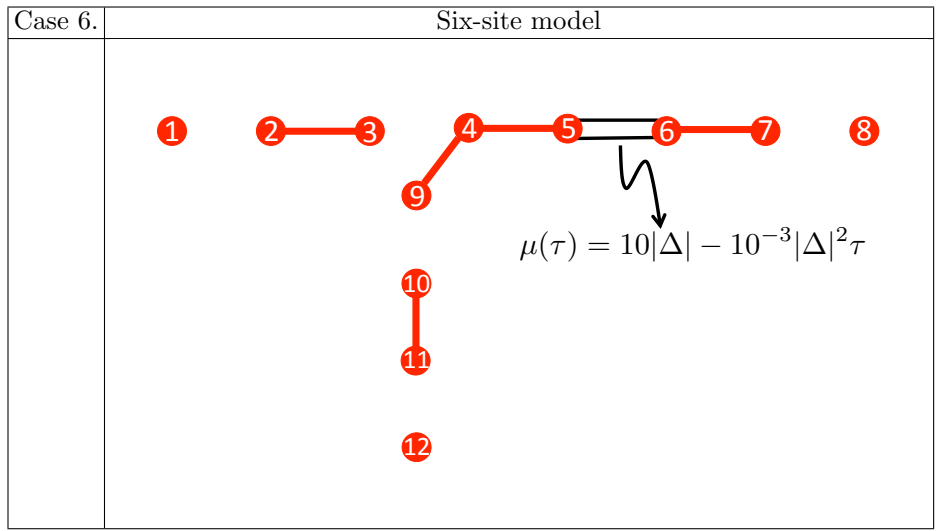

$$\mu(\tau) = 10|\Delta| - 10^{-3}|\Delta|^2\tau$$

| Case 6. | Unitary evolution |
|---|---|

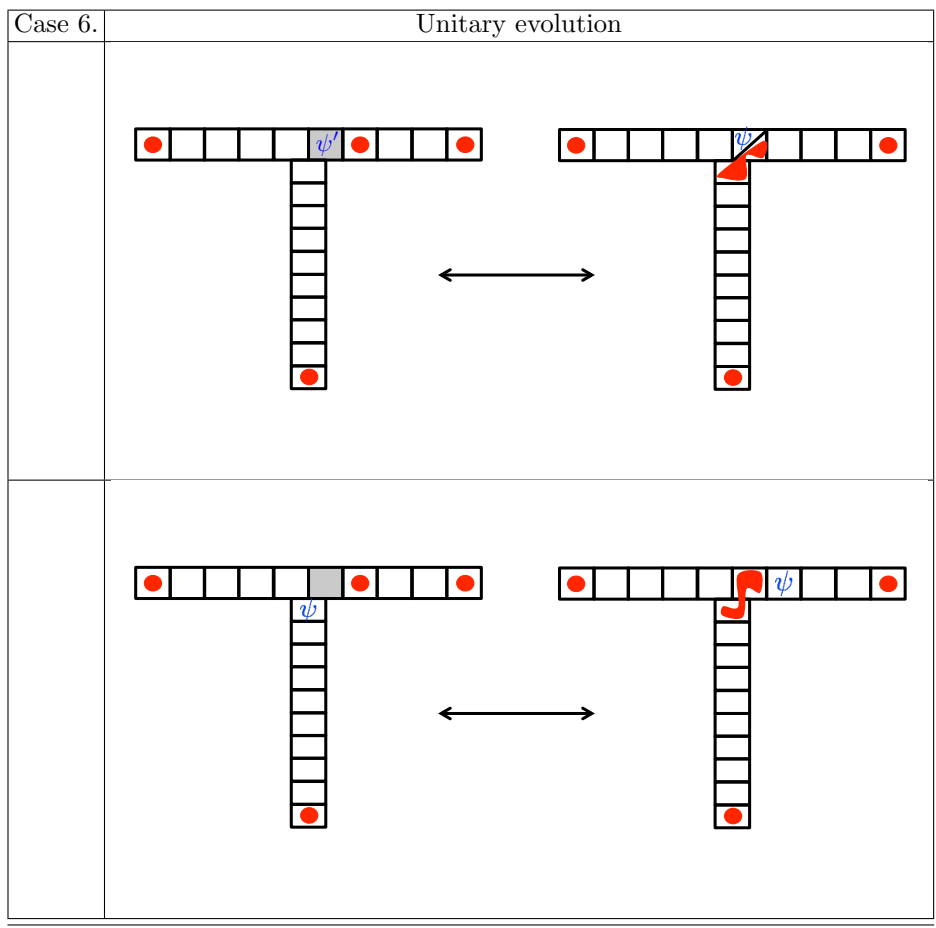

| Case 6. | Error processes supported by the bath |
|---------|----------------------------------------|



$\gamma_5\gamma_6$



$\gamma_5\gamma_6$



$\gamma_5\gamma_6$



$\gamma_5\gamma_6$

Case 6. Error processes supported by the bath

$\gamma_5\gamma_6$

$\gamma_5\gamma_6$

$\gamma_5\gamma_6$

$\gamma_5\gamma_6$

| Case 6. | Error processes supported by the bath |
|---------|---------------------------------------|



$\gamma_5\gamma_6$



$\gamma_5\gamma_6$



$\gamma_5\gamma_6$



$\gamma_5\gamma_6$

| Case 6. | Error processes supported by the bath |
|---------|---------------------------------------|
| $\gamma_5\gamma_6$ |  |
| $\gamma_5\gamma_6$ |  |
| $\gamma_5\gamma_6$ |  |
| $\gamma_5\gamma_6$ |  |

| Case 6. | Error processes supported by the bath |
|---|---|

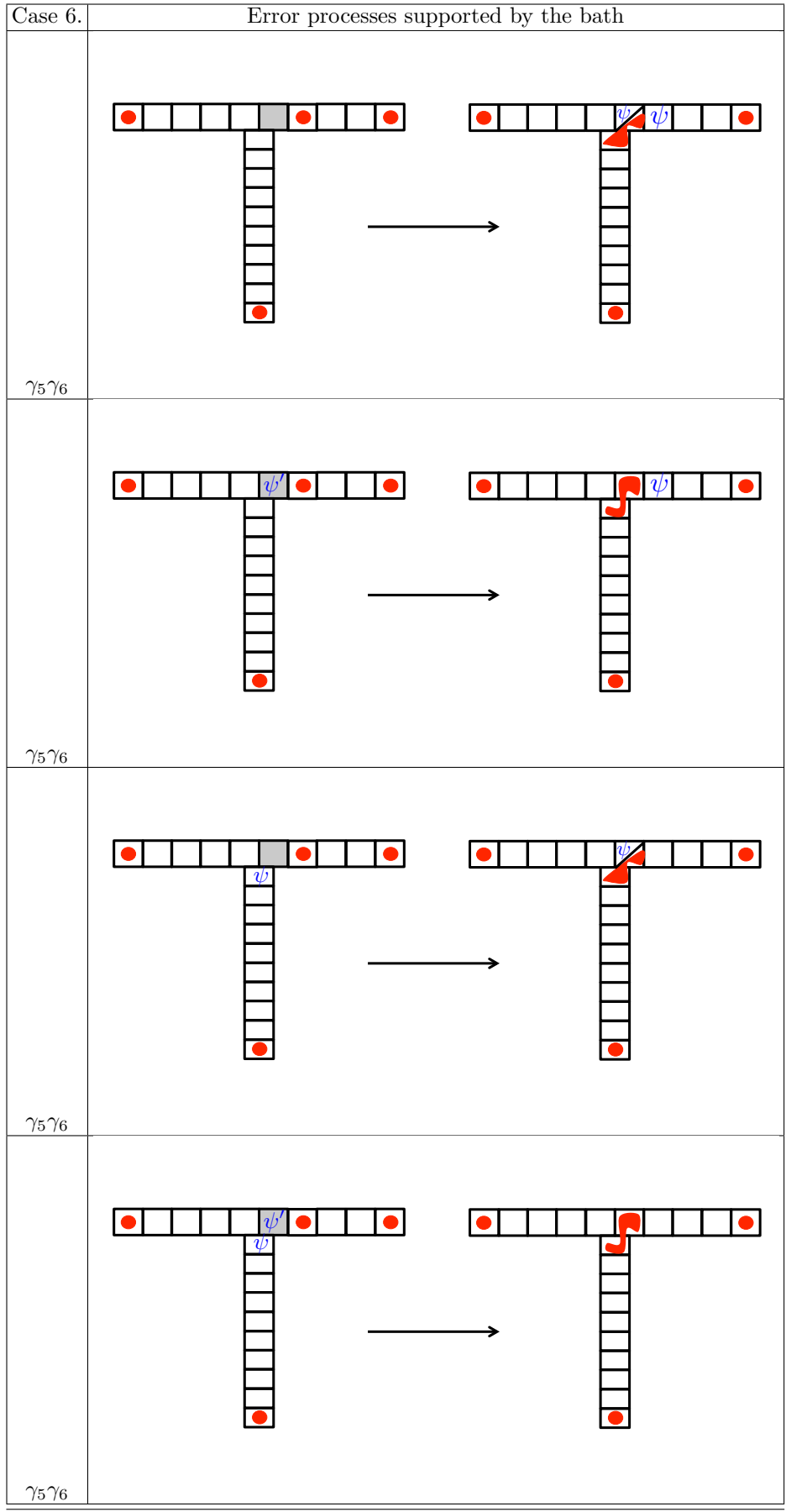

$\gamma_5\gamma_6$

$\gamma_5\gamma_6$

$\gamma_5\gamma_6$

$\gamma_5\gamma_6$

| Case 6. | Error processes supported by the bath |
|---|---|

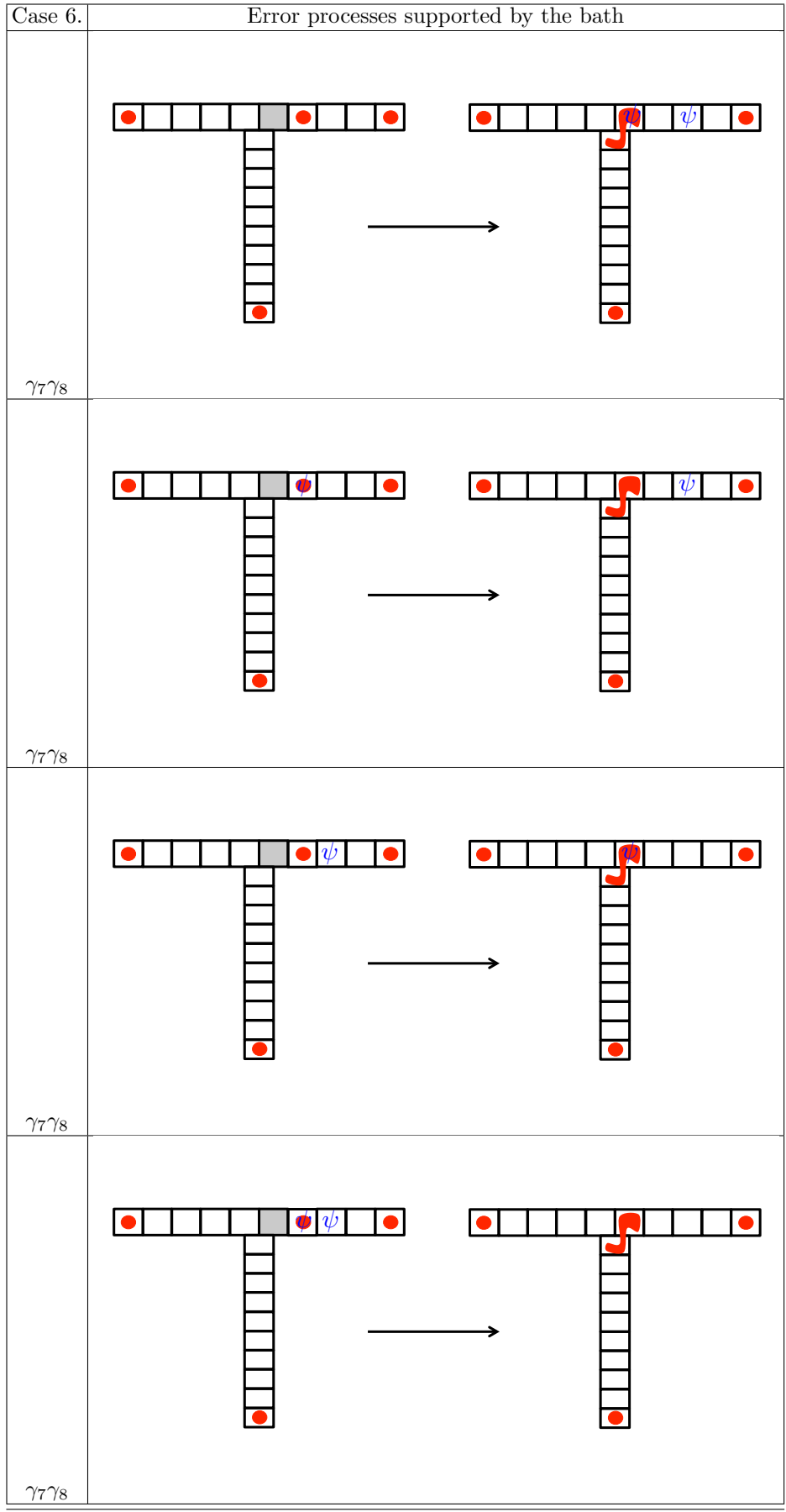

48

| Case 6. | Error processes supported by the bath |
|---|---|
| $\gamma_7\gamma_8$ |  |
| $\gamma_7\gamma_8$ |  |
| $\gamma_7\gamma_8$ |  |
| $\gamma_7\gamma_8$ |  |

| Case 6. | Error processes supported by the bath |
|---|---|
| | 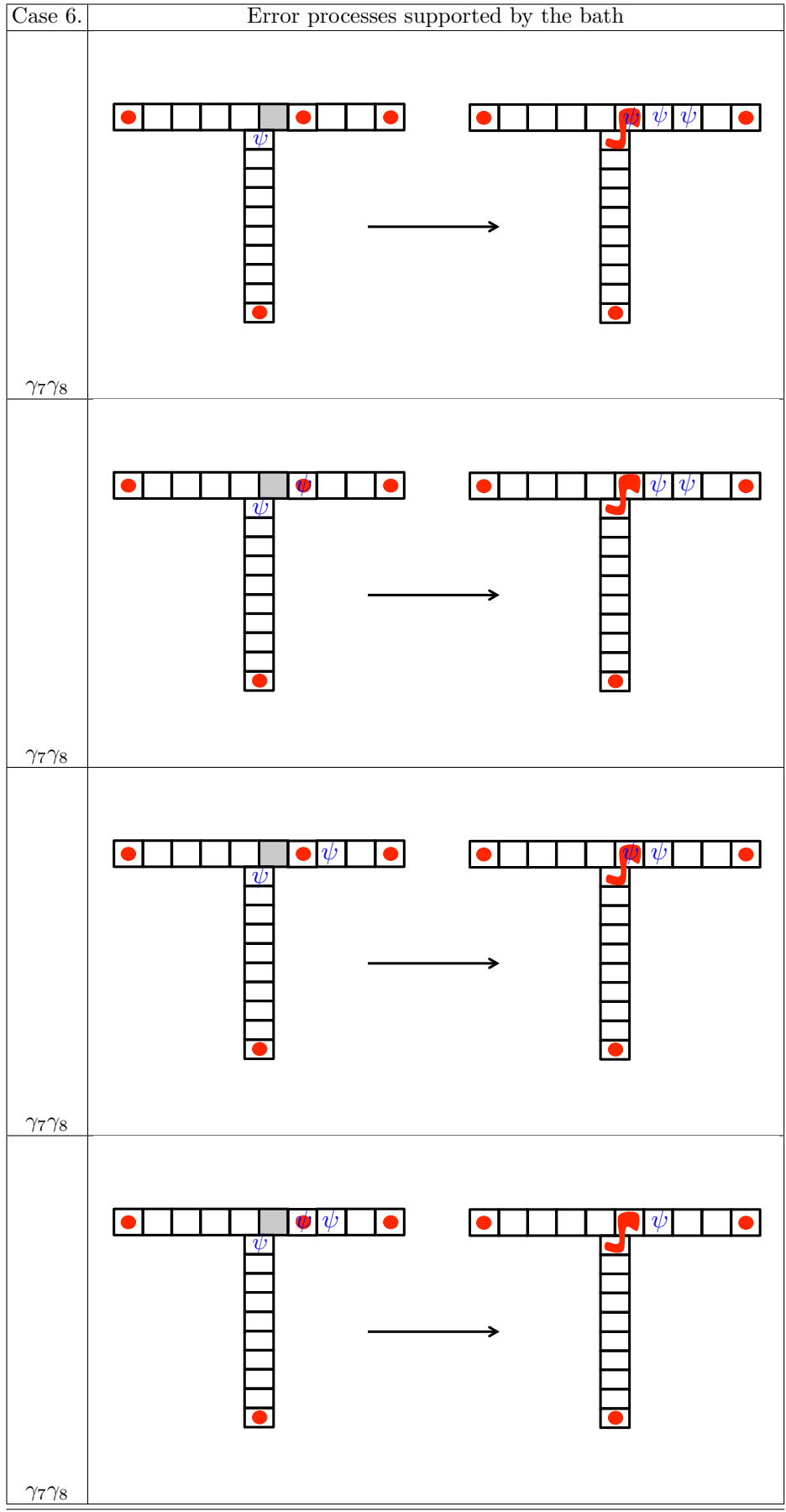 |
| $\gamma_7\gamma_8$ | |
| $\gamma_7\gamma_8$ | |
| $\gamma_7\gamma_8$ | |
| $\gamma_7\gamma_8$ | |

| Case 6. | Error processes supported by the bath |
|---|---|

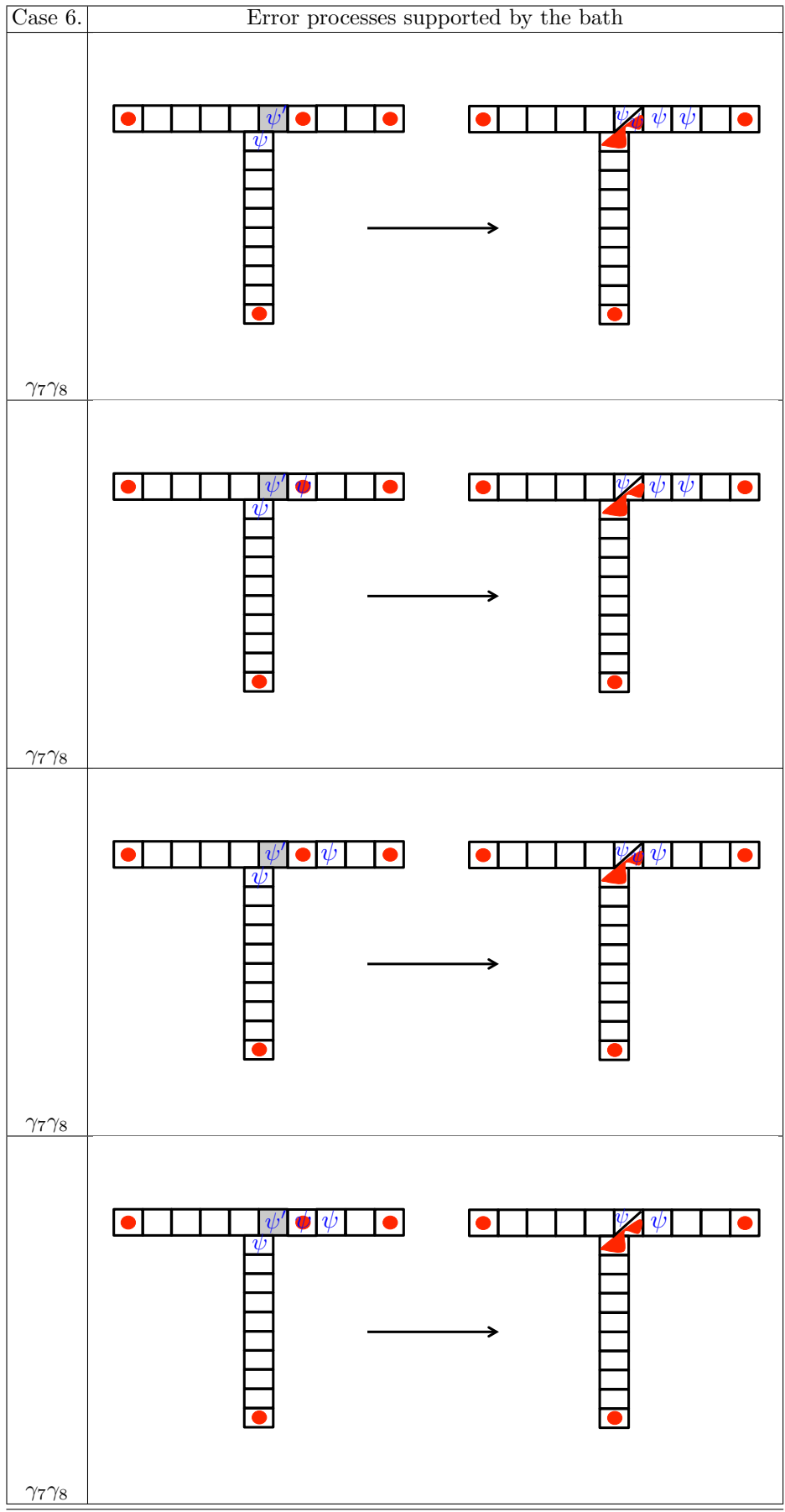

51

| Case 6. | Error processes supported by the bath |
|---|---|

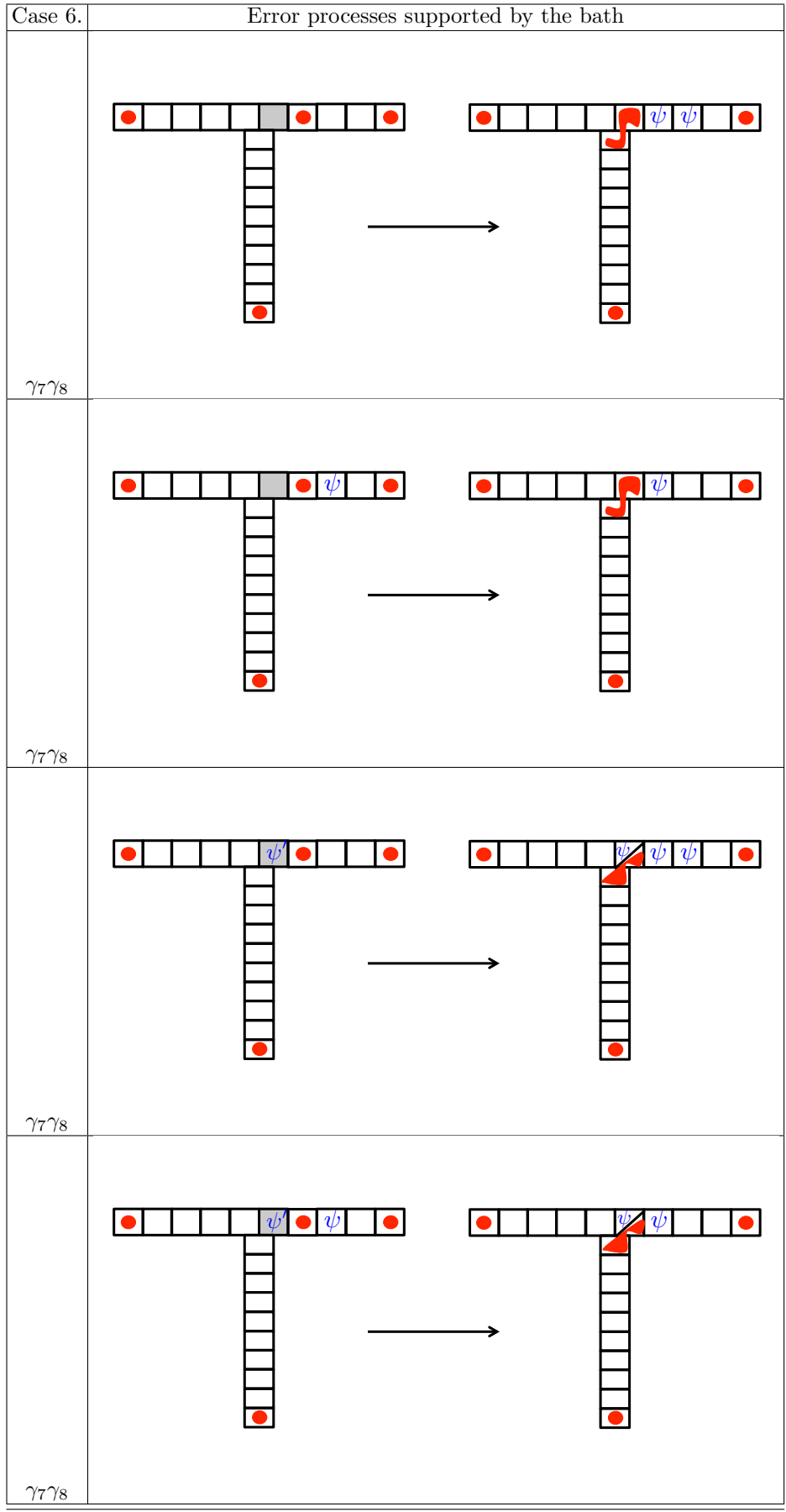

| Case 6. | Error processes supported by the bath |
|---|---|

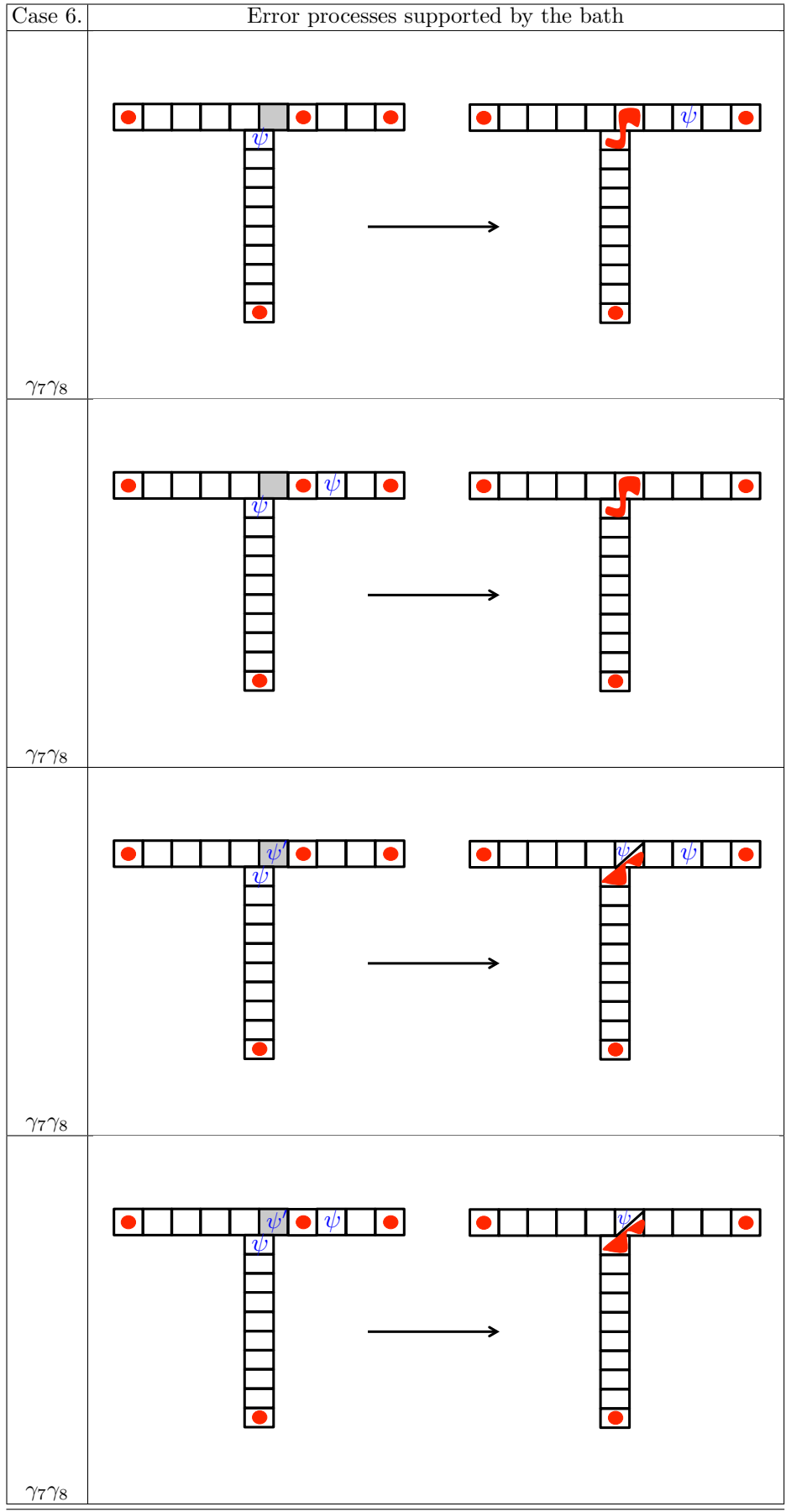

| Case 6. | Error processes supported by the bath |
|---|---|

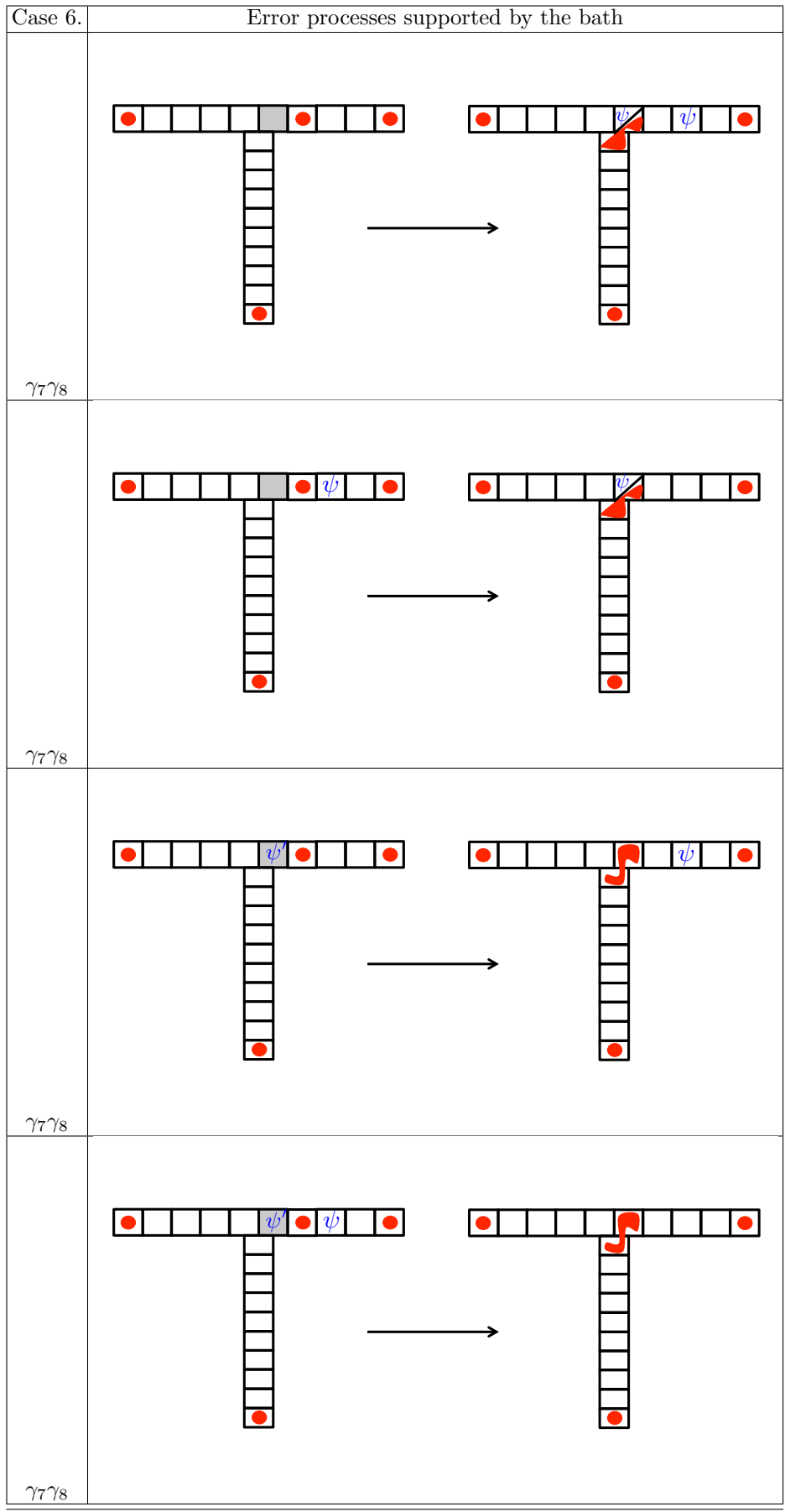

| Case 6. | Error processes supported by the bath |
|---|---|

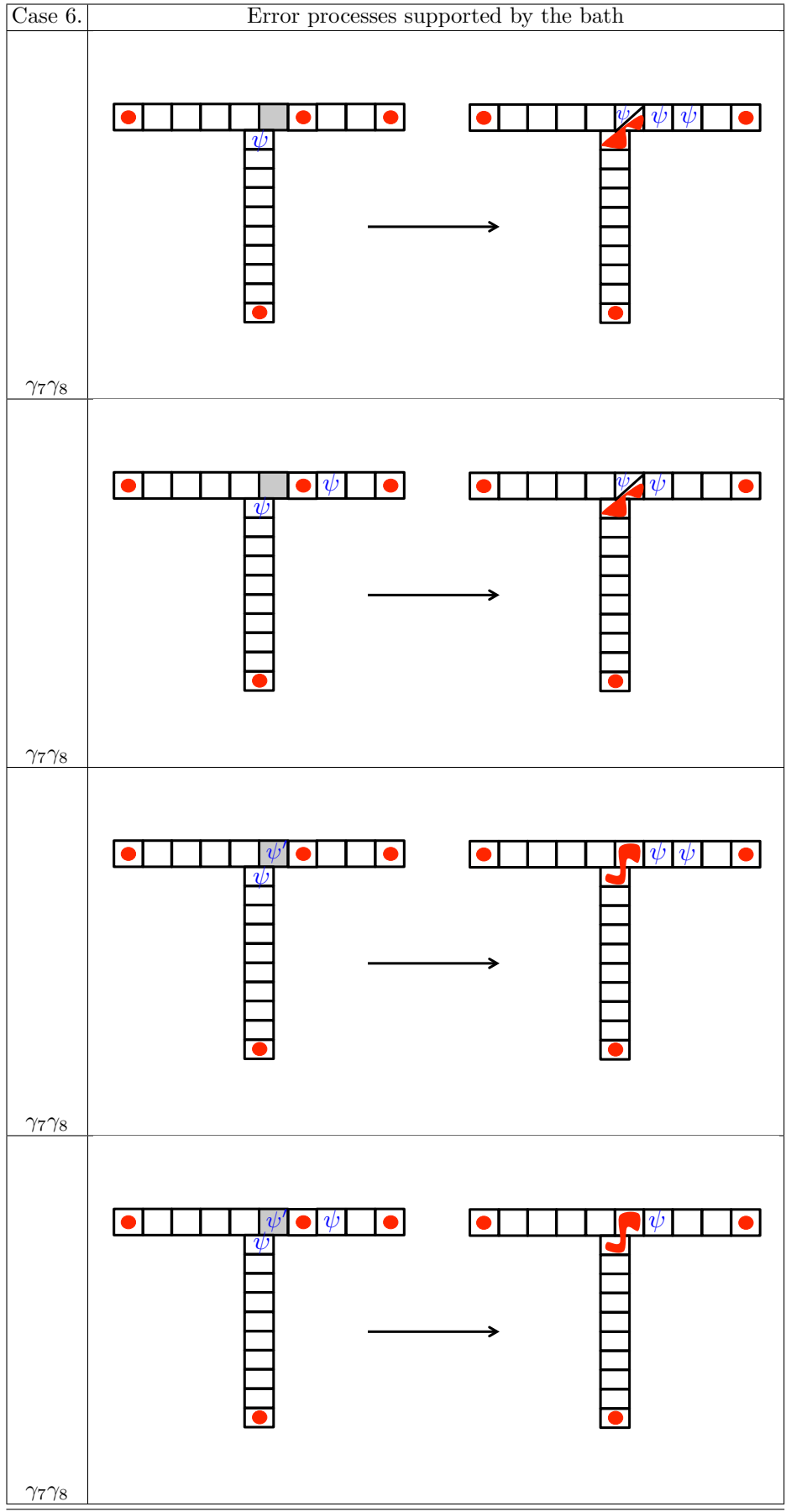

| Case 6. | Error processes supported by the bath |
|---|---|

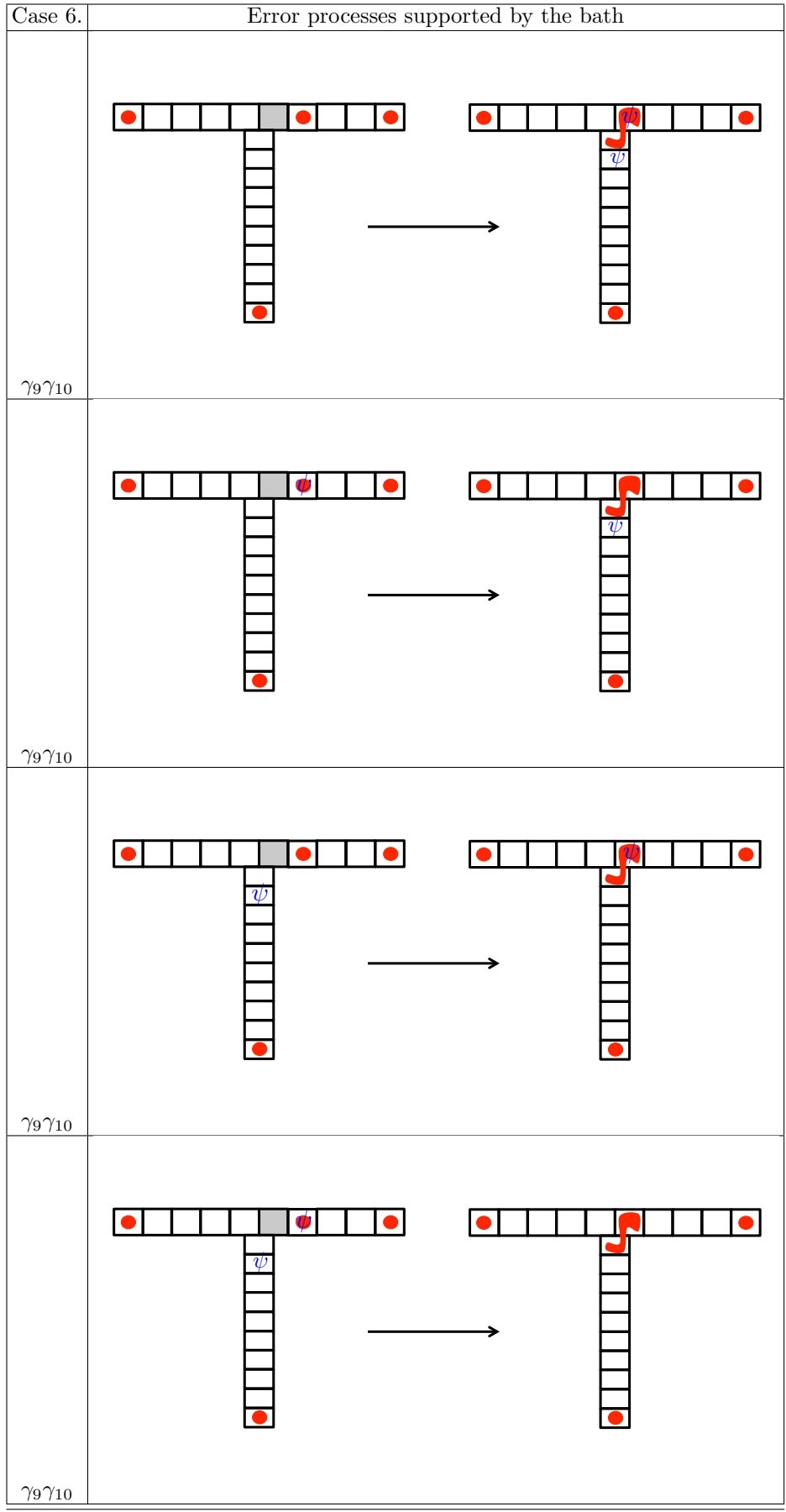

$\gamma_9\gamma_{10}$

$\gamma_9\gamma_{10}$

$\gamma_9\gamma_{10}$

$\gamma_9\gamma_{10}$

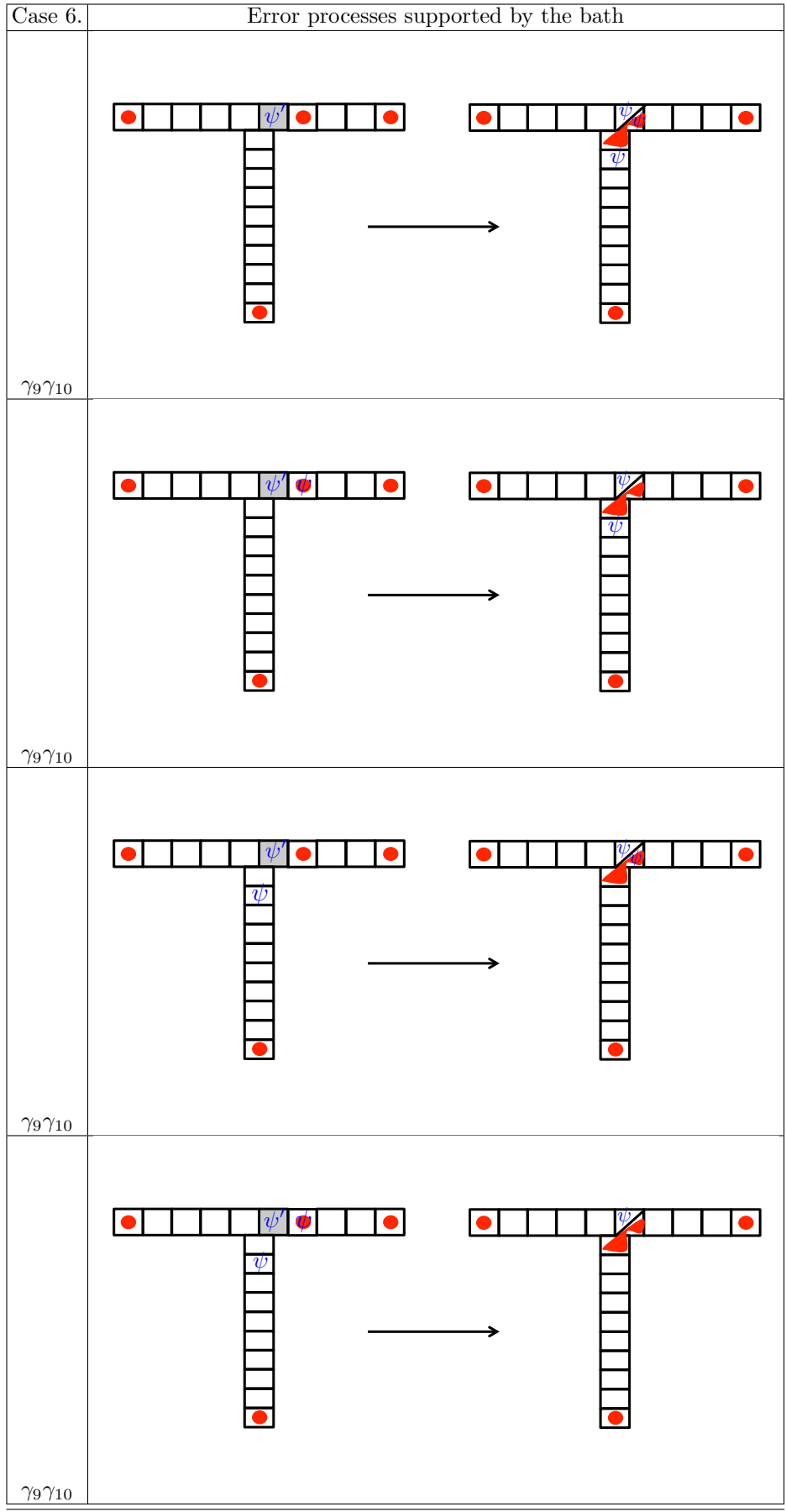

Case 6. | Error processes supported by the bath

$\gamma_9\gamma_{10}$

$\gamma_9\gamma_{10}$

$\gamma_9\gamma_{10}$

$\gamma_9\gamma_{10}$

| Case 6. | Error processes supported by the bath |
|---|---|

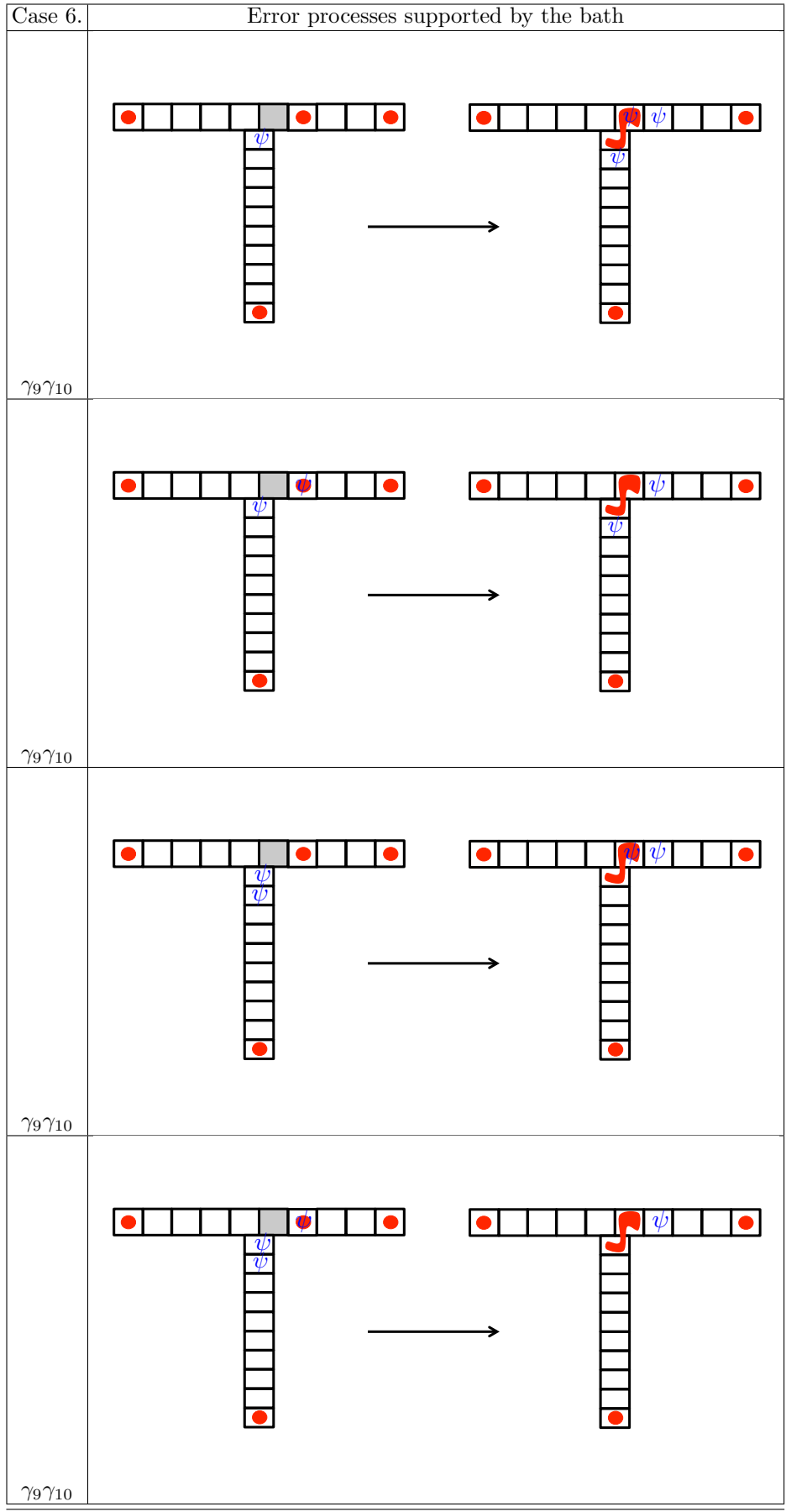

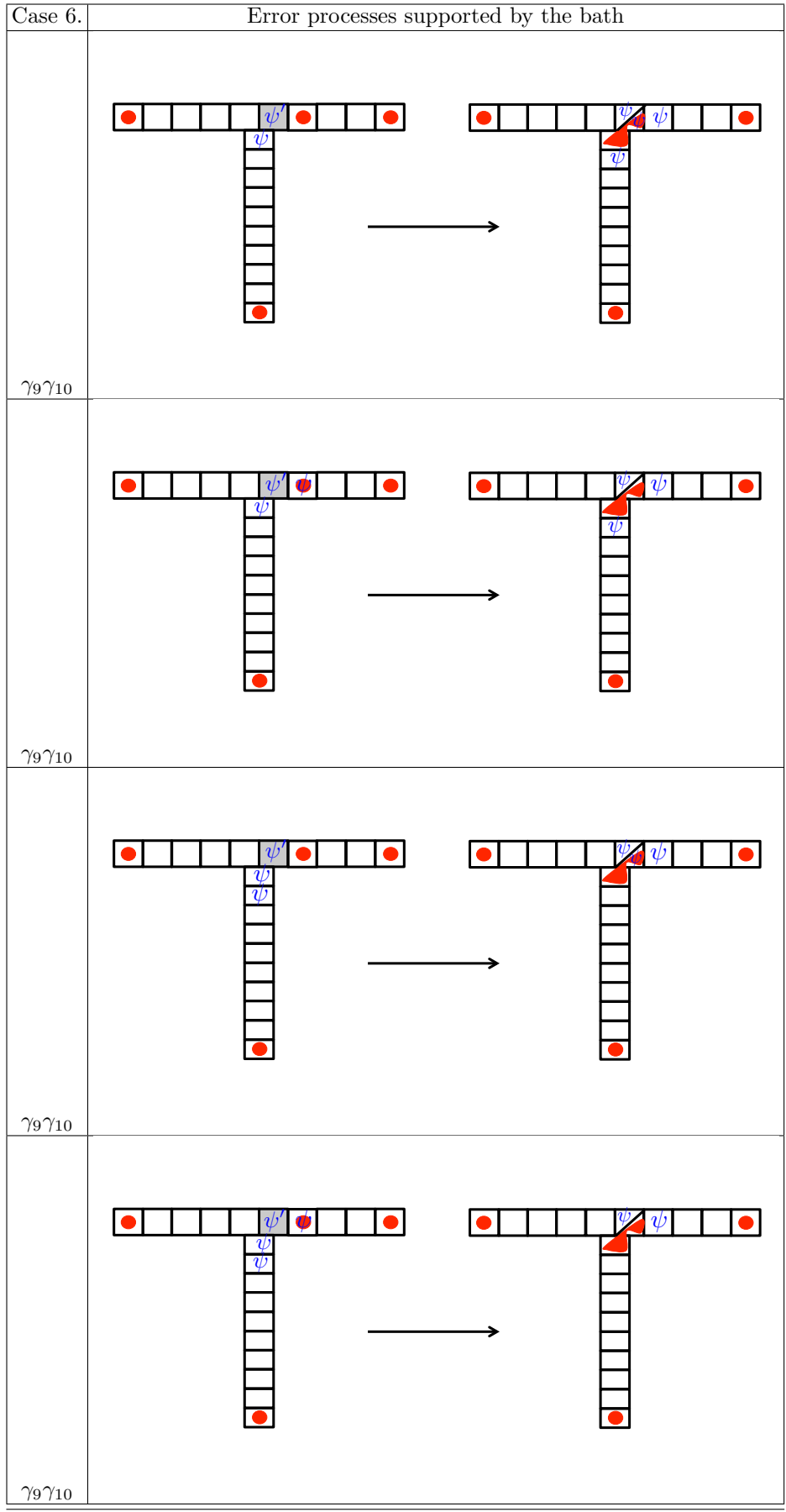

| Case 6. | Error processes supported by the bath |
|---------|---------------------------------------|
| $\gamma_9\gamma_{10}$ | |
| $\gamma_9\gamma_{10}$ | |
| $\gamma_9\gamma_{10}$ | |
| $\gamma_9\gamma_{10}$ | |

| Case 6. | Error processes supported by the bath |
|---------|----------------------------------------|

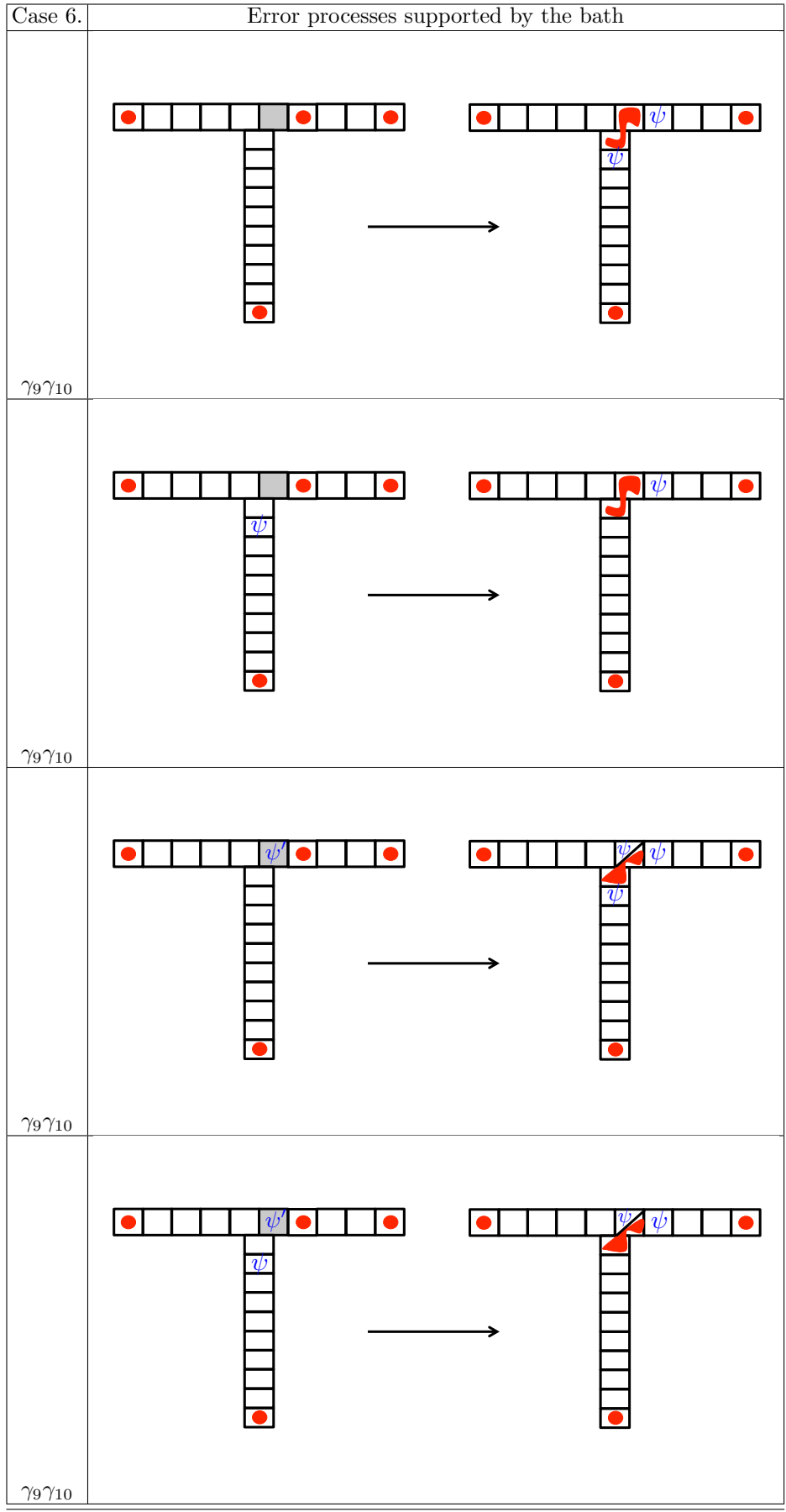

$\gamma_9\gamma_{10}$

$\gamma_9\gamma_{10}$

$\gamma_9\gamma_{10}$

$\gamma_9\gamma_{10}$

| Case 6. | Error processes supported by the bath |
|---|---|
| 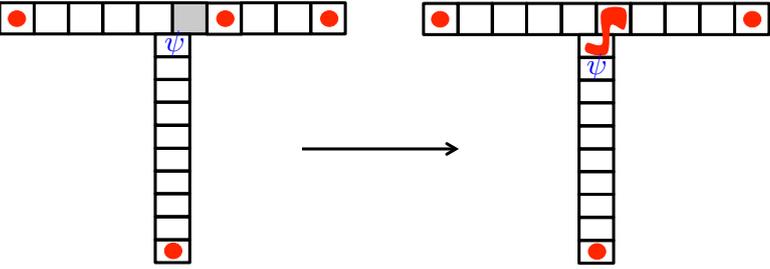 | |
| $\gamma_9\gamma_{10}$ | |
| 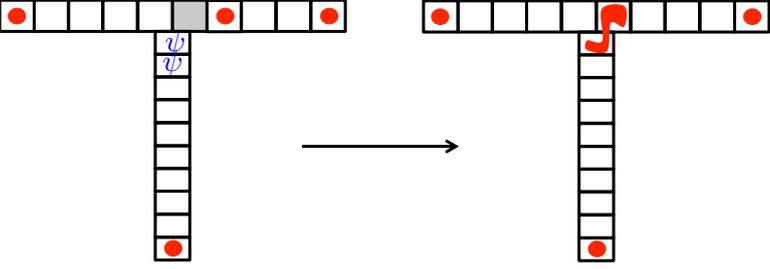 | |
| $\gamma_9\gamma_{10}$ | |
| 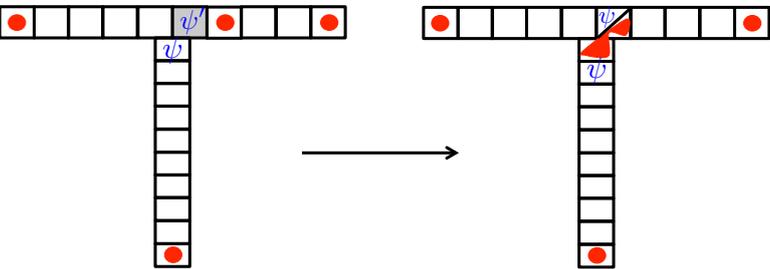 | |
| $\gamma_9\gamma_{10}$ | |
| 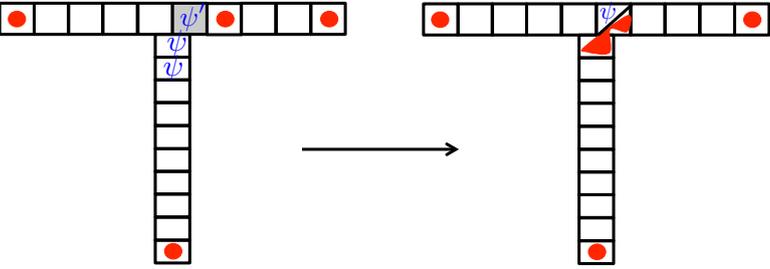 | |
| $\gamma_9\gamma_{10}$ | |

| Case 6. | Error processes supported by the bath |
|---|---|

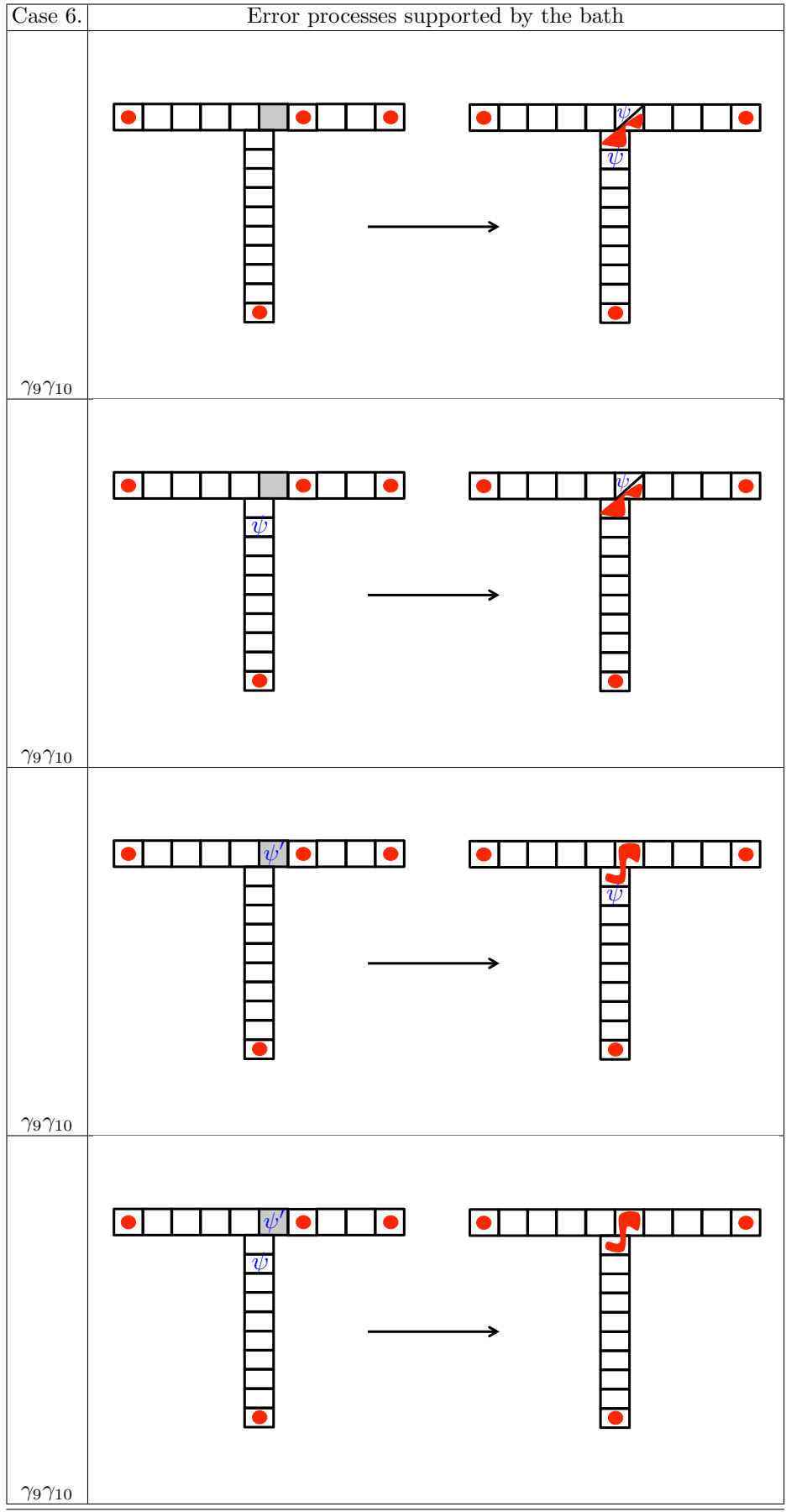

$\gamma_9\gamma_{10}$

$\gamma_9\gamma_{10}$

$\gamma_9\gamma_{10}$

$\gamma_9\gamma_{10}$

62

| Case 6. | Error processes supported by the bath |
|---|---|

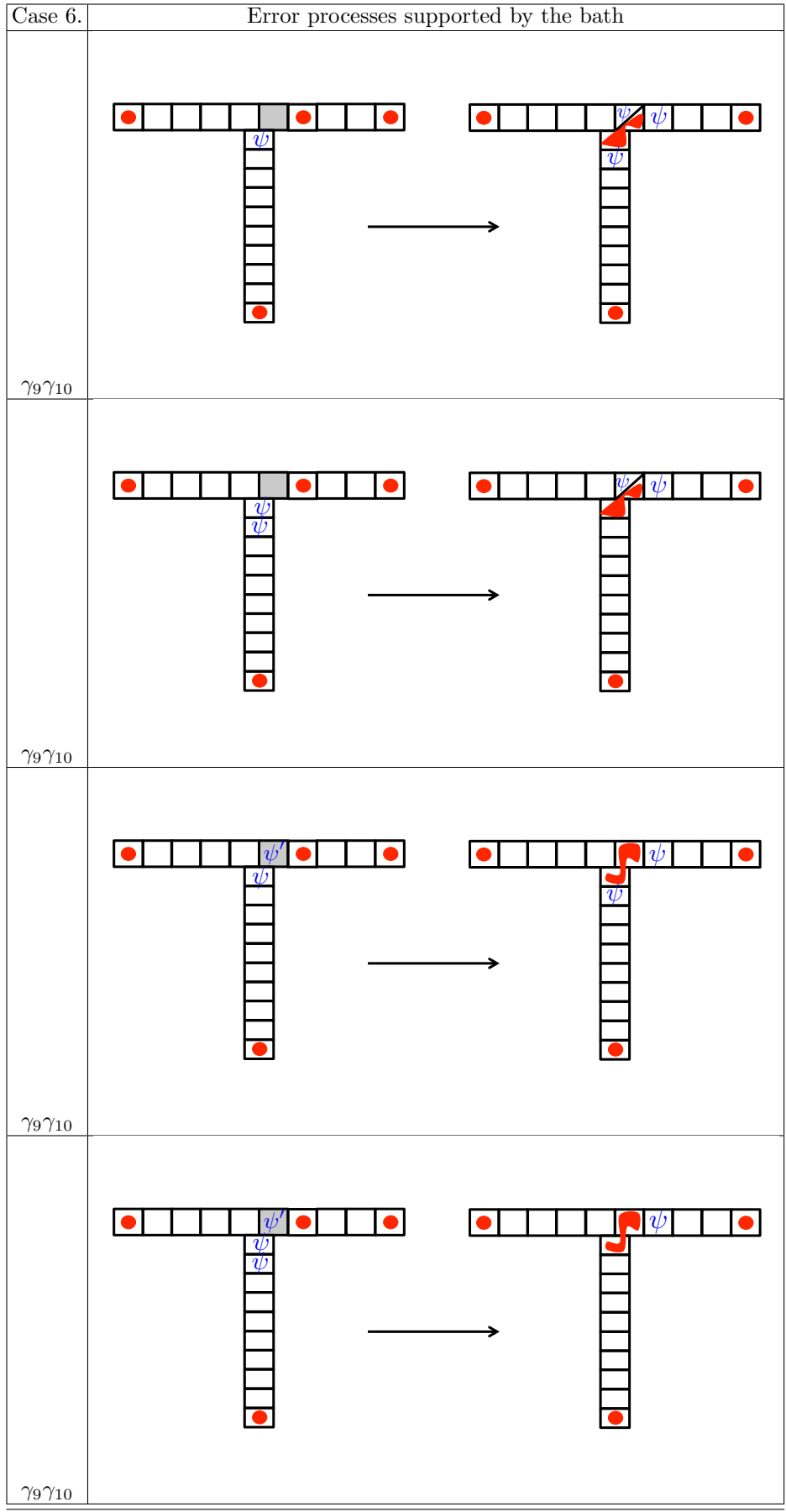

| Case 6. | Error processes supported by the bath |
|---|---|

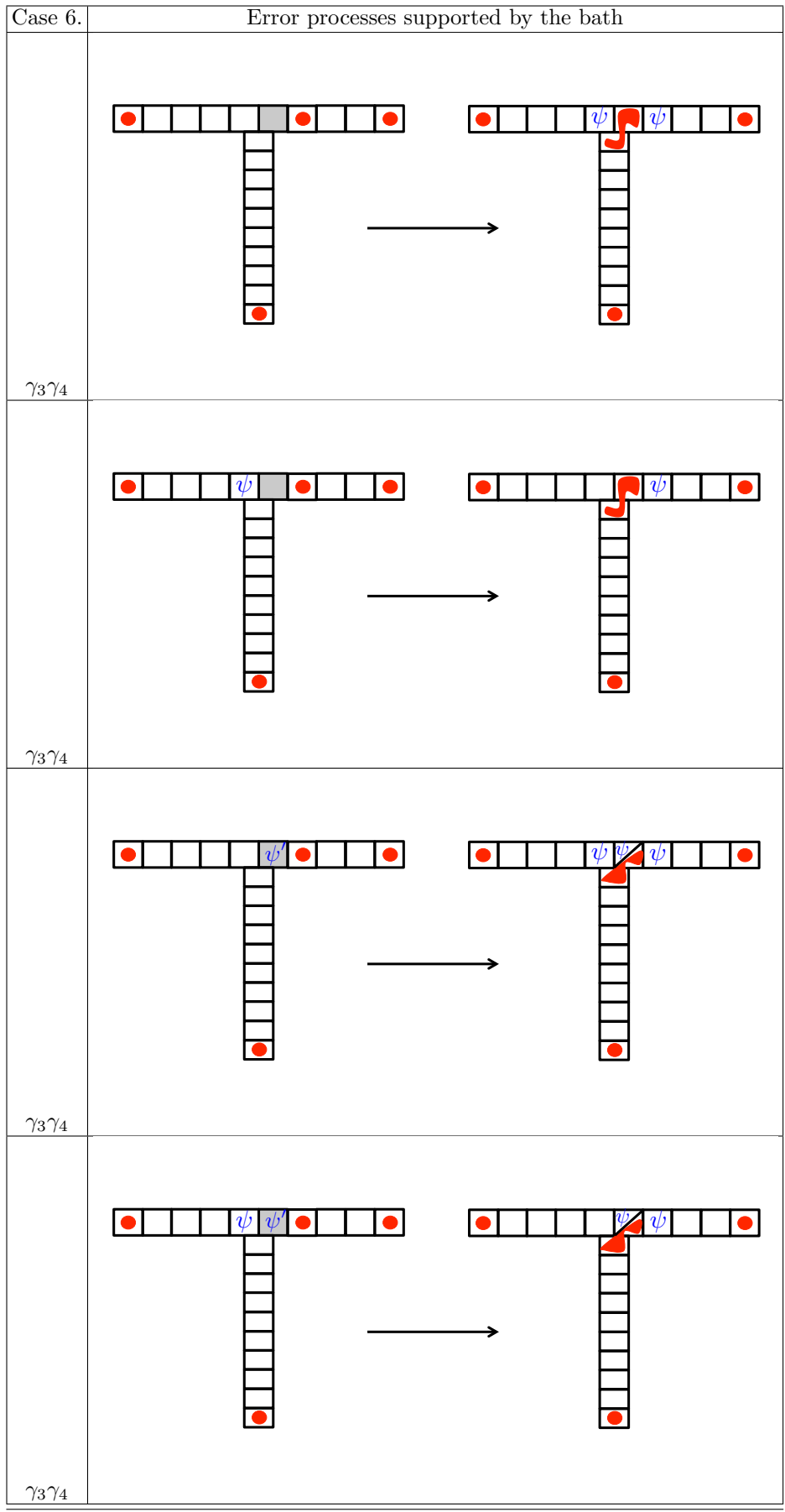

| Case 6. | Error processes supported by the bath |

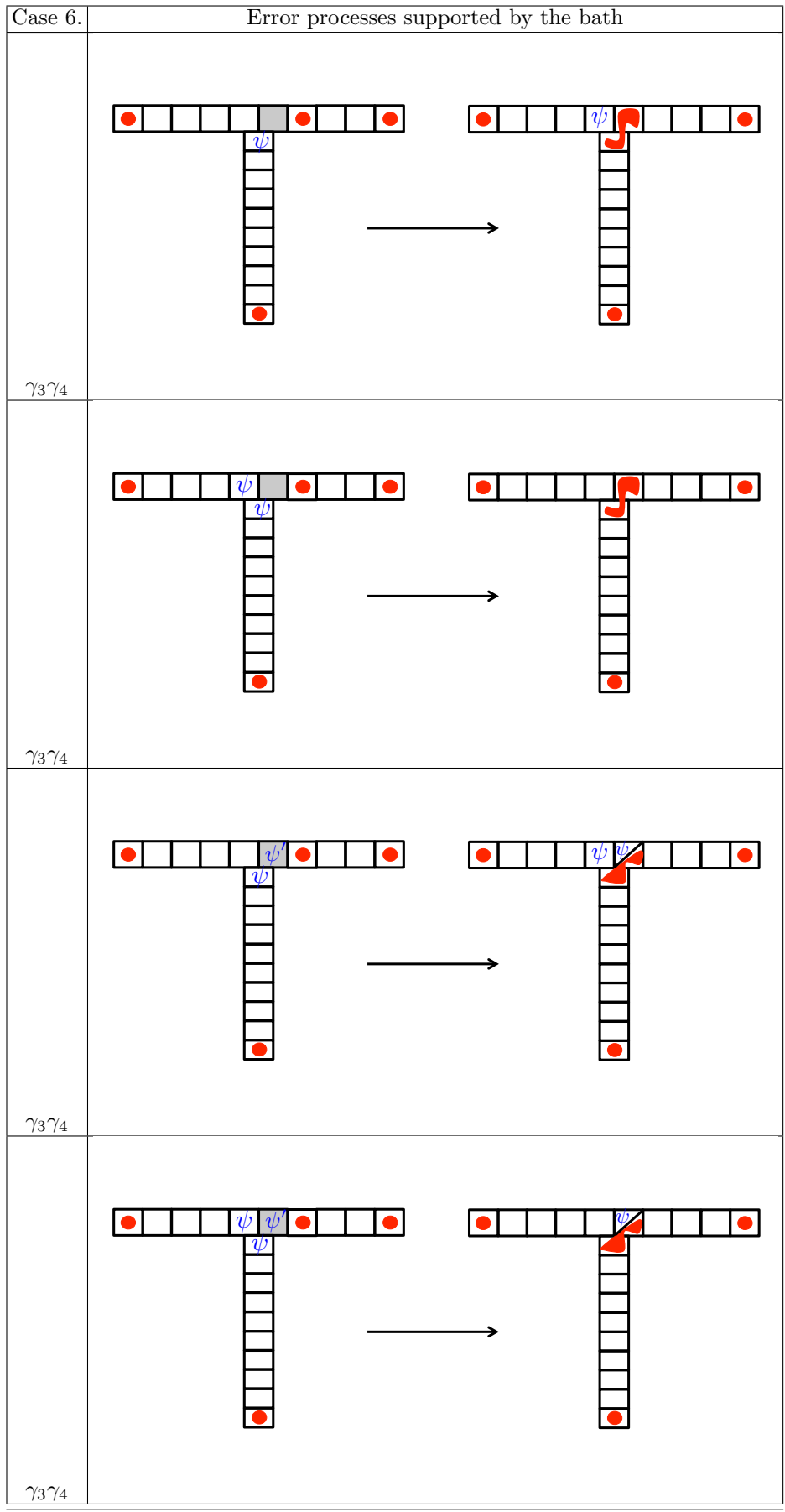

$\gamma_3\gamma_4$

$\gamma_3\gamma_4$

$\gamma_3\gamma_4$

$\gamma_3\gamma_4$

| Case 6. | Error processes supported by the bath |
|---|---|

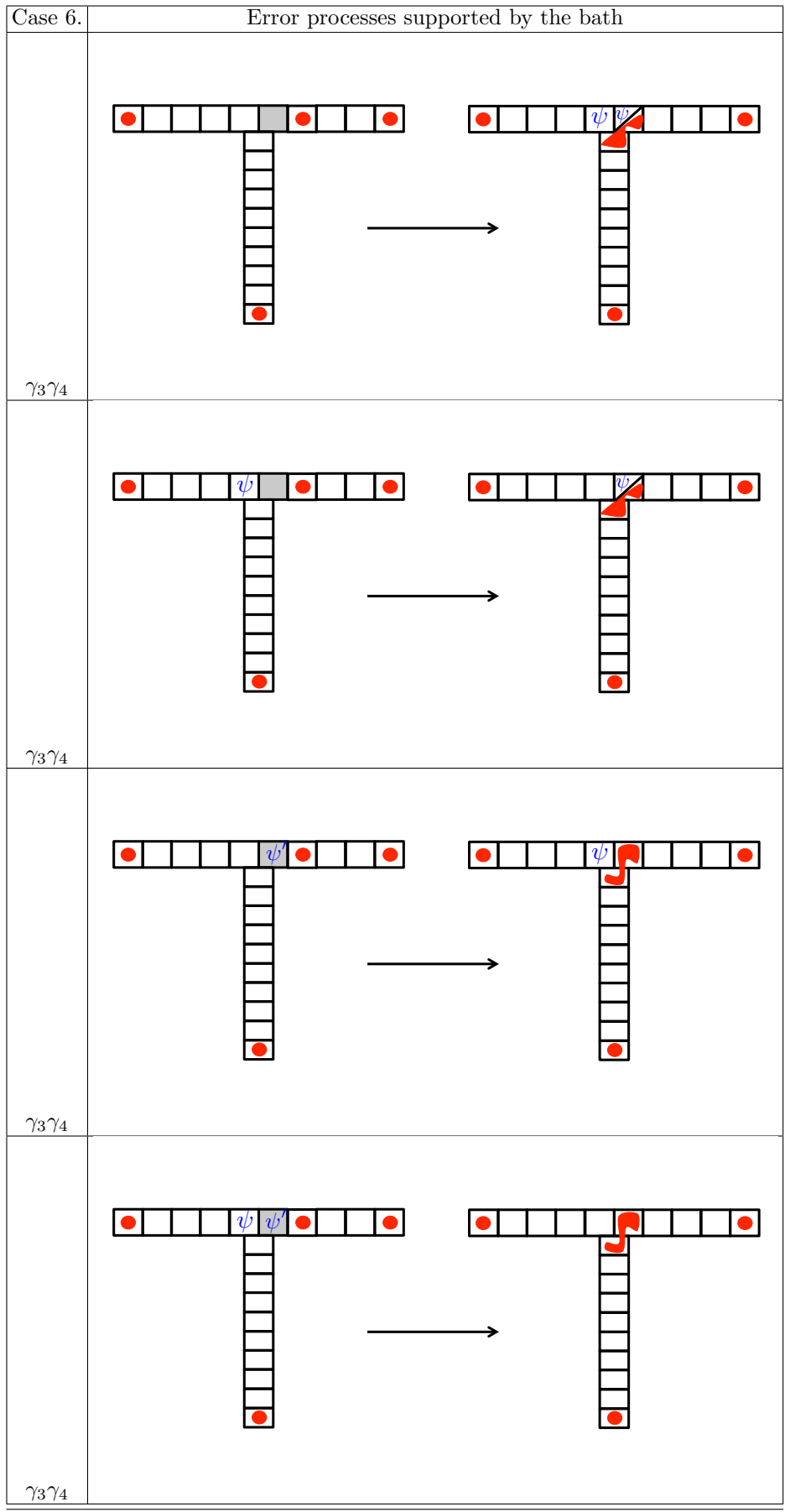

| Case 6. | Error processes supported by the bath |
|---|---|

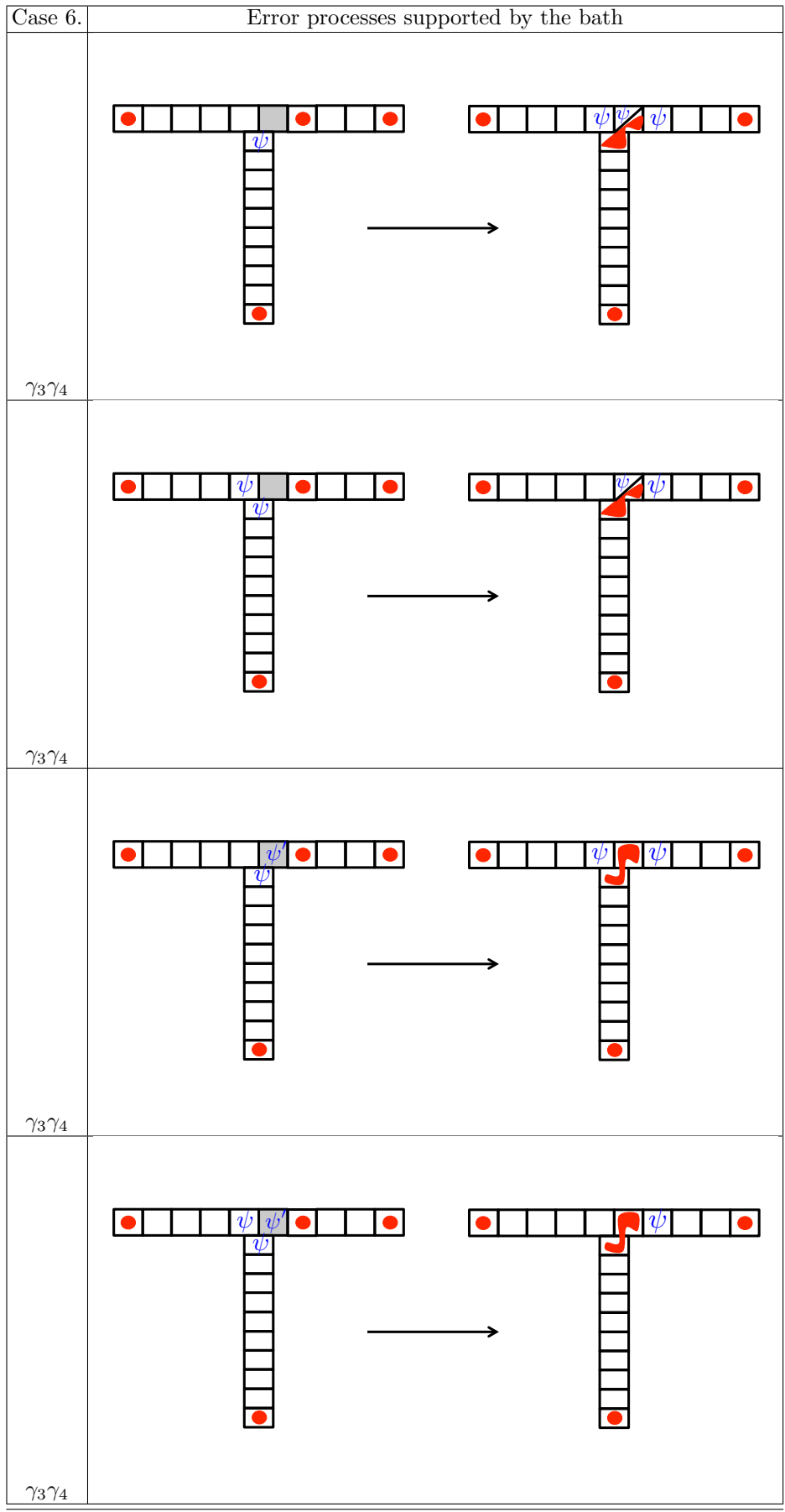



\end{document}